\documentclass[11pt]{article}
 \usepackage{color}
\usepackage{amsmath}
\usepackage{amssymb}
\usepackage{amsthm}
\usepackage{epstopdf}
\usepackage{graphicx}
\usepackage{mathtools}
\usepackage[usenames,dvipsnames, table]{xcolor}
\usepackage[hang]{caption}
\usepackage{hyperref}
\usepackage{oldgerm}  
\usepackage{float}
\usepackage{marginnote}

\def\subsubsubsection#1{\medskip{\bf #1}}

\newcommand{\nco}{\newcommand}
\nco{\one}{\ensuremath{\,\,\mathrm{l}\!\!\!1}} 
\nco{\ZZ}{\mathbb{Z}}
\nco{\CC}{\mathbb{C}}
\def\CX{{\mathcal X}}
\def\CP{{\mathcal P}}

\nco{\yellow}{\color{yellow}}
\nco{\green}{\color{green}}
\nco{\red}{\color{red}}
\nco{\cyan}{\color{cyan}}
\nco{\blue}{\color{blue}}
\nco{\violett}{\color{violet}}
\nco{\magenta}{\color{magenta}}
\nco{\redend}{\normalcolor}
\nco{\0}{\underline{0}}

\definecolor{violet}{rgb}{1,0,1}
 
\def\ie{{\rm i.e.,\/}\ }
\def\etc{{\rm etc.\/}\ }
\def\be{\begin{equation}}\def\ee{\end{equation}}
\def\bea{\begin{eqnarray}}\def\eea{\end{eqnarray}}
\def\bee{\begin{enumerate}}\def\eee{\end{enumerate}}
\def\bei{\begin{itemize}}\def\eei{\end{itemize}}

\def\ommit#1{{}}
\def\su{{\rm su}}
\def\SU{{\rm SU}}
\def\inv#1{\frac{1}{#1}}

\def\Tr{\mathrm{Tr}}
\def\p{\lambda_1}
\def\q{\lambda_2}
\def\r{\mu_1} 
\def\s{\mu_2}

 \def\CC{{\cal C}}
 
 \def\u{\lambda} 
 \def\eq=#1{\buildrel #1 \over{=}}
 
 \nco{\rnc}{\renewcommand}
\rnc{\title}[1]{{\Large\bf\mbox{}\\\medskip#1\bigskip\medskip\\}}
\rnc{\author}[1]{{\large #1\smallskip\\}}
\nco{\address}[1]{{\em #1\medskip\\}}

\newtheorem{theorem}{Theorem}

\newtheorem{corollary}{Corollary}

\newtheorem{conjecture}{Conjecture}

\begin{document}
\begin{titlepage}
\begin{center}
\title{{On  some properties of SU(3) Fusion Coefficients}}
\medskip
\author{Robert Coquereaux} 
\address{Aix Marseille Universit\'e, Universit\'e de Toulon, CNRS, CPT, UMR 7332, 13288 Marseille, France\\
Centre de Physique Th\'eorique (CPT)}
\vspace*{6mm}
\centerline{and}
\author{Jean-Bernard Zuber}
\address{
 Sorbonne Universit\'es, UPMC Univ Paris 06, UMR 7589, LPTHE, F-75005, 
Paris, France\\
\& CNRS, UMR 7589, LPTHE, F-75005, Paris, France
 }
\bigskip\medskip


\begin{abstract}
\noindent {Three aspects of the SU(3) fusion coefficients are revisited: the generating polynomials of fusion 
coefficients are written explicitly; some curious identities generalizing the classical Freudenthal-de Vries formula
are derived; and the properties of the fusion coefficients under conjugation of one of the factors, previously analysed
in the classical case, are extended to the affine algebra $\widehat{su}(3)$ at finite level.}
\end{abstract}
\end{center}

\vspace*{60mm}


\end{titlepage}


\section{Introduction}

The aim of this paper is threefold:\\
1) Write generating functions for \SU(3) fusion matrices and coefficients.\\
2) Use them to get general formulae for dimensions of spaces of essential paths on fusion graphs and derive identities generalizing the classical Freudenthal-de Vries formula. \\
3) Compare multiplicities for $\lambda \otimes \mu$ and $\lambda \otimes \overline \mu$ (provide a proof that was missing in our paper \cite{RCJBZ2014}).\\
Along the way we discuss several other properties of fusion coefficients that do not seem to have been discussed elsewhere (for instance in section \ref{miscellanea}).
\\
But the purpose of this paper is also 
to make a modest hommage to the memory of our distinguished colleague and great friend Raymond Stora. 
In scientific discussions Raymond was a listener second to none, with 
unsurpassable insight, critical sharpness and good humour. \dots We hope that he 
would have found entertaining the 
following  mix of algebraic, geometric and group theoretical  considerations, but surely, he would have 
stimulated us with witty comments and inspiring suggestions.

\section{Notations}
\label{notations}
{{When dealing with  $\widehat{\su}(3)_k$, the \su(3) affine algebra  at level $k$, 
the highest weights (h.w.) will be restricted to the Weyl alcove of level $k$, \ie  all dominant weights of
level $\|\lambda\| :=\lambda_1+\lambda_2\le k$,
thus \be\label{Walcove} {{\cal P}}^{(k)}:= \{\lambda|\p\ge 0\,,\ \q\ge 0\,, \p+\q\le k\}\,. \ee
Fusion matrices describe the multiplication, denoted $\lambda\star \mu$, 
{and sometimes $\lambda\star_k \mu$ if needed,}  of the $(k+1)(k+2)/2!$ simple objects (irreps) in the fusion category 
${\mathcal A}_k(\su(3))$. 
They are  denoted $N_{\lambda}$, with $\lambda=(\p,\q)$ in the Weyl alcove{,}
and satisfy the following  conjugation property: $N_\lambda^T=N_{(\p,\q)}^T=N_{(\q,\p)}=N_{\bar\lambda}$,
with $\bar\lambda=(\q,\p)$.
 \\
  In SU(3) or $\widehat{\su}(3)_k$, there is  a $\ZZ_3$ grading $\tau$ (``triality'') on irreps, 
stemming from the fact that this discrete group is the center of SU(3). We set  
for the two fundamental weights
$\tau(N_{(1,0)})=~1,\ \tau(N_{(0,1)})=2\; \textrm {mod} \; 3$
and more generally 
\be \label{triality} \tau(\lambda)\equiv\tau(N_{(\p,\q)}):=\p+2\q \; \textrm{mod} \; 3\,. \ee
For tensor product or fusion, this triality is conserved (mod 3): $N_{\lambda\mu}^\nu\ne 0$ only if
 \be\label{cons-trial}\tau(\lambda)+\tau(\mu)-\tau(\nu) =0\ \mod 3\,.\ee}}

\section{Generating functions of fusion matrices}
Generating series giving SU(3) fusion coefficients have been discussed and obtained in \cite{BCM-tensorpro}, \cite{BCM-fusionpro}. What we do here is to provide generating formulae for the fusion matrices themselves.
This material is presumably known to many {\it affinociados} of fusion algebras, but has never appeared 
in print to the best of our knowldege, and we think it may be
helpful to recall its essentials  for the convenience of the reader.

In this section we write down formulae for the generating function of  
fusion matrices. As the finite set ${{\cal P}}^{(k)}$ is closed under fusion, this generating function 
turns out to be a polynomial, see below (\ref{defF}). This property will be taken as granted in the following.\\
For reference and comparison, the same problem in the case of $\SU(2)$ is well known \cite{DFMS}. 

For the affine algebra $\widehat{su}(2)_k$, consider the matrices 
obtained by the (Chebyshev) recursion formula  
\be\label{chebySU2}  N_{\lambda+1} =G\cdot N_\lambda -N_{\lambda-1} \qquad \lambda \in \ZZ\,, \ee
with $N_{(0)}=\one$ and $G =N_{(1)}$, the latter being taken as the adjacency matrix of the Dynkin diagram $A_{k+1}$.
Notice that this implies $N_{(-1)}=0$. The obtained sequence is periodic of period $2h$ with $h=k+2$ (use Cayley-Hamilton for $G$);
actually the $N_\lambda$ matrices obey a Weyl group symmetry: $N_{\lambda+2h}=N_{\lambda}$, $N_{\lambda+h}=-N_{h-2-\lambda}$. 
This infinite set of matrices\footnote{In the non-affine case, the multiplicities for the decomposition of tensor product of irreps are encoded by matrices obtained by taking the large $k$ limit of the $N_\lambda$ matrices.} 
 can be obtained from the generating function $\tfrac{\one}{\one - s\, G + s^2}=\sum_{\lambda=0}^\infty s^\lambda N_\lambda$.
When $0\le \lambda\le k$ (Weyl alcove),  the $N_\lambda$ matrices have non-negative integer entries and are the fusion matrices  of  the affine algebra $\widehat{su}(2)_k$.
Introducing a term $+ s^h\, P$, where $P=N_{(k)}$ is an involution ($P.P=\one$), in the numerator of the previous generating function, has the effect of truncating the series stemming from the denominator to a finite polynomial, 
so that the generating function of fusion matrices at level $k$ reads 
\be X(s)=\sum_{\lambda=0}^k s^\lambda N_\lambda=\frac{\one + s^h P }{\one - s\, G + s^2}\,, \ee

%
\subsection{SU(3): definition of $X$ and recursion formulae}
\label{SU3genfct}
The generators of the $\widehat{\su}(3)_k$ fusion algebra
are $G =N_{(1,0)}$ and $G^T =N_{(0,1)}$, while again $ N_{(0,0)}=\one$.
We define the generating polynomial 
\be\label{defF} X(s,t)=\sum_{\lambda=(\p,\q)\in {{\cal P}}^{(k)}} s^{\p} t^{\q} N_{(\p,\q)}\,. \ee
It satisfies the conjugation property
$$ X(s,t)^T=X(t,s)\,.$$
We then write the recursion formulae stemming from fusion by $G$ or $G^T$ 
$$ G\cdot N_{(\p,\q)} = N_{(\p+1,\q)}+N_{(\p-1,\q+1)}+N_{(\p,\q-1)}$$
where each of the three terms in the rhs is present only if respectively, $\p\le k-1$, $\p\ge 1$ and $\q\le k-1$, 
$\q\ge 1$. We  have a similar expression for $G^T.N_{(\p,\q)}$. We finally translate the latter formulae 
into  identities on $X$
$$ G\cdot X(s,t)= \inv{s} ( X(s,t)- \Lambda_2(t)) +\frac{s}{t}( X(s,t)- \Lambda_1(s))+t( X(s,t)- \Lambda_3(s,t))$$
where 
\be\label{defL3} \Lambda_1(s):=X(s,0)\,,\  \Lambda_2(t):=X(0,t)\,,\ 
\Lambda_3(s,t):=\sum_{\p+\q=k} s^{\p} t^{\q} N_{(\p,\q)}\,, \ee
from which we obtain
\be\label{eqn1} (s^2 + t +s t^2 - s t\, G)\cdot X(s,t)= s^2 \Lambda_1(s) +t \Lambda_2(t) +s t^2 \Lambda_3(s,t)\ee
and likewise, from the fusion with $G^T$, 
\be\label{eqn2}  (t^2 + s +s^2 t - s t\, G^T)\cdot X(s,t)= s \Lambda_1(s) +t^2 \Lambda_2(t) +s^2 t \Lambda_3(s,t)\,.\ee
Subtracting equ. (\ref{eqn2}) multiplied by $t$ from equ. (\ref{eqn1}) multiplied by $s$, we eliminate
the unwanted $\Lambda_3$ and get
\be\label{Fequation} (s^3 \one - s^2 t G + s t^2 G^T - t^3 \one)\cdot X(s,t)= (s^3-s t) \Lambda_1(s)+(s t -t^3) \Lambda_2(t) \,.\ee

Thus, if we can  determine the boundary generating polynomials $\Lambda_1(s)=X(s,0)$ and $\Lambda_2(t)=X(0,t)$
we find
\be\label{diffeqforX}
X(s,t)=\frac{(s^3-s t) \Lambda_1(s)+(s t -t^3) \Lambda_2(t) }{s^3 \one - s^2 t G + s t^2 G^T - t^3 \one}
\ee

Here the inverse of ${s^3 \one - s^2 t G + s t^2 G^T - t^3 \one}$ may be understood in two alternative ways.
Either as the actual matrix inverse, which exists for generic values\footnote{For example, for $s\ne 0$ and $t$ 
small enough, the matrix is certainly invertible.}  of $s$ and $t$; calculating that inverse  requires the use of a computer but leads,
for each value of the level $k$, to an explicit expression. Or as a
formal series in, say, $t/s$, which actually truncates to a finite degree polynomial.  
Taking for example $s\ne 0$ in equ. (\ref{Fequation}),
we write $(s^3 \one - s^2 t G + s t^2 G^T - t^3 \one) =s^3 (\one - \frac{t}{s} G+\frac{t^2}{s^2} G^T -\frac{t^3}{s^3} \one)$.
The latter factor may then be inverted as a formal series%
\footnote{Explicitly, we may write $1/(\one-\CX)=\sum_p \CX^p$, here $\CX=\frac{t}{s} G -\frac{t^2}{s^2} G^T +\frac{t^3}{s^3} \one$,
and we expand $\CX^p$ in terms of multinomial coefficients, hence $\CX^p=\sum_{q,r\ge 0 \atop q+r \le p} 
(-1)^r \frac{p!}{q!r!(p-q-r)!} \big(\frac{t}{s}\big)^{q+2r+3(p-q-r)} G^q (G^T)^r$; identifying the term of degree $\ell$ in $t/s$ gives
 $X_\ell = \sum_{p,q,r\ge 0\atop q+2r +3(p-q-r )=\ell} (-1)^r \frac{p!}{q!r!(p-q-r)!} \big(\frac{t}{s}\big)^{q+2r+3(p-q-r)} G^q (G^T)^r$,
which may be restricted to $\ell\le k$.}
$(\one - \frac{t}{s} G+\frac{t^2}{s^2} G^T -\frac{t^3}{s^3} \one)^{-1}=\sum_\ell  \big(\frac{t}{s}\big)^\ell X_\ell$
and equ. (\ref{diffeqforX}) reads
$X(s,t)=\sum_\ell  \big(\frac{t}{s}\big)^\ell X_\ell\cdot \big( (1-\frac{t}{s^2}) X(s,0) +(\frac{t}{s^2} -\frac{t^3}{s^3}) X(0,t)\big)$.
By consistency, this expression is such that the  summation over $\ell$  truncates to a finite order so as to make $X(s,t)$ a 
polynomial in $s$ and $t$ of total degree $k$. 

The two issues of 
determining the boundary terms will be addressed in the following subsections. 
Here, we just observe that because of the
commutativity of the fusion algebra, we may then write the rational fraction  in (\ref{diffeqforX}) without 
worrying about the order of terms. 


\subsection{Boundary generating polynomials}
  We may again write recursion formulae for the boundary generating polynomials $\Lambda_1(s)= X(s,0)$   
  and $\Lambda_2(t)=X(0,t)$
$$ G\cdot\Lambda_1(s)= \sum_{0\le \p \le k-1} s^\p N_{(\p+1,0)} + \sum_{1\le \p \le k} s^\p N_{(\p-1,1)} = \inv{s} (\Lambda_1(s)-\one) +L(s)$$
  where $L(s):= \sum_{\p=1}^k s^{\p} N_{(\p-1,1)}$, and likewise
  $$G^T\cdot \Lambda_1(s)=s(\Lambda_1(s)-N_{(k,0)} s^k) + \inv{s} L((s)\,.$$
  Eliminating $L(s)$ between these two relations yields
  \be (\one - s G + s^2 G^T - s^3 \one)\cdot \Lambda_1(s) = (\one - s^h N_{(k,0)})\, \ee
  where for SU(3), $h=k+3$. 
  
  Call $P = N_{(k,0)}$. This permutation matrix  describes a rotation of angle   $ 2 \pi/3$ 
around the center of the Weyl alcove, as 
\be\label{Pppty} P\cdot N_{(\p\q)}=N_{k-\p-\q,\p}\,.\ee
  In particular, for all $\lambda: 0 \leq \lambda \leq k$,  we have $P\cdot N_{(\lambda,k-\lambda)} = N_{(0,\lambda)}$, and for $\lambda=k$ we have $N_{(0,k)}=P^2$. 
  As $P^3=\one$ we notice that  $\{\one, P, P^2 \}$ is isomorphic with the $\ZZ_3$ group.
Note that this action is {\it not} related to the $\ZZ_3$ grading (``triality'') on irreps described in section \ref{notations}. 

Now, it is clear that the matrix $(\one - s G + s^2 G^T - s^3 \one)$ may be inverted as a formal
series in $s$, thus giving the generating function $\Lambda_1(s)= X(s,0)$
\be\label{L1formula} \Lambda_1(s)= X(s,0) =\frac{\one - s^h P}{\one - s G + s^2 G^T - s^3 \one}\,.\ee
The effect of the numerator $\one -s^h P$ is to truncate these generating series to a polynomial of degree 
$k$. In a similar way, we have 
\be\label{L2formula} \Lambda_2(t)= X(0,t) =\frac{\one - t^h P^2}{\one - t G^T + t^2 G - t^3 \one}\,,\ee
which satisfies $\Lambda_2(t)=(\Lambda_1(t))^T$ as anticipated,  but also, using (\ref{Pppty}), 
\be\label{L1L2} \Lambda_2(t)=t^k P^2 \Lambda_1(\inv{t})\,.\ee

But we may also now determine the third boundary matrix, namely $\Lambda_3(s,t)$ defined in (\ref{defL3}). 
From (\ref{Pppty}), it follows that 
$$ \Lambda_3(s,t)= \sum_{\p+\q=k} s^{\p} t^{\q} N_{(\p,\q)} =  \sum_{0\le \q \le k}s^k \Big(\frac{t}{s}\Big )^{\q}  P\cdot N_{(\q,0)} =
s^k P\cdot X(\frac{t}{s},0)$$
which, according to what we already know  about $X(s,0)=\Lambda_1(s)$, is indeed a polynomial in $s$ and $t$ of total degree $\le k$. 


\subsection{The $ X $ generating polynomial}
\label{genpolsu3}
We  return to equ. (\ref{diffeqforX}), into which the expressions (\ref{L1formula}) and (\ref{L2formula}) 
for $\Lambda_1(s)$ and $\Lambda_2(t)$ may now be inserted. After some algebra, the result may be recast in the following form  \begin{eqnarray}
 \label{symeqforF}
  X(s,t) & = &
\frac{(1-st) \one}{(\one - s G + s^2 G^T - s^3 \one)(\one - t G^T + t^2 G - t^3 \one)}  \\
& {} &
+ \frac{s^h (st-s^3) P}{(\one - s G + s^2 G^T - s^3 \one)(s^3 \one - s^2 t G + s t^2 G^T - t^3 \one)}    \nonumber \\
& {} &
+ \frac{t^h (st -t^3) P^2}{(\one - t G^T + t^2 G - t^3 \one)(t^3 \one - s t^2  G^T +  s^2 t  G - s^3 \one) \nonumber }\,.
 \end{eqnarray} 
 which is clearly symmetric under conjugation (interchange $P$ and $P^2$, $s$ and $t$, $G$ and $G^T$).
 
 Taking the level $k$, hence $h$,     
 arbitrarily large in (\ref{symeqforF}), \ie dropping the last two terms and taking for $G$ the (infinite size) 
 adjacency matrix of the set of SU(3) dominant weights, gives 
\be  X_\infty(s,t) =
\frac{(1-st) \one}{(\one - s G + s^2 G^T - s^3 \one)(\one - t G^T + t^2 G - t^3 \one)}  \ee
which is the generating function (now an infinite series !) 
for the tensor product of SU(3) irreps.

There are alternative, different looking, expressions for the generating polynomial $X(s,t)$. 
Adding (rather than subtracting as we did in order to obtain equ. (\ref{diffeqforX}))
equ. (\ref{eqn2}) multiplied by $t$ to equ. (\ref{eqn1}) multiplied by $s$ gives 
\begin{equation}
\label{sumeqforX}
(s^3 + t^3 + 2 s t + 2 s^2 t^2) \one - s t (s G + t G^T)) \cdot X(s,t) = (s^3+ s t)  \Lambda_1(s) + (t^3 + s t) \Lambda_2(t) + 2 s^2 t^2 \Lambda_3(s,t)\,.
\end{equation}  
There, $X(s,t)$ appears as the inverse of the matrix $(s^3 + t^3 + 2 s t + 2 s^2 t^2) \one - s t (s G + t G^T))$ times
the rhs, the solution of which being given by (\ref{symeqforF}). Note that evaluated at $s=t=1$, the former matrix
reads $ K^{-1} = (6 \one - (G+G^T))$, generalizing to SU(3) what would be $(2\one -G)$ for $\widehat{\su}(2)_k$, namely
 the Cartan matrix of the  $A_{k+1}$ algebra, and the last relation reads $X = K \, \Lambda$, with $X=X(1,1)$ and where $\Lambda=2 (\Lambda_1+\Lambda_2+\Lambda_3)$ is the boundary matrix evaluated at $s=t=1$.



\section{Generalized Freudenthal - de Vries formulae and dimensions of spaces of paths}
From a fusion algebra point of view, all Dynkin diagrams are manifestations of SU(2), as they can be introduced as graphs encoding the action of the fusion ring of the affine $\widehat{su}(2)$ on appropriate modules (nimreps). Strictly speaking this is only true for simply laced diagrams but there is a way to accommodate the non-ADE's in the same common framework. 
From this point of view, the classical Freudenthal-de Vries formulae, that hold for all simple Lie groups, are also a manifestation of the underlying SU(2) theory. 
In turn, those formulae are related to the counting of a particular kind of paths on appropriate Dynkin diagrams.
Moving from an underlying SU(2) to an underlying SU(3) framework leads to several intriguing formulae that we present in this section.


\subsection{SU(2) :   Dimensions of spaces of essential paths on Dynkin diagrams}  
Call $A$ the Cartan matrix associated with some chosen Dynkin diagram of rank $r$ (the number of vertices), and $\langle \;, \; \rangle$ the corresponding inner product in the space of roots. Call $s$ the vector of scaling coefficients, with components defined as  $\langle \alpha, \alpha \rangle/2$ where $\alpha$ runs in a basis of simple roots, the squared norm being $2$ if $\alpha$ is a long root. Call also $\bold{s}$ the corresponding diagonal matrix. It is the identity matrix if the Dynkin diagram is simply laced (ADE cases). Call $K = A^{-1} \bold{s}$ the quadratic form matrix (it gives the scalar products between fundamental weights).  Call $g$ the Coxeter number, $g^\vee$ the dual Coxeter number (they are equal for simply laced cases) and $k=g-2$ the (SU(2)) level to be used below.
Call $\rho$ the Weyl vector (by definition its components are all equal to $1$ on the basis of fundamental weights). Its squared norm is then  $\langle \rho,  \rho\rangle = \sum_{a,b} K_{ab}$ and it is given by the Freudenthal - de Vries formula : $\langle \rho, \rho \rangle = g^\vee (g +1) r /12$. \\
\def\F{F}
Call $\F_\lambda = (\F_\lambda)_{a,b}$ the $r\times r$ matrices defined by the SU(2) Chebyshev recurrence relation (\ref{chebySU2}), with $G = 2-A$. In a conformal field theory context, and for the ADE cases, 
they are called the ``nimrep" matrices,
(for {\sl non negative integer valued matrix representation} of the fusion algebra) and they describe boundary conditions.\\
Call $X = \sum_\lambda \F_\lambda$, the path matrix, and $u$, with components\footnote{these components are also equal to twice the components of the dual Weyl vector on the basis of simple coroots} 
$u_a = \sum_b (X^T)_{b,a}=\sum_b (X)_{a,b}$
the height vector (as defined by Dynkin \cite{Dyn}, see also 
ref. \cite{DFMS}).  In this general setup it is also useful to define the vector $v$, with components\footnote{these components are also equal to twice the components of the Weyl vector on the basis of simple roots} $v_a = \sum_{b} (X)_{b,a}$. 
\\
Given $\lambda \in \{0, 1,\ldots, g-2\}$ and two vertices $a$ and $b$ in the chosen Dynkin diagram, the vector space of dimension $(\F_\lambda)_{a,b}$ is called the space of essential paths of length $\lambda-1$ from the vertex $a$ to the vertex $b$; for the purpose of the present paper we don't need to explain how these spaces are realized in terms of actual paths -- see \cite{Ocneanu:paths}.
With a slight terminological abuse {(dimension versus cardinality)}
we may say that there are $X_{a,b}$ essential paths (of any length) from $a$ to $b$ and that the component $v_b$ of $v$ gives the total number of essential paths (of arbitrary origin and length)
reaching $b$.  Then, $d_\lambda = \sum_{a,b} (F_\lambda)_{a,b}$ gives the dimension of the space of essential paths of length $\lambda$, and $d_H = \sum_{\lambda \in \{0, 1,\ldots, g-2\}} d_\lambda$ is the total dimension of the space of essential paths, but the reader should only take these two equalities as mere definitions for the integers $d_\lambda$ and $d_H$. Now, obviously, $d_H = \sum_a u_a = \sum_a v_a = \sum_{a,b} X_{a,b}$. It is also of interest to consider the integer $d_B = \sum_{\lambda} d_\lambda^2$ that can be interpreted (at least in the ADE cases, since simply laced Dynkin diagrams classify module-categories over ${\mathcal A}_k(\SU(2))$, \cite{CIZ}, \cite{Ostrik}) as the dimension of a weak Hopf algebra \cite{PZ}.\\
A last relation of interest, for us, relates the path matrix $X$ to the quadratic form matrix $K$.  It can be established as we did in the SU(3) case (see the end of section \ref{genpolsu3}).
For an arbitrary Dynkin diagram, it reads again : $X^T = K \bold{s}^{-1} \Lambda$, where $\Lambda$ is the ``boundary matrix'' $\Lambda = F_0 + F_{r-1}$.

If the diagram is simply laced then $g^\vee=g$, $\bold{s}$ is the identity, and $A$, $K$, $X$ are symmetric. 
For those cases the former relation reads $X = K \Lambda$, but  $F_0$ is the identity  and a detailed analysis of all cases shows that $F_{r-1}$ is a permutation matrix, so that $\sum_{a,b} X_{ab} = 2 \sum_{a,b} K_{a,b}$. The Freudenthal - de Vries formula then implies\footnote{In the general case (non necessarily ADE) one would obtain $d_H=2 \sum_{a,b} A^{-1}_{a, b}$}${}^,$\footnote{This number also gives the dimension of the vector space underlying the Gelfand-Ponomarev preprojective algebra associated with the corresponding unoriented quiver \cite{MOV}.}
\be d_H = \sum_{a,b} X_{a,b} =  2 \sum_{a,b} K_{a,b} = g (g +1) r /6 \qquad \text{(ADE cases)}\,.\ee
One may notice that the Freudenthal - de Vries formula (giving  $\langle \rho, \rho \rangle$) and the expression giving the dimension $d_H$ of the space of essential paths only differ by a factor $2$.
In particular, if the diagram is $A_r$, which in particular encodes fusion by the fundamental representation in fusion categories of type $\mathcal{A}_k(SU(2))$, then $g=r+1=k+2$ and 
\be d_H=(k+1)(k+2)(k+3)/6 \qquad \text{($A_{k+1}$ cases)}\,.\ee


\subsection{Generalization to SU(3)}
The purpose of this section is to show how the above results generalize in the case of graphs (McKay graphs) describing fusion by the fundamental irreps in the case $\mathcal{A}_k(SU(3))$.
We will show that the following two formulae hold:

 \begin{equation}
 \label{eqfordK}
\sum_{a,b} K_{a,b} = \frac{1}{2} \, \frac{(k+1)(k+2)}{2} \, \frac{(k+4)(k+5)}{60}
\end{equation}

 \begin{equation}
 \label{eqfordX}
d_H=\sum_{a,b} X_{a,b} = \frac{(k+1)(k+2)(k+3)(k+4)(k+5)((k+3)^2+5)}{1680}
\end{equation}

These formulae 
were already announced in  \cite{CoquereauxSchieberRio} and \cite{CoquereauxSchieberJMP} where one can find tables containing several other ``characteristic numbers'' describing the geometry of graphs related to fusion categories of type SU(2), \ie simply laced Dynkin diagrams,  and of type SU(3).  However, for these two formulae, no proof was given. 
 The  one that we shall give below relies on a crucial property (a theorem that we recall below in sect. \ref{summaryonmutiplicities}) that was actually obtained much later \cite{RCJBZ2011}.

The sum $d_H$ of matrix elements of $X$, for SU(3) at level $k$, equal to the sum of all multiplicies for all  possible fusion products up to level $k$, can again be interpreted as counting essential ``paths'' although the ``length'' is no longer a non-negative integer (\ie an irrep of $\SU(2)$) but a  weight of $\SU(3)$ (\ie a pair of non-negative integers) belonging to the Weyl alcove. The notion of path should therefore be generalized in a way that is appropriate, but we do not intend to enter this discussion and shall stay at the level of combinatorics.


\subsection{Proof of the relation (\ref{eqfordX})} 

To ease the writing, we introduce, for an arbitrary  square matrix $M$,  the notation $\Sigma M$ that denotes the sum of its matrix elements,
 thus $\Sigma M=\Tr MU$, where $U$ is the square matrix of same dimensions that has all its coefficients equal to $1$.

$X(s,t)$ being the SU(3) generating functional introduced in section \ref{SU3genfct}, we call $X = X(1,1)$.
From the definition of $U$, we have $d_H = \Sigma X = \Tr(X U)$. 
One important step of the proof is to show that   $\Tr(G - G^T)  X(s,t)  U= 0$  for arbitrary values of $s$ and $t$. 
This property is obvious for $s=t$ as $X(s,s)$ is symmetric, but in general we have only $X(s,t)^T = X(t,s)$.

{\bf Lemma 1.}  {\sl One has $U X(s,t) = U X(t,s)$. There is no transpose sign here.}\\
Proof. The matrix element $(i,j)$ of $U X(s,t)$, which is obviously independent of $i$ is the sum of all matrix elements of the column $j$ of $X(s,t)$.
The matrix element $(i,j)$ of $U X(t,s)$, is the sum of all matrix elements of the column $j$ of $X(t,s) = X(s,t)^T$.
Since $X(s,t)^T = \sum_{p,q} N_{(p,q)}^T s^p t^q$, the equality of these two matrix elements results from 
the Theorem \ref{CZthm} proved in \cite{RCJBZ2011} and recalled below.

{\bf Lemma 2.} {\sl  One has $\Tr( G^T X(s,t) U) =  \Tr(G X(s,t) U)$.}\\
Proof. Using the previous lemma, and the fact that $G$ commutes with all fusion matrices, and therefore also with $X(s,t)$, one gets:
$\Tr( G^T X(s,t) U) = \Tr(U^T  X(s,t)^T G) = \Tr(U X(t,s) G) = \Tr(U X(s,t) G) = \Tr(U G X(s,t)) = \Tr(G X(s,t) U)$.
In contrast,  the trace of $(G^T - G)  X(s,t)$ has no reason to vanish.

The previous lemma implies:\\
 $\Tr((s G - t G^T ) X(s,t)  U =  (s-t) \Tr(G  X(s,t)  U)$ and $\Tr((s G + t G^T) X(s,t) U =  (s+t) \Tr(G X(s,t) U)$

 Multiplying both equations (\ref{sumeqforX}) and (\ref{diffeqforX}) on both sides by $U$, using the previous lemma, calling $\Sigma X(s,t) = \Tr(X(s,t) U)$, $\Sigma \Lambda_1(s)=\Tr(\Lambda_1(s) U)$, $\Sigma(GX(s,t))= \Tr(G.X(s,t) U)$, and taking the trace gives: 
 {\small
\begin{eqnarray}
 \label{coupledequationsfordX}
\!\!\!\!\!\!\!\!
(s^3 + t^3 + 2 s t + 2 s^2 t^2) \Sigma X(s,t)  - s t (s+t )  \Sigma(GX(s,t))  &=& (s^3+ s t)  \Sigma \Lambda_1(s) + (t^3 + s t) t^k  \Sigma \Lambda_1(1/t)  + 2 s^2 t^2 s^k \Sigma \Lambda_1(t/s) \nonumber \\
((s^3 - t^3) \Sigma X(s,t) - st (s-t) \Sigma(GX(s,t))  &=& (s^3 - st ) \Sigma \Lambda_1(s)   - (t^3 - s t) t^k \Sigma \Lambda_1(1/t) 
 \end{eqnarray}
 }
 where we have used the fact that in that calculation $P$  can be replaced by $\one$ 
 since $P\cdot U=U\cdot P=U$, so 
 $\Tr(P^2 \Lambda_1(1/t) U)= \Tr(\Lambda_1(1/t) U)$ and $\Tr(P \Lambda_1(t/s)) = \Tr(\Lambda_1(t/s) U)$.
As it is easy enough to determine separately the value of $\Sigma \Lambda_1(s)$  (see below), 
 the above (\ref{coupledequationsfordX}) is a system of two linear equations for the two unknown $\Sigma X(s,t)$ and $\Sigma(GX(s,t))$.
 One finds~:
 {\small
 \begin{eqnarray}
  \label{couplesolutionsfordX}
 \Sigma X(s,t)&=&\big(\Sigma \Lambda_1(s) (1-s) s+\Sigma \Lambda_1(t/s) s^{(k+1)} t (s-t)+\Sigma \Lambda_1(1/t)(t^{(k+1)}) (t-1) \big)/((-1+s) (s-t) (-1+t))\nonumber  \\
\Sigma(GX(s,t))&=&\big(\Sigma \Lambda_1(s)(1- s^3)+\Sigma \Lambda_1(1/t) t^k (-1+t^3)+\Sigma \Lambda_1(t/s) s^k (s^3- t^3)\big)/((-1+s) (s-t) (-1+t)) \,.\nonumber
   \end{eqnarray}
   }
 It is known, and in any case it is easy to show, that the sum of matrix elements of $N_{(p, 0)}$ is a product of two triangular numbers:  $(k + 2 - p) (k + 1- p) (1 + p) (2 + p)/4$.
 The common value of $\Sigma \Lambda_1 = \Sigma \Lambda_1(1)$ and $\Sigma \Lambda_i= \Tr(\Lambda_i[1].U)$ for $i=1,2,3$
 is obtained by summing the previous quantity along an edge: $\Sigma \Lambda_1=\Tr(\Lambda_1(1).U)= \sum_{p=0}^k \, (k + 2 - p) (k + 1- p) (1 + p) (2 + p)/4 = \binom{k+5}{5}$.
 Unfortunately, this last equality is of little help since one cannot solve the previous system (\ref{coupledequationsfordX}) while setting $s=t=1$.
 So, we plug the value of the polynomial $\Sigma \Lambda_1(s) = \sum_{p=0}^k \, (k + 2 - p) (k + 1- p) (1 + p) (2 + p)/4 \;   s^p$ in the above solutions (\ref{couplesolutionsfordX}) for 
 $\Sigma X(s,t)$ and $\Sigma(GX(s,t))$, and calculate the first term of their Taylor series around $s=t=1$. Calling $X = X(1,1)$ and $GX = GX(1,1)$ one finally gets:
  \begin{eqnarray}
  \label{dXanddGX}
 \Sigma X &=& ((k + 1) (k + 2) (k + 3) (k + 4) (k + 5) ((k + 3)^2 + 5))/1680 \nonumber \\
 \Sigma(GX) &=& 
 (k+6)!/(k-1)!/560 
  \nonumber
  \end{eqnarray}
  hence the result  (\ref{eqfordX}).

As a side result, we also obtain the value of $\Sigma \Lambda$. Indeed, \\
$\Tr(6 K  \Lambda U -   G K \Lambda. U - G^T K  \Lambda U) =  \Tr(( 6 \one - G - G^T) K  \Lambda U ) =  \Tr[(A K \Lambda U ) = \Tr( \Lambda  U )$\\
Evaluating the sum $6 \Sigma X - \Sigma(GX) - \Sigma(G^TX)$ using the two previous formulae gives immediately
\be \Sigma \Lambda = (k+1) (k+2 ) (k+3 ) (k+4) (k+5)/20\,.\ee


\subsection{Proof of the relation (\ref{eqfordK})}

Let $M$ be an element of the matrix algebra generated by the commuting family of the fusion matrices $\F_\lambda$.
By Verlinde formula \cite{DFMS}, all these elements (in particular $X$, $A$, $K$, $\Lambda$, $\ldots$) are diagonalized by the (symmetric) modular matrix $S$.
We call $\Delta_M$ the diagonal matrix of eigenvalues of $M$, \ie $\Delta_M = S M S^{-1}$.

{\bf Lemma 3.} {\sl The element $(i,j)$ of $S U S^{-1}$ vanishes whenever $i$ and $j$ do not both label real representations.}\\
 Proof. 
By definition of $U$, the element $(i,j)$ of $S U S^{-1}$ is equal to $\sum_x S_{ix} \sum_y S^{-1}_{yj}$ but $S^{-1}=S^3=C S$, where $C$ is the conjugation matrix, so that this element is $\sum_x S_{ix} \sum_y S_{\overline{y}j}$ = $(\sum_x S_{ix})(\sum_y S_{jy})$. Because of theorem  3 of ref. \cite{RCJBZ2011}, the sum $\sum_x S_{ix}$ vanishes if $x$ is not of real type,
hence the result. In the present case of SU(3), real irreps have highest weight $\{\mu, \mu\}$, with $\mu\in \{0,\cdots,\lfloor k/2\rfloor\}$.

The $S$-matrix elements for those real irreps read  \cite{KacPeterson},\cite{DFMS}
$$ S_{\{\lambda_1,\lambda_2\},\{\mu, \mu\}}= \mathrm{const.} \Big(\sin \frac{2\pi (\mu+1)(\lambda_1+\lambda_2+2)}{k+3}
-\sin \frac{2\pi(\mu+1)(\lambda_1+1)}{k+3}-\sin \frac{2\pi (\mu+1)(\lambda_2+1)}{k+3}\Big)$$
where the $(\lambda, \mu)$-independent constant is of no concern to us.
A straightforward but tedious calculation leads then to the corresponding eigenvalue of $\Lambda_1$
 $$\varpi_\mu(\Lambda_1) = \sum_{0\le \lambda_1\le k} \frac{S_{\{\lambda_1,0\},\{\mu,\mu\} }} {S_{\{0,0\},\{\mu,\mu\}}}
 = \frac{(k + 3)}{4 \sin^2(\pi (\mu+1)/(k + 3))}= \frac{2(k+3)}{6- 2(1+2 \cos(2\pi(\mu+1)/(k+3)))}$$
\\
Since $P$ acts as a permutation, and $P^3 = 1$,  we get immediately the corresponding eigenvalue of  
$\Lambda = 2(\Lambda_1+\Lambda_2+\Lambda_3) = 2(1+P+P^2)\Lambda_1$, namely $\varpi_\mu(\Lambda) = 6 \,\varpi_\mu(\Lambda_1)$.
\\
By the same token, the corresponding eigenvalue of $A = 6 - (G+G^T)$ is 
$$ \varpi_\mu(A)= 6 -2 \frac{S_{\{1,0\},\{\mu,\mu\} }} {S_{\{0,0\},\{\mu,\mu\}}}=6-2(1+2 \cos(2\pi (\mu+1)/(k+3)) )$$
and we conclude that the ``real" eigenvalues of $K=A^{-1}$ are proportional to those of $\Lambda$
$$
12(k+3) \varpi_\mu(K) =  \varpi_\mu(\Lambda)  = 12(k+3) \times \frac{1}{6 - 2(1+2 \cos(\frac{2\pi (\mu+1)}{k+3}))}\,. 
$$
\\
By Lemma 3, in the calculation of $S A \Lambda U S^{-1}= \Delta_A \Delta_\Lambda S U S^{-1}$, 
only real irreps contribute and we have found that
$S A S^{-1} S \Lambda S^{-1} S U S^{-1}  = 12 (k+3) \times  S U S^{-1}$, \ie $\Lambda U = 12(k+3) K U$.
It follows that $\Sigma K= \inv{12(k+3)}\Sigma \Lambda$, thus establishing (\ref{eqfordK}).

Note that in the SU(2) case, \ie for Dynkin diagrams, 
we have seen that the proportionality factor  $\Sigma X / \Sigma K$ was just equal to $2$.
Equations  (\ref{eqfordK}) and  (\ref{eqfordX}) illustrates a non-trivial effect of the boundary terms in this SU(3) situation since the proportionality factor is now a $k$-dependent polynomial.



\section{Multiplicity properties for fusion products : from $\lambda\otimes\mu$ to $\lambda\otimes\overline{\mu}$ }

\subsection{Known results} 
\subsubsection{Known results about multiplicities} 
\label{summaryonmutiplicities}

The multiplicity of the trivial representation in the decomposition into irreducibles of a tensor product (resp. the fusion product) of three irreducible representations of a group (resp. an affine Lie algebra) is invariant under an arbitrary permutation of the three factors. This property, sometimes called Frobenius reciprocity, is well known. Assuming that the notion of conjugation is defined in the case under study, the multiplicity stays also constant if we conjugate simultaneously the three factors.\\
Invariance of the multiplicity of  $\lambda \otimes \lambda^\prime \otimes \lambda^{\prime\prime} \rightarrow \one$ under the above transformations implies, for example, that this integer is also equal to the multiplicity of $\lambda \otimes \lambda^\prime \rightarrow \overline{\lambda^{\prime\prime}}$, of $\overline{\lambda} \otimes \overline{\lambda^\prime} \rightarrow\lambda^{\prime\prime}$, of $\lambda^{\prime\prime} \otimes \lambda  \rightarrow \overline{\lambda^{\prime}}$,  of $\lambda^{\prime\prime} \otimes \lambda^\prime \rightarrow \overline{\lambda}$, \etc
\\
These properties implies that the total multiplicity in the decomposition into irreducibles of the product $\lambda\otimes \mu$ of two irreducible representations is trivially invariant if we conjugate {\sl both} of them.\\
The following theorem was recently proven \cite{RCJBZ2011} 
{\theorem\label{CZthm} 
The total multiplicity in the decomposition into irreducibles of the tensor
product of two irreducible representations of a simple\footnote{The simplicity requirement can be lifted, see \cite{ZeierZimboras}} Lie algebra stays constant
if we conjugate only {\sl one} of them.  At a given level, this  property also holds for the fusion multiplicities of affine algebras.
} 
\\
Another sum rule, which is much easier to prove (see for instance \cite{RCJBZ2014}, section 1.1), using for instance Verlinde formula or its finite group analog, and which holds at least for finite groups, semi-simple Lie groups and affine algebras, states that the sum of squares of multiplicities is the same for $\lambda \otimes \mu$ and $\lambda \otimes \overline{\mu}$.
There are no such properties, in general, for higher powers of multiplicities.
\\
Very recently, cf \cite{RCJBZ2014}, it was further shown that 
{\theorem\label{CZ2014} In the special case of SU(3),  the {\rm lists} of multiplicities, in the tensor products  $\lambda \otimes \mu$ and $\lambda \otimes \overline{\mu}$, are identical up to permutations.}\\
This does not hold in general for other Lie algebras.
It was conjectured that the same property should hold for the fusion product of representations of the affine algebra of su(3) at finite levels, but this stronger result,
{that we shall call  property $\mathfrak P$ in the following}, was not proven in the quoted paper.
{\theorem\label{ppteP}{\bf (property $\mathfrak P$). }{At any finite level $k$ and for any integrable highest 
weights $\lambda$ and $\mu$ of 
$\widehat{su}(3)_k$, the lists of multiplicities $N_{\lambda\mu}^\nu$ and $N_{\lambda\bar\mu}^{\nu'}$ are identical up to permutations.
}}\\[2pt]
The purpose of the present section is to discuss and prove this property.\\[3pt]
A trivial corollary of Theorem \ref{ppteP} is 
{\begin{corollary}\label{coroll0}{At any finite level $k$ and for any integrable highest weights $\lambda$ and $\mu$ of 
$\widehat{su}(3)_k$, the number of {\it distinct} irreps $\nu$, resp. $\nu'$, appearing in $\lambda\star_k\mu$, resp
$\lambda\star_k\bar\mu$ is the same.}\end{corollary}}
\vskip-5mm
Strangely, whereas for SU(4) or $\widehat{su}(4)_k$ Theorem \ref{ppteP} is invalid  (for a counter-example, see 
\cite{RCJBZ2014}), Corollary \ref{coroll0} seems to hold.
{\conjecture\label{conj0}{At any finite level $k$ and for any integrable weights $\lambda$ and $\mu$ of 
$\widehat{su}(4)_k$, the number of {\it distinct} irreps $\nu$, resp. $\nu'$, appearing in $\lambda\star_k\mu$, resp
$\lambda\star_k\bar\mu$ is the same.}}\\[2pt]
Of that we have no proof, only some evidence from the computation of all fusion products up to level 15.\\
In contrast, this property does not hold in general for higher SU($N$) or $\widehat{su}(N)_k$. For example, in 
$\widehat{su}(5)_5$, and for  $\lambda=( 1,1,1,0), \mu=(1, 1, 0, 1)$, the line $\mu$ of the matrix $N_\lambda$
has 21 non vanishing entries, whereas the line $\bar \mu$ has 22.


\subsubsection{A short summary on  couplings, intertwiners, thresholds, and pictographs}

In the case of groups or Lie algebras, intertwiners, thought of as equivariant linear maps from a tensor product of three representations to the scalars, are  Ò3J-operatorsÓ that, when evaluated on a triple of vectors, become 3J symbols (Clebsh-Gordan coefficients).   We shall sometimes refer to a particular space of intertwiners, associated to a particular term $\nu$ in the decomposition into irreducibles of $\lambda \otimes \mu$ as a ``branching'', and denote it by $\lambda \otimes \mu \rightarrow \nu$ (and indeed, in the case of a group $G$, it also corresponds to a branching from $G \times G$ to its diagonal subgroup $G$). The multiplicity of $\nu$ in $\lambda \otimes \mu$ is also the dimension $d$ of the associated space of intertwiners. From the point of view of representation theory, all the irreps $\nu$ that appear on the rhs of the decomposition of $\lambda \otimes \mu$ are equivalent; nevertheless, it is convenient to consider each of them as a different ``coupling'' of the chosen representations (this is actually what is done in conformal field theory where such a multiplicity $d$ is interpreted as the number of distinct couplings of the associated primary fields). In the same way combinatorial models typically associate to a space of intertwiners of dimension $d$, and therefore to a given triple or irreducible representations, a set of $d$ combinatorial and graphical objects, that we call generically pictographs: the reader may look at \cite{RCJBZ2014} for a discussion of three kinds of pictographs (KT-honeycombs, BZ-triangles, O-blades), in the framework of the Lie group SU(3).
Finally, let us mention that the choice of a basis in a space of intertwiners allows one to decompose a coupling along so-called elementary couplings \cite{BMW} (equivalently, decompose pictographs along fundamental pictographs, see again \cite{RCJBZ2014} and sect. \ref{applOblade} of the present paper).

As in the first part, we denote an irreducible representation of a Lie algebra, or the integrable irrep of the corresponding affine algebra at level $k$ (which may not exist if $k$ is too small) by its associated highest weight, and therefore by the same symbol. Multiplicities $N_{\lambda \mu}^\nu$, also called  Littlewood-Richardson coefficients,  or  fusion coefficients in the present context of affine Lie algebras, are the matrix elements of the fusion matrices considered in the first section but here we have to make the level explicit in the notation and call $N_{\lambda \mu}^{(k)\;\nu}$ the multiplicity at level $k$ and keep the notation $N_{\lambda \mu}^{\nu}$ for the classical multiplicity (infinite level).
Beginning with $k=0$ and letting $k$ increase, it is usually so that the three irreps $\lambda, \mu, \nu$ start to exist at possibly distinct levels; then, even if the three irreps exist at level $k$, it may be that $N_{\lambda \mu}^{(k)\;\nu}$ is still $0$,  or has a value smaller than the classical one (\ie of infinite level). 
For instance, one has $\{1, 1\} \otimes \{0, 2\} = \{0, 2\} \oplus \{1, 0\}\oplus \{1, 3\}\oplus \{2, 1\}$, but at level $k = 2$, although the irrep $\{0, 2\}$ exists, it does not appear in the fusion product, which is $\{1, 1\} \star_ 2 \{0, 2\} = \{1, 0\}$; the other irreps $\{2,1\}, \{0,2\}$ and $\{1,3\}$ appear respectively in the fusion product $\star_k$ at levels $k=3, 3,4$.\\[3pt]
{\bf Definition.} A triple $(\lambda,\mu;\nu)$ is called {\it admissible}, or {\it classically admissible}, if $N_{\lambda\mu}^\nu\ne 0$; it is called {\it admissible at level $k$} if $N_{\lambda\mu}^{(k)\;\nu}\ne 0$.

{The following results can be gathered from \cite{BMW}.}

The threshold level (or threshold, for short) $k_0^{min}$ of a triple (or {\sl of a branching} $\lambda \otimes \mu \rightarrow \nu$) is the smallest value
of $k$ for which the fusion coefficient $N_{\lambda \mu}^{(k)\;\nu}$ is non-zero.
It is known that, for any given triple of irreps, the fusion coefficient is an increasing function of the level, and that it becomes equal to its classical value $N_{\lambda \mu}^\nu$ when $k$ reaches a value $k_0^{max}$, then it stays constant. 
The integers $k_0^{min}$ and $k_0^{max}$ are functions of $\lambda$, $\mu$ and $\nu$, but for given $\lambda$ and $\mu$ one calls $k^{min}$ the infimum of the $k_0^{min}$
over the set $\nu$,  and  $k^{max}$ the supremum of the $k_0^{max}$ over the set $\nu$.

In terms of couplings (or pictographs),  the discussion goes as follows: for fixed $\lambda$, $\mu$ and $\nu$ and a given $k$, we have a set of $N_{\lambda \mu}^{(k)\;\nu}$ distinct couplings; these couplings still exist when the level becomes $k+1$, but new couplings may appear, since  $N_{\lambda \mu}^{(k+1)\nu} \geq  N_{\lambda \mu}^{(k)\;\nu}$.
The sets $I(k)$ of couplings are therefore ordered by inclusion: $I(k) \subset I(k+1)$. 
The threshold  $k_0(i)$ {\sl of a coupling} $i$ is, by definition,  the smallest value of $k$ at which it appears: the coupling $i$ belongs to the set $I(k_0(i))$ but it does not belong to the set  $I(k_0(i)-1)$.

The threshold
 $k_0^{min}$ of a triple (or of a branching) is therefore equal to the value of $k$ for which the first associated couplings appear (such couplings have a threshold  equal to $k_0^{min}$).
Conversely the value $k_0^{max}$ of a branching is equal to the value of $k$ above which no new couplings appear.\\
We have $I(k_0^{min}) \subset I(k_0^{min}+1)  \subset I(k_0^{min}+2) \subset \ldots I(k) \subset \ldots \subset I(k_0^{max})$.  The set $I(k)$ has cardinality $N_{\lambda \mu}^{(k)\;\nu}$, and the largest set $ I(k_0^{max})$ can be considered as classical since it contains all the couplings.

{\sl From now on, we consider the special case of SU(3)}. In all cases, as we saw, for any given triple of irreps, the fusion coefficient is equal to $0$ when $k < k_0^{min}$, equal to the constant $N_{\lambda \mu}^\nu$ when $k \geq k_0^{max}$, and is an increasing function of $k$ when $k_0^{min} \leq k \leq k_0^{max}$. In the special case of the affine algebra su(3), 
 the multiplicity of the branching $\lambda \otimes \mu \rightarrow \nu$ is a piecewise affine function of $k$,
 first equal to $0$, then it starts to increase with $k$, taking values respectively equal to $1,2,3,\ldots,N_{\lambda \mu}^{\nu}$ for successive values of $k=k_0^{min}, k_0^{min}+1, \ldots, k_0^{max}$. Then it stays constant. 
 At level $k$ the multiplicity (the number of couplings) is therefore $k-k_0^{min}+1$:  one new coupling appears for every value of $k$ between $k_0^{min}$ and $k_0^{max}$.  
 
Adapting the results of  \cite{BMW} to our own notations\footnote{When discussing
general couplings $\lambda \otimes \mu \rightarrow \nu$, it happens that most formulae are more simply expressed in terms of $\lambda$, $\mu$, and $\overline{\nu}$, 
than in terms of $\lambda$, $\mu$ and $\nu$, because of Frobenius reciprocity.
Admittedly, we should have chosen notations where the roles of $\nu$ and $\overline{\nu}$ are exchanged. However we shall not do that, in order to be consistent with the choices made in the companion paper \cite{RCJBZ2014}.} (see also \cite{DFMS}  {and the discussion at the end of our section 5.1.3}), we have\footnote{In order to be interpreted as a minimal or maximal threshold, the argument $(\lambda,\mu;\nu)$ appearing in eqs (\ref{k0min}, \ref{k0max}), should refer to a classically admissible triple, although the rhs of these two equations make sense for arbitrary arguments. We hope that there should be no confusion. } 
\bea
\nonumber   k_0^{\min}(\lambda,\mu;\nu)
 &=&\max\Big( \lambda_1 + \lambda_2,\;   \frac{\lambda_1 - \lambda_2 +  \mu_1 + 2\mu_2 + 2 \nu_1 + \nu_2}{3}, \;  \frac{-\lambda_1 +  \lambda_2 + 2\mu_1 + \mu_2 + \nu_1 +  2\nu_2}{3},
 \\ \nonumber & &   
  \mu_1 + \mu_2, \;  \frac{2 \lambda_1 + \lambda_2 - \mu_1 + \mu_2 +   \nu_1 + 2\nu_2}{3}, \   \frac{\lambda_1 + 2 \lambda_2 + \mu_1 - \mu_2 + 2\nu_1 +  \nu_2}{3},  
  \\ \label{k0min} & &     
   \nu_1 + \nu_2,\;   \frac{\lambda_1 + 2 \lambda_2 + \mu_1 + 2 \mu_2 - \nu_1 + \nu_2}{3},\;    \frac{2 \lambda_1 + \lambda_2 + 2 \mu_1 + \mu_2 + \nu_1 - \nu_2}{3}   \Big)\\
    \label{k0max}
   \!\!\!\!\!\! k_0^{\max}(\lambda,\mu;\nu) \!\!&\!\!\!=\!\!\!&\!\! \min\Big(\frac{2 \lambda_1 + \lambda_2 +2 \mu_1 + \mu_2 +   \nu_1 + 2\nu_2}{3},
    \frac{\lambda_1 + 2 \lambda_2 + \mu_1 +2\mu_2 + 2\nu_1 +  \nu_2}{3}\Big)\qquad
\eea
 and 
\begin{eqnarray}\label{defskmin-kmax}
k^{min} &=& \max(\lambda_1+\lambda_2, \mu_1+\mu_2)\\ \nonumber
k^{max} &=& \lambda_1+\lambda_2+ \mu_1+\mu_2
\end{eqnarray}
The threshold  $k_0(i)$ of a given coupling $(i)$ can be read from one of its associated pictographs. Using BZ-triangles for example, this level is obtained\footnote{This property was proven, in terms of BZ-triangles, by \cite{KMSW}, \cite{LK}, see also \cite{BMW}.} as follows:  $k_0(i)$ is the maximum of $\lambda_1+\lambda_2+ \alpha,  \mu_1+\mu_2+ \beta, \nu_1+\u_2+ \gamma$, where $\alpha, \beta, \gamma$ are the values of the vertices respectively opposite to the three sides  $\lambda, \mu, \nu$.
Using O-blades, they are the values  $(c,e,a)$ of the internal edges opposite to the three sides  in Fig.\;\ref{permutedforks}. 
The same integers can also be easily read from the
SU(3)-honeycombs, since they are dual to O-blades (see Fig. 17 of  \cite{RCJBZ2014}). The translation in terms of  
KT-honeycombs (which are actually GL(3)-honeycombs) or in terms of the hive model can be done by using methods explained in \cite{RCJBZ2014}.
We shall illustrate the above considerations on one example, at the end of the next section.

 
\subsubsection{Formulae for multiplicities}
\label{formulaeformultiplicities}

Explicit formulae for SU(3) multiplicities have been known for quite a while, the first known reference going back to 
\cite{Mandeltsveig}, in 1965 (see also \cite{RCJBZ2014} and references therein). 
For the affine case, they were obtained in \cite{BMW}.
Assuming that the triplet $(\lambda,\mu;\nu)$ is admissible, we have
\be\label{Nklmn}N_{\lambda \mu}^{(k)\;\nu}=\begin{cases}  0 & {\rm if}\ \ k <  k_0^{min}(\lambda,\mu;\nu) \\
k-k_0^{min}(\lambda,\mu;\nu)+1 & {\rm if}\ \  k_0^{min}(\lambda,\mu;\nu)\le k\le k_0^{max}(\lambda,\mu;\nu)\\
k_0^{max}(\lambda,\mu;\nu) -k_0^{min}(\lambda,\mu;\nu)+1 =N_{\lambda\mu}^\nu & {\rm if}\ \   k\ge k_0^{max}(\lambda,\mu;\nu)
\end{cases}\ee
This expression entails 
a recursion formula on fusion coefficients under a shift of all h.w. by $\rho=(1,1)$, the Weyl vector, and of 
$k$ by $3$. Assume that $\lambda,\mu,\nu$ are h.w. vectors such that $\lambda-\rho,  \mu-\rho, \nu-\rho$ are also integrable h.w.
at level $k-3$,
\ie, $\lambda_i,\mu_i,\nu_i\ge 1$, $i=1,2$, and $\lambda_1+\lambda_2\le k+1$ etc. 
Then\footnote{\label{k0shifts} More generally,  for all integers $u,v$, one shows using  eq.\;(\ref{k0min}) that
$k_0^{min}(\lambda + (u,v), \mu + (u,v); \nu + (v,u)) = k_0^{min}(\lambda, \mu; \nu) + (u+v)$, and
using eq.\;(\ref{k0max}), 
that $k_0^{max}(\lambda+(u,v), \mu+(u,v), \nu+(v,u)) - k_0^{max}(\lambda, \mu, \nu) = (u+2v)$, (resp $(2u+v)$),  if  $\lambda_1+\mu_1+\nu_2 \geq   \lambda_2+\mu_2+\nu_1$ and $u \geq v$,  (resp if $\lambda_1+\mu_1+\nu_2 \leq   \lambda_2+\mu_2+\nu_1$ and $u \leq v$).
}
$k_0^{min}(\lambda-\rho,\mu-\rho;\; \nu-\rho)=
k_0^{min}(\lambda,\mu;\;\nu)-2$, whereas 
$k_0^{max}(\lambda-\rho,\mu-\rho;\; \nu-\rho)= k_0^{max}(\lambda,\mu;\;\nu)-3$,
 thus according to (\ref{Nklmn}),  if $k\ge 3$ 
\begin{eqnarray*} \!\!\!\!\!\!\!\!\!\!\!\!\!\!\!\!\!\!
N_{\lambda-\rho\; \mu-\rho}^{(k-3)\nu-\rho}&=&\begin{cases}\scriptstyle  0 &\scriptstyle {\rm if}\ \ k-3 <  k_0^{min}(\lambda-\rho,\mu-\rho;\nu-\rho)=
k_0^{min}(\lambda,\mu;\;\nu)-2 \\
\scriptstyle (k-3)-(k_0^{min}(\lambda,\mu;\nu)-2)+1 &\scriptstyle {\rm if}\ \  k_0^{min}(\lambda,\mu;\nu)-2\le k-3\le k_0^{max}(\lambda,\mu;\nu)-3\\
\scriptstyle (k_0^{max}(\lambda,\mu;\nu)-3) -(k_0^{min}(\lambda,\mu;\nu)-2)+1&\scriptstyle {\rm if} \ \   k-3\ge k_0^{max}(\lambda,\mu;\nu)-3
\end{cases}\\
&=&\begin{cases}  0 & {\rm if}\ \ k <  k_0^{min}(\lambda,\mu;\;\nu) +1 \\
k-k_0^{min}(\lambda,\mu;\nu)=N_{\lambda \mu}^{(k)\;\nu} -1  & {\rm if}\ \  k_0^{min}(\lambda,\mu;\nu) +1\le k \le k_0^{max}(\lambda,\mu;\nu) \\
k_0^{max}(\lambda,\mu;\nu) - k_0^{min}(\lambda,\mu;\nu) =N_{\lambda\mu}^\nu -1 & {\rm if}\ \   k \ge k_0^{max}(\lambda,\mu;\nu)
\end{cases}
\end{eqnarray*}
In other words
\be\label{recurN} N_{\lambda-\rho\; \mu-\rho}^{(k-3)\nu-\rho}=\begin{cases} N_{\lambda \mu}^{(k)\;\nu} -1
     & {\rm if}\ \ k\ge \max(k_0^{min}(\lambda,\mu;\nu),3)  \\
  0 &  {\rm if}\ \ k\le  k_0^{min}(\lambda,\mu;\nu) \ {\rm or}\ k<3\end{cases}\,. \ee
We will make use of this relation in our proof of property $\mathfrak{P}$.
%

\subsubsection{A simple algorithm}
 In order to simply determine the multiplicity of the branching $\lambda \otimes \mu \rightarrow \nu$, or equivalently of $\lambda \otimes \mu \otimes  \overline \nu \rightarrow \one$, at level $k$, one may proceed as follows, see also  \cite{Suciu}.
 We use the following notations, and drop the label $k$ for classical multiplicities:
$$\mathfrak{m}_k=mult_k(\lambda, \mu; \nu) =  mult_k(\lambda \otimes \mu \otimes \overline{\nu}) = N_{\lambda \mu}^{(k)\;\nu}$$ 
{\sl Classical case}. Call $S_1 = \lambda_1 + \mu_1 + \overline \nu_1$, $S_2 = \lambda_2 + \mu_2 + \overline \nu_2$. If $S_1-S_2$ is not a multiple of $3$, the multiplicity vanishes \cite{BMW}.
Assuming $S_1>S_2$, we define three new irreps $\lambda^\prime=(\lambda_1-x, \lambda_2)$, $\mu^\prime=(\mu_1-x, \mu_2)$, $\overline \nu^\prime=(\overline \nu_1 -x, \overline \nu_2)$ where  $x = (S_1-S_2)/3$. Then\footnote{Indeed, from eq.\;(\ref{k0min}) and footnote \ref{k0shifts}, for an arbitrary integer $x$, we have  $k_0^{min}(\lambda - (x,0), \mu - (x,0); \nu - (0,x)) - k_0^{min}(\lambda,\mu; \nu) =-x$; moreover, if $S_1\geq S_2$, from eq.\;(\ref{k0max}) and for the  particular value $x = (S_1-S_2)/3$, we have
 $k_0^{max}(\lambda - (x,0), \mu - (x,0); \nu - (0,x)) - k_0^{max}(\lambda,\mu; \nu) =-x$.}  $mult(\lambda \otimes \mu \otimes \overline{\nu}) =  mult(\lambda^\prime \otimes \mu^\prime \otimes \overline{\nu^\prime})$ and the new sums $S_1^\prime$ and $S_2^\prime$ are both equal to $S = min(S_1, S_2)$.
If $S_1<S_2$ one shifts the second components by $x$ instead. 
Define $\ell=S-\sum_i \lambda^\prime_i$, $m=S-\sum_i \mu^\prime_i$, $n= S- \sum_i \nu^\prime_i$ and consider the tableau of order $3$ with lines 
$(\lambda^\prime_1, \mu^\prime_1, \overline\nu^\prime_1)$, $(\lambda^\prime_2, \mu^\prime_2, \overline\nu^\prime_2)$ and $(\ell, m, n)$. 
By construction, this tableau is a semimagic square of order $3$ with magic constant $S$; it is not necessarily magic because the sums along the diagonals do not add to $S$ in general. The classical multiplicity $\mathfrak{m}$ is equal to $c+1$ where $c$ is the minimum of the entries of the tableau; indeed, by construction, $c=k_0^{max} - k_0^{min}$.

{\sl Affine case}. Start from $k=0$ and and let $k$ increase: the irrep $\nu$ appears for the first time, with multiplicity $\mathfrak{m}_k=1$, in the tensor product $\lambda \otimes \mu$ when $k$ reaches\footnote{Remember that $k_{\min}(\lambda, \mu) = \inf_\nu \; k_{0}^{min}(\lambda, \mu, \nu)$ and that $k_{\max}(\lambda, \mu) = \sup_\nu \; k_{0}^{max}(\lambda, \mu,\nu)$.} 
the threshold  $k_{0}^{min}=S+x-c$.
The multiplicity then increases with the level, each time by one unit, so  $\mathfrak{m}_k = k - k_{0}^{min} +1$, 
until $k$ reaches the value $k_{0}^{max}=S+x=k_{0}^{min}+ (\mathfrak{m}-1)$ for which the multiplicity becomes equal to the classical one, $\mathfrak{m}$.
 It then stays constant when $k$ increases beyond $k_{0}^{max}$.\\
Warning. We have $mult(\lambda \otimes \mu \otimes \overline{\nu}) =  mult(\lambda^\prime \otimes \mu^\prime \otimes \overline{\nu^\prime})$ when the level is infinite (classical case), but their thresholds $k_{0}^{min}$ are distinct: it is $S+x-c$ for the first but $S-c$ for the next since $x$ vanishes in the latter case.

Example: let us determine $\mathfrak{m}=mult( (9,5), (6,2); (8,6)) = mult( (9,5) \otimes (6,2)  \otimes (6,8) $. We have $S_1=21, S_2=15$, so $S =15 $ and $x =(21-15)/3=2$.
Then\footnote{The thresholds of $((9,5), (6,2); (8,6))$ and $((7,5), (4,2); (8,4))$ are distinct:  $15$ for the first and $13$ for the next.} $\mathfrak{m}= mult((7,5) \otimes (4,2)  \otimes (4,8))$ and  $(\ell, m, n)=(3,9,3)$, the associated semimagic square being 
{\footnotesize $\begin{array}{ccc}
7&4&4 \\
5&2&8 \\
3&9&3 \\
\end{array}$}.
Therefore $c =2$ and $\mathfrak{m}=3$. The irrep $(8,6)$ appears with multiplicity $1$ in $(9,5) \otimes (6,2)$ when $k$ reaches the threshold value $k_{0}^{min}=S+x-c=15+2-2=15$. 
Therefore ${\mathfrak m}_{15}=1, {\mathfrak m}_{16}=2, {\mathfrak m}_{17}=3$, then it stays constant.


\subsubsection{One example}

\begin{figure}[]
\centering{\includegraphics[width=25pc]{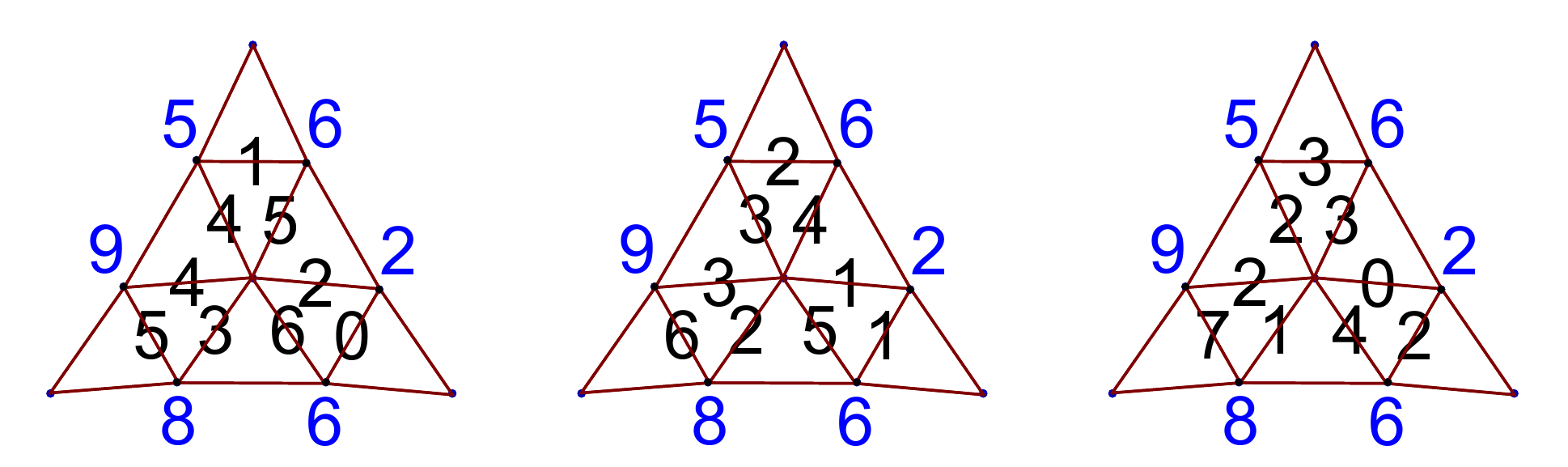}}
\caption{\label{oblades956286} {The three couplings of $(9,5)\otimes(6,2)\rightarrow(8,6)$ using the O-blade model}}
\end{figure}

\begin{figure}[]
\centering{\includegraphics[width=25pc]{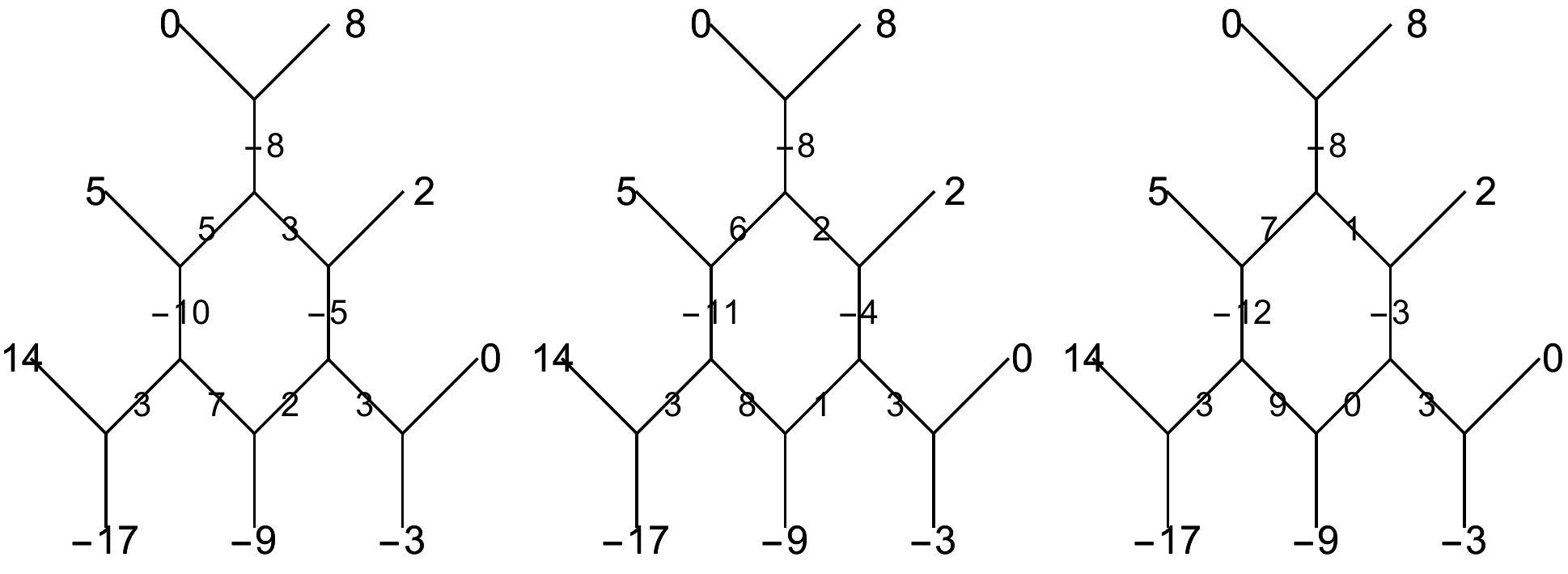}}
\caption{\label{honeycomb956286} {The three couplings of $(9,5)\otimes(6,2)\rightarrow(8,6)$ using the KT-honeycomb model}}
\end{figure}

{\footnotesize
\begin{eqnarray*}
\!\!\!\!\!\!\!\!\!\!\!\!
(9,5)\otimes(6,2)&=&\{(6, 1) _ {14}, (7, 2) _ {14, 15}, (5, 3) _ {14}, (8, 3) _ {14, 15, 
   16}, (6, 4) _ {14, 15}, (4, 5) _ {14}, (7, 5) _ {14, 15, 16}, (5, 
   6) _ {14, 15}, (3, 7) _ {14}, (6, 7) _ {14, 15, 16}, \\
 & & (4, 8) _ {14}, (2, 9) _ {14}, (5, 9) _ {14, 15, 16}, (3, 10) _ {14, 
   15}, (1, 11) _ {14},
 (8, 0) _ {15}, (9, 1) _ {15, 16}, (10, 2) _ {15, 16, 17}, (9, 
   4) _ {15, 16, 17}, \\
   & & 
   (8, 6) _ {15, 16, 17}, (4, 8) _ {15}, (7, 
   8) _ {15, 16, 17}, (4, 11) _ {15, 16}, (2, 12) _ {15}, (11, 
   0) _ {16}, (12, 1) _ {16, 17}, (11, 3) _ {16, 17, 18}, (10, 
   5) _ {16, 17, 18},\\
   & &  (9, 7) _ {16, 17, 18}, (6, 10) _ {16, 17}, (3, 
   13) _ {16}, (14, 0) _ {17}, (13, 2) _ {17, 18}, (12, 4) _ {17, 18, 
   19}, (11, 6) _ {17, 18, 19}, (8, 9) _ {17, 18}, (5, 
   12) _ {17},\\
   & &  (15, 1) _ {18}, (14, 3) _ {18, 19}, (13, 5) _ {18, 19, 
   20}, (10, 8) _ {18, 19}, (7, 11) _ {18},
 (16, 2) _ {19}, (15, 4) _ {19, 20}, (12, 7) _ {19, 20}, (9, 
   10) _ {19}, (17, 3) _ {20},\\
   & & (14, 6) _ {20, 21}, (11, 
   9) _ {20}, (16, 5) _ {21}, (13, 8) _ {21}, (15, 7) _ {22}\}
\end{eqnarray*}
}
The subindices refer to the levels at which the branching rule occurs. Let us consider the branching to $(8,6)$ that we already analysed in the previous subsection. The term $(8,6)_{15,16,17}$, for instance, means that the first coupling associated with the branching $(9,5)\otimes(6,2)\rightarrow(8,6)$ appears at level 15 (so the multiplicity is $1$), another at level 16 (the multiplicity is $2$), and a last one at level 17 (the multiplicity is $3$). The tensor (\ie classical) multiplicity of this branching is $3$. Note that  $k_0^{min}((9,5),(6,2),(8,6))=15,\,  k_0^{max}((9,5),(6,2),(8,6))=17$ but $k^{min}((9,5),(6,2))=14$ and $k^{max}((9,5),(6,2))=22$. 
The three couplings are illustrated by their respective pictographs, using O-blades and KT-honeycombs\footnote{These are GL(3) honeycombs, one could prefer to draw instead the corresponding SU(3) honeycombs, where the lengths of all edges are non-negative, see \cite{RCJBZ2014}.}, on Figs. \ref{oblades956286}, \ref{honeycomb956286}. 
Notice that the threshold  $k_0^{(i)}$, $i=1,2,3$, for each of these couplings can be read from the pictures, as explained previously.
The associated BZ-triangles can be immediately obtained from the correspondence explicited on Fig. 16 of \cite{RCJBZ2014}.

\begin{figure}[htbp]
\includegraphics[width=35pc]{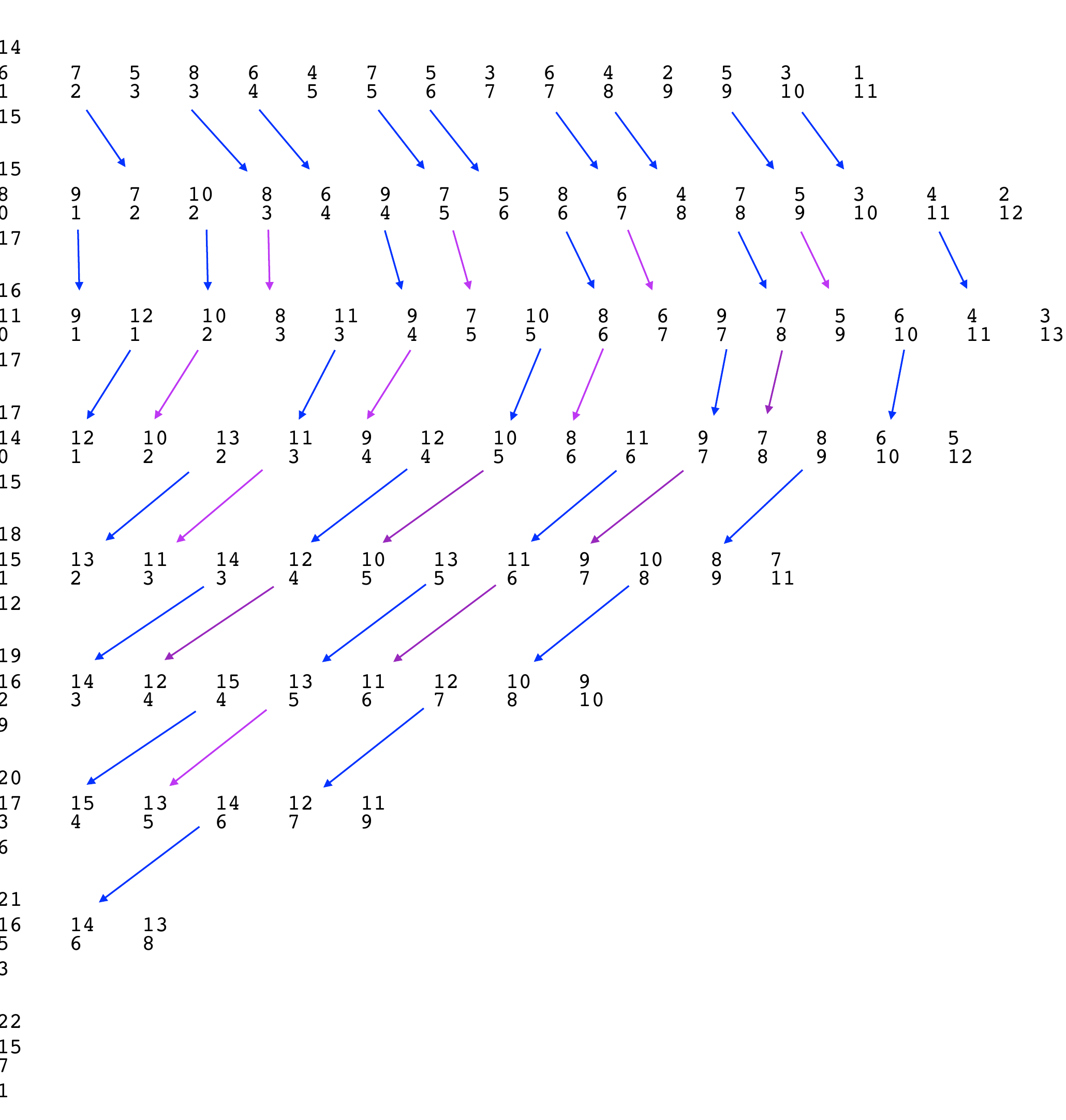}
\caption{Thresholds for intertwiners $(9,5) \otimes (6,2) \rightarrow (\nu_1, \nu_2)$.
\small{
Each line (levels $k^{min}=14$ to $k^{max}=22$) lists those irreps $\nu$ (components given vertically) for which the multiplicity increases at the chosen level, \ie for which a new coupling appears. 
The total number of new couplings at a given level is also given on the left.
Same irreps for which there is a multiplicity increase are connected by arrows.  The number of arrows with the same colour gives the  $u_{pj}$ of sect. \ref{proofofP}: at level $k=18$, \ie $p=k-k^{min}+1=5$ there 
are $5$ irreps that are not hit by any arrowhead ($u_{50}=5$), $4$ hit by a blue arrow ($u_{51}=4$), and ($u_{52}=3$) hit by a red arrow, the total multiplicity increasing of $5+4+3=12$.}}
\label{9562}
\end{figure}

 Another way of displaying the results is given on Fig.\;\ref{9562}. Rather than giving the irreps $\nu$ obtained 
 in the fusion of $(9,5)$ and $(6,2)$, together with the level at which they appear, we display, for each level, the {\sl couplings} that  show up  at that level. The highest weight of a particular irrep therefore appears several times on the figure (for instance $(8,6)$ appears on the three lines labelled by $k=15, 16, 17$).

\begin{figure}[htbp]
\includegraphics[width=35pc]{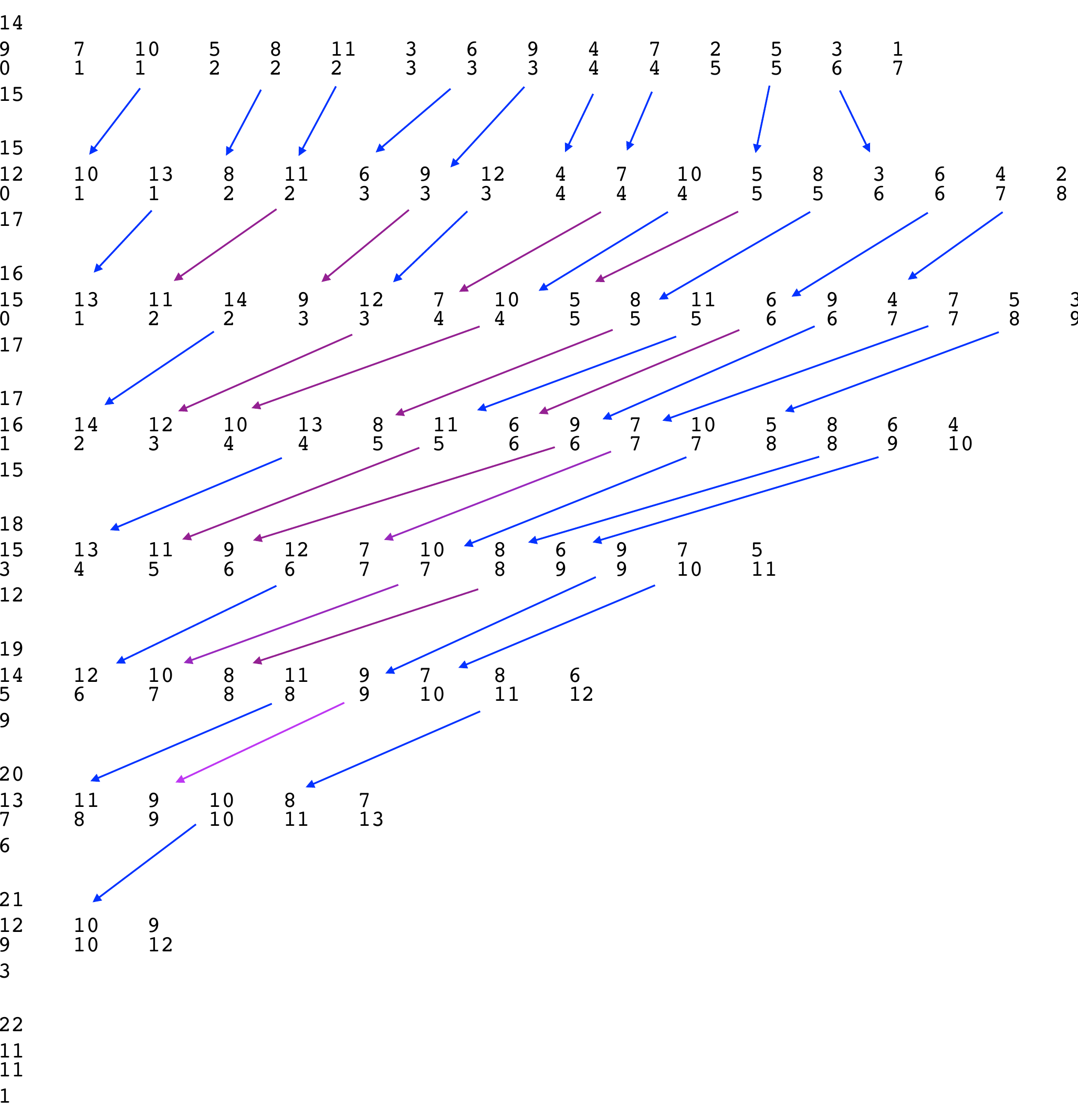} 
\caption{Thresholds for intertwiners $(9,5) \otimes (2,6) \rightarrow (\nu_1^\prime, \nu_2^\prime)$} 
\label{9526}
\end{figure}

What we did for the fusion product $(9,5)\otimes(6,2)$ can be repeated for $(9,5)\otimes(2,6)$.
We leave this as an exercise for the reader. Fig.\;\ref{9526}, that should be compared with Fig.\;\ref{9562}, summarizes the results.

As it was announced, the two lists of multiplicities at the classical level are the same, up to permutations (we find $21$, $16$ and $3$ irreps with respective multiplicities $1, 2, 3$, and therefore a total multiplicity 
 equal to $95$), something that we already knew from \cite{RCJBZ2014}, but they are also equal for all values of the level.  
 In the next section, we turn to a proof of this property.


\subsection{From $\lambda\otimes\mu$ to $\lambda\otimes\overline{\mu}$ : proof of property $\mathfrak P$ using Wessl\'en polygons}

\subsubsection{New inequalities and the convex polygonal domain}
Recall from \cite{Wesslen, RCJBZ2014}
the inequalities satisfied by $\nu=(\nu_1,\nu_2)$, $\nu_1,\nu_2\ge 0$,  in the intertwiner $\lambda\otimes \mu \to \nu$
{\bea  \scriptstyle
 \label{W1} 
 \scriptstyle \max(2\mu_1+\mu_2-\lambda_1-2\lambda_2, 2\lambda_1+\lambda_2-\mu_1-2\mu_2,\lambda_2-\lambda_1+\mu_2-\mu_1) \le  &\!\!\!  2\nu_1+\nu_2\!\!\!  & \scriptstyle \le 2\lambda_1+\lambda_2+2\mu_1+\mu_2 \\
 \label{W2} 
 \scriptstyle\max(\lambda_1+2\lambda_2-2\mu_1-\mu_2,\mu_1+2\mu_2-2\lambda_1-\lambda_2,\lambda_1-\lambda_2+\mu_1-\mu_2) \le &\!\!\!  \nu_1+2\nu_2 \!\!\! & \scriptstyle \le  \lambda_1+2\lambda_2+\mu_1+2\mu_2\\
\label{W3}  \scriptstyle\max(\mu_1-\mu_2-2\lambda_1-\lambda_2,\lambda_1-\lambda_2-2\mu_1-\mu_2) 
 \le &\!\!\!  \nu_1-\nu_2\!\!\! &\le \scriptstyle \min(\lambda_1-\lambda_2+\mu_1+2\mu_2,\lambda_1+2\lambda_2+\mu_1-\mu_2)
\eea }
and from \cite{BMW} the additional {``BMW"} ones satisfied in the fusion product  $\lambda\star_k \mu \to \nu$ at level $k$
\bea  \label{BMW1} { \nu_1+\nu_2}&\le& k  \\ 
\label{BMW2}  { 2\nu_1+\nu_2}&\le& \min(3k-\lambda_1+\lambda_2-\mu_1-2\mu_2,\, 3k-\lambda_1-2\lambda_2-\mu_1+\mu_2) \\
\label{BMW3} {\nu_1+2\nu_2} &\le&\min(3k-2\lambda_1-\lambda_2+\mu_1-\mu_2,\, 3k +\lambda_1-\lambda_2-2\mu_1-\mu_2) \\
\label{BMW4} -3k+\lambda_1+2\lambda_2+\mu_1+2\mu_2 &\le& { \nu_1-\nu_2}\ \le\  3k -2\lambda_1-\lambda_2-2\mu_1-\mu_2  
\eea 
which are simple consequences of $k\ge k_0^{min}(\lambda,\mu;\nu)$ with the explicit expression given in (\ref{k0min}).
Using standard symmetries of fusion coefficients, we may always assume that $\lambda_1 \ge \max(\lambda_2,\mu_1, \mu_2)$.
Then note that the second inequality (\ref{BMW3}) follows from the first since $\mu_1\le \lambda_1$.
As discussed above, the issue of finite level is relevant only in the range $k^{min}\le k\le k^{max}$,
with $k^{min}$ and $ k^{max}$ given in (\ref{defskmin-kmax}).
 Below $k^{min}$, one of the weights $\lambda$ or $\mu$ is not integrable, 
 above $k^{max}$ we are in the classical, ``$k=\infty$", regime.

For a given $k\ge k^{min}$ the set of inequalities (\ref{W1}-\ref{BMW4}) defines in the $\nu$-plane a 
convex polygonal domain. 
For $k\ge k^{max}$ this domain stabilizes to its ``classical" shape defined by the only
Wessl\'en's inequalities (\ref{W1}-\ref{W3}), with vertices $V_i$ defined on Fig.\;\ref{wessl-tronqu}. In particular the 
``highest highest weight" is $H=V_1=(\p+\r,\q+\s)$, while the ``lowest {highest} weight" $h=V_5$ is given below in (\ref{h-V5}).
For $k^{min}\le k < k^{max}$, we refer to the polygonal domain 
as the truncated Wessl\'en's domain.

By the same type of recursive arguments as in sect. 5.1.3  and eq.\;(\ref{recurN}), 
we may also find the polygons that bound the domain of  multiplicity 2. 
By (\ref{Nklmn}), they are obtained
by imposing that $k = 1
+ k_0^{min}(\lambda,\mu;\nu)$, and result in inequalities of the form (\ref{BMW1}-\ref{BMW4}) in which $k$ is substituted for 
$k-1$. But the transformation $k\to k-3$, 
$\lambda\to \lambda- \rho$, $\mu\to \mu- \rho$, $\nu\to \nu-\rho$ in (\ref{BMW1}-\ref{BMW4})
has the same effect. We conclude that the set of $\nu$ of level $k$ and multiplicity $N_{\lambda\mu}^{(k)\;\nu}=2$ 
is the translate by $\rho$ 
of the set of $\nu'$ of level $k-3$ 
and multiplicity  $N_{\lambda-\rho\,\mu-\rho}^{(k-3)\,\nu'}=1$. 
It would seem
that this argument requires the $\lambda_i,\mu_i,\nu_i$ to be $\ge 1$ in order for $\lambda- \rho$, $\mu- \rho$, $\nu-\rho$ to be
integrable weights, and $k-3\ge 0$. In fact  
as the recursion relation is used  for quadruplets $(\lambda,\mu,\nu;k)$ for which $N_{\lambda \mu}^{(k)\;\nu}= 2$, these
conditions  
are automatically enforced.\\ 
The recursion extends to higher values of the multiplicity, see below  sect. 5.2.3.

\begin{figure}
   \mbox{ 
  \setlength{\unitlength}{1.3pt}
\begin{picture}(100,60)
\put (180,0){}
\put(205,18){$H=V_1=(\lambda_1+\mu_1,\lambda_2+\mu_2)$}\put(200,18){\vector(1,-2){25}}\put(210,0){$\scriptstyle \min(\q,\s)\, (1,-2)$} 
\put(225,-32){\vector(-1,-1){25}}\put(220,-45){$\scriptstyle |\min(\q-\s,\r)|\, (-1,-1)$}\put(225,-35){$V_2$}
\put(210,-57){$V_3$}
\put(200,-57){\vector(-3,0){20}}
\put(210,-70){\vector(-2,1){20}}
\put(210,-70){$\scriptstyle \max[\min(\p,\q,\r,\s,\r+\s-\q),0] \, (-3,0)$}
\put(180,-57){\vector(-2,1){20}}
\put(145,-63){\vector(2,1){20}}
\put(170,-66){$V_4$}
\put(15,-73){$\scriptstyle
\begin{rcases}
\scriptstyle \min(\p-\s,\r-\q)(-2,1)&\scriptstyle {\rm if\ } \r> \q \\
\scriptstyle |\min(\q-\r,\s)|(-2,1)&\scriptstyle{\rm if\ }\q\ge \r
\end{rcases} $}
\put(125,-50){$h=V_5$}
\put(160,-47){\vector(-1,2){20}}\put(5,-35)
{$\scriptstyle |\min(\p-\r-\s,0)+\min(\q,\r)| 
 (-1,2)$}
\put(128,-13){$V_6$}
\put(140,-7){\vector(0,3){25}}\put(20,2){$\scriptstyle \max[\min(\q,\r+\s-\p),0] \, (0,3)$}
\put(128,18){$V_7$}
\put(140,18){\vector(1,1){20}}\put(65,30){$\scriptstyle |\min(\p-\r,\s)|\, (1,1)$}
\put(160,38){\vector(2,-1){40}}\put(180,30){$\scriptstyle \r 
\, (2,-1)$}
\put(155,40){$V_8$}
 \put(130,30){\red\line(1,1){20}}
  \put(210,-60){\red\line(1,1){27}}
   \put(220,0){\green\line(1,-2){17}}
   \put(190,30){\violett\line(1,-1){30}}
      \put(150,50){\blue\line(2,-1){40}}
   \end{picture}}
\vskip45mm
\caption{The oriented sides of {a} tensor polygon (with the assumption that  $\p\ge \max(\q,\r,\s)$). 
\footnotesize(We have used orthogonal coordinate axes,  although it would be more correct to have them at a $\pi/3$ angle.)
The violet,  green, blue and red lines represent respectively the inequalities (\ref{BMW1}),  (\ref{BMW2}), (\ref{BMW3})  and (\ref{BMW4}). As $k$ decreases from $k^{max}$ they start cutting the classical domain, hence
restricting the allowed values of $\nu$.
}\label{wessl-tronqu}
\end{figure}

{A useful remark is that the point $V_5$ (the ``lowest highest weight") 
always satisfies the BMW inequalities for any value of $k^{min}\le k\le k^{max}$ and thus always belongs to the 
truncated Wessl\'en's domain. \\
Proof : a tedious case by case verification using 
\be\label{h-V5} V_5=\begin{cases}  (\r+\s-\p-\q,\p-\s) &{\rm if\ } 0\le \p-\s\le \r-\q \\ 
                                                           (\p+\q-\r-\s, \r-\q) & {\rm if \ } 0\le \r-\q\le \p-\s \\ 
                                                            (\p-\s,\q-\r) &{\rm if\ } \r-\q<0\,.
   \end{cases}\ee
Intuitively, $V_5$   
``far away" from the colored lines of Fig.\ref{wessl-tronqu}. \\
In contrast all the other vertices of the original (``classical") tensor polygon may be removed  in some cases
and for some value of $k$. }


\subsubsection{The truncated Wessl\'en's domain for $k^{min} \le  k < k^{max}$}
As $k$ is incremented from $k^{min}$ by one unit, 
new points $\nu$ appear until $k=k^{max}$, the 
new points having multiplicity 1, {see \cite{BMW}}. The points $\nu$ organize themselves in a pattern of 
polygons included into one another with the 
multiplicity increasing from 1 on the outside polygon\footnote{When $k=k^{min}$ all the inner points have multiplicity $1$ too.} to a $k$-dependent maximum inside. 
At a given point $\nu$ inside one of the polygons, the multiplicity increases as $k$ grows, 
until $k$ reaches the value $k_0^{max}(\lambda,\mu;\,\nu)$.\\
This is illustrated on Table~\ref{table:polygons} for our favorite examples of $(9,5)\star (6,2)$ and $(9,5)\star (2,6)$. \\
%
\begin{table}[H]
\noindent
\begin{tabular}{l|c|c|c|}
\raisebox{25pt}{$k=14$}&\includegraphics[width=0.26\textwidth]{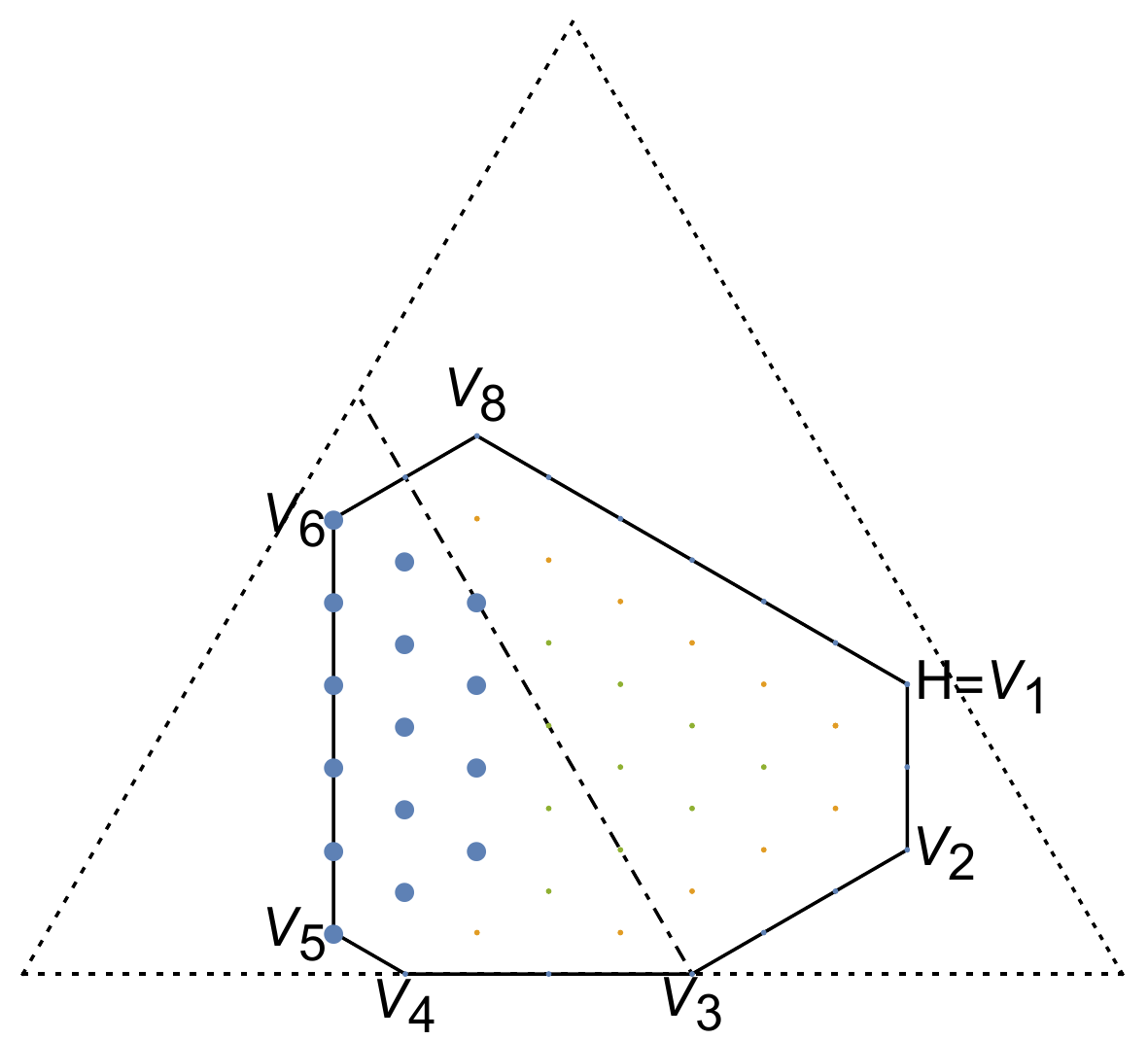}&\includegraphics[width=0.26\textwidth]{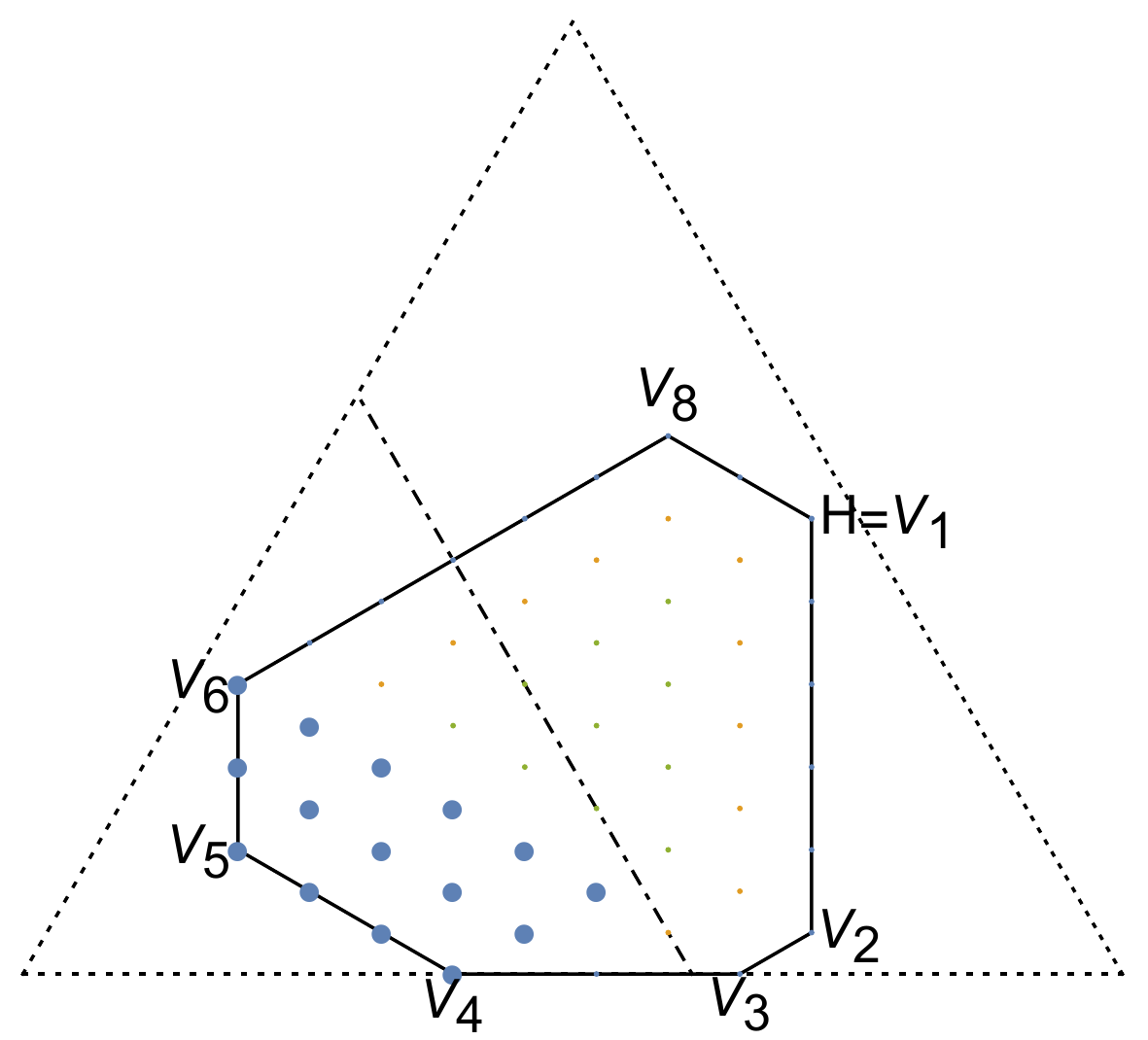}& 
\raisebox{25pt}{$\scriptstyle u_{00}=15$}\\
\raisebox{25pt}{$k=15$}&\includegraphics[width=0.26\textwidth]{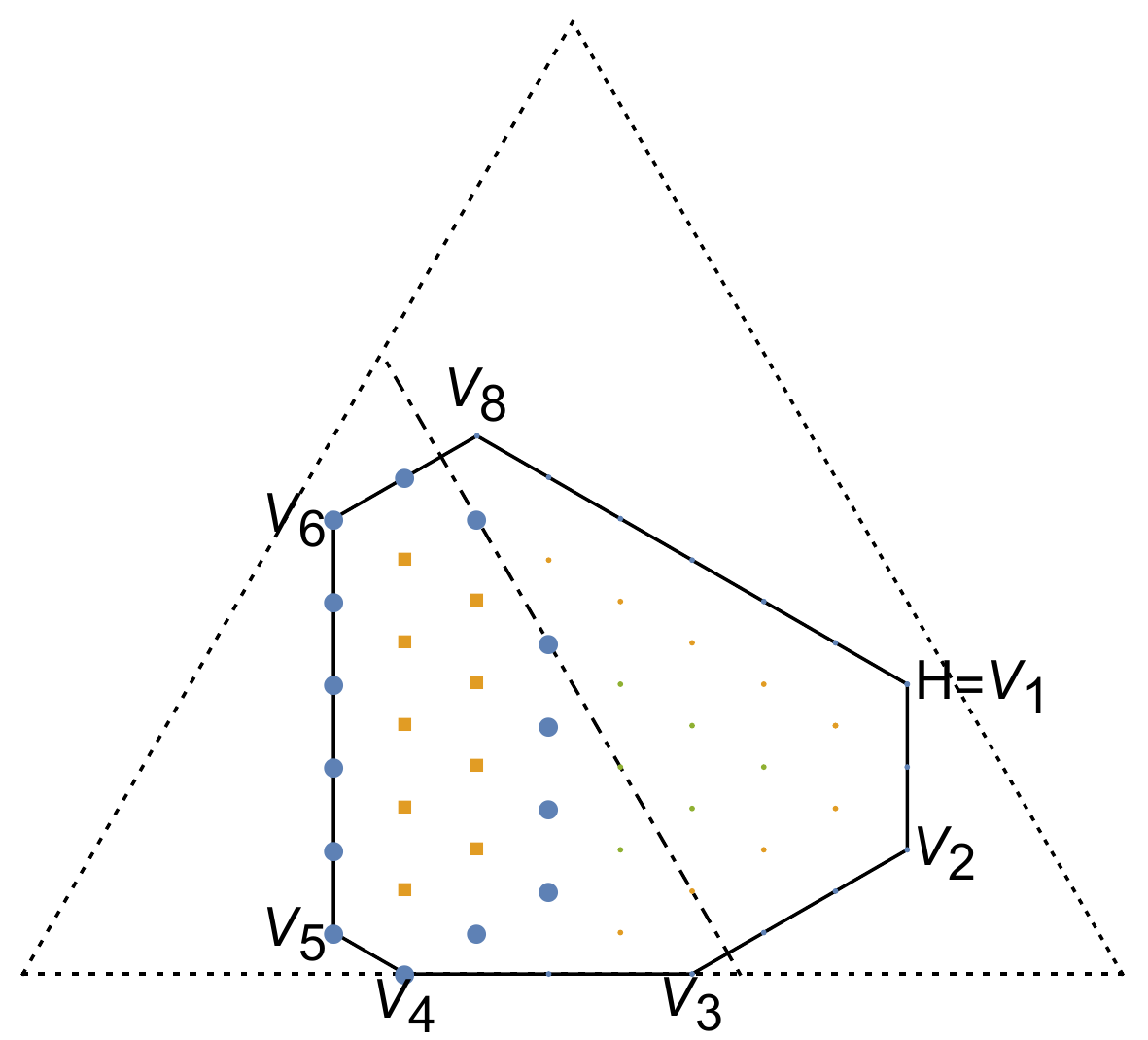}&\includegraphics[width=0.26\textwidth]{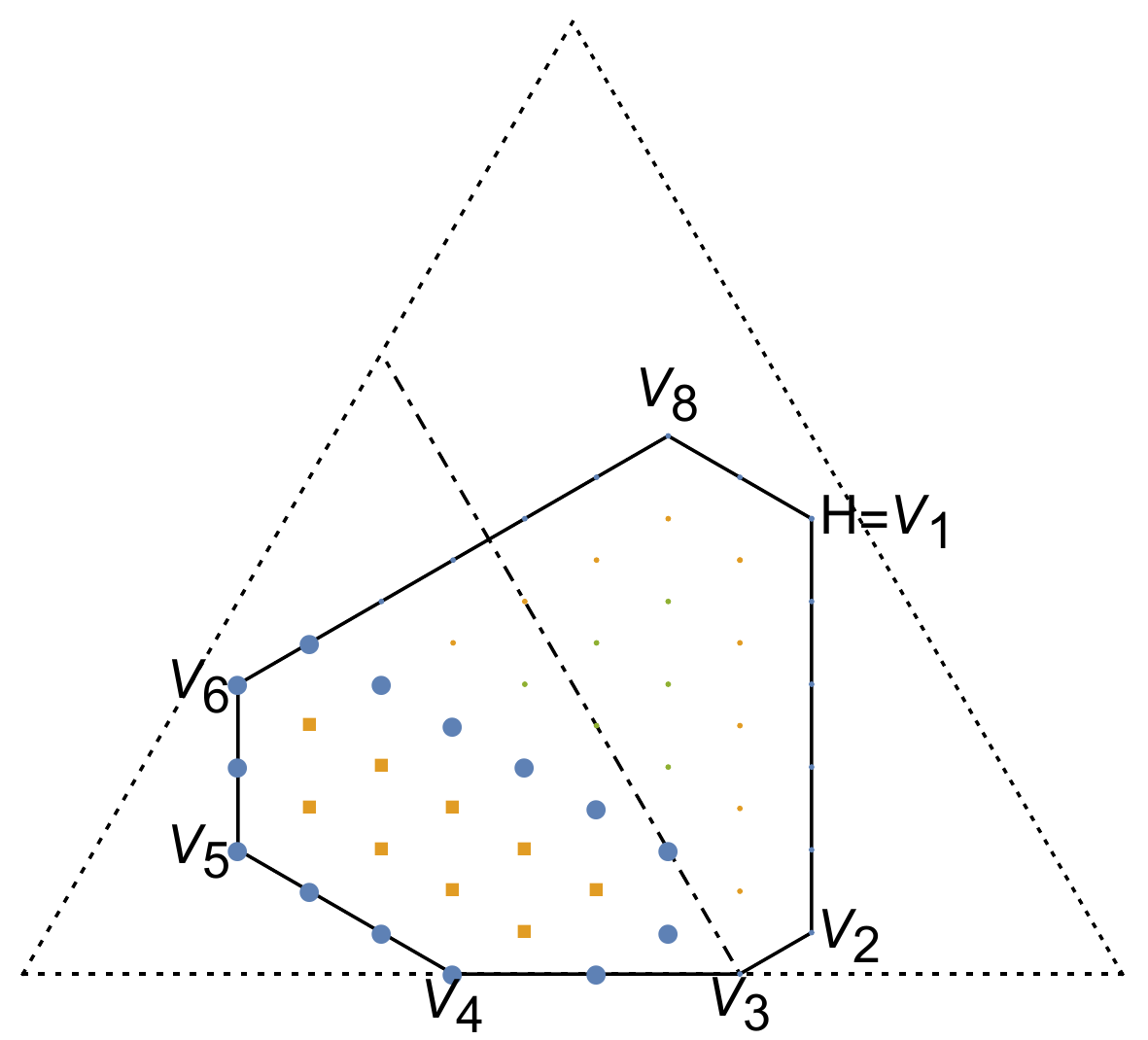}& \raisebox{25pt}{$\scriptstyle u_{10}=8\ u_{11}=9$}\\
\raisebox{25pt}{$k=16$}&\includegraphics[width=0.26\textwidth]{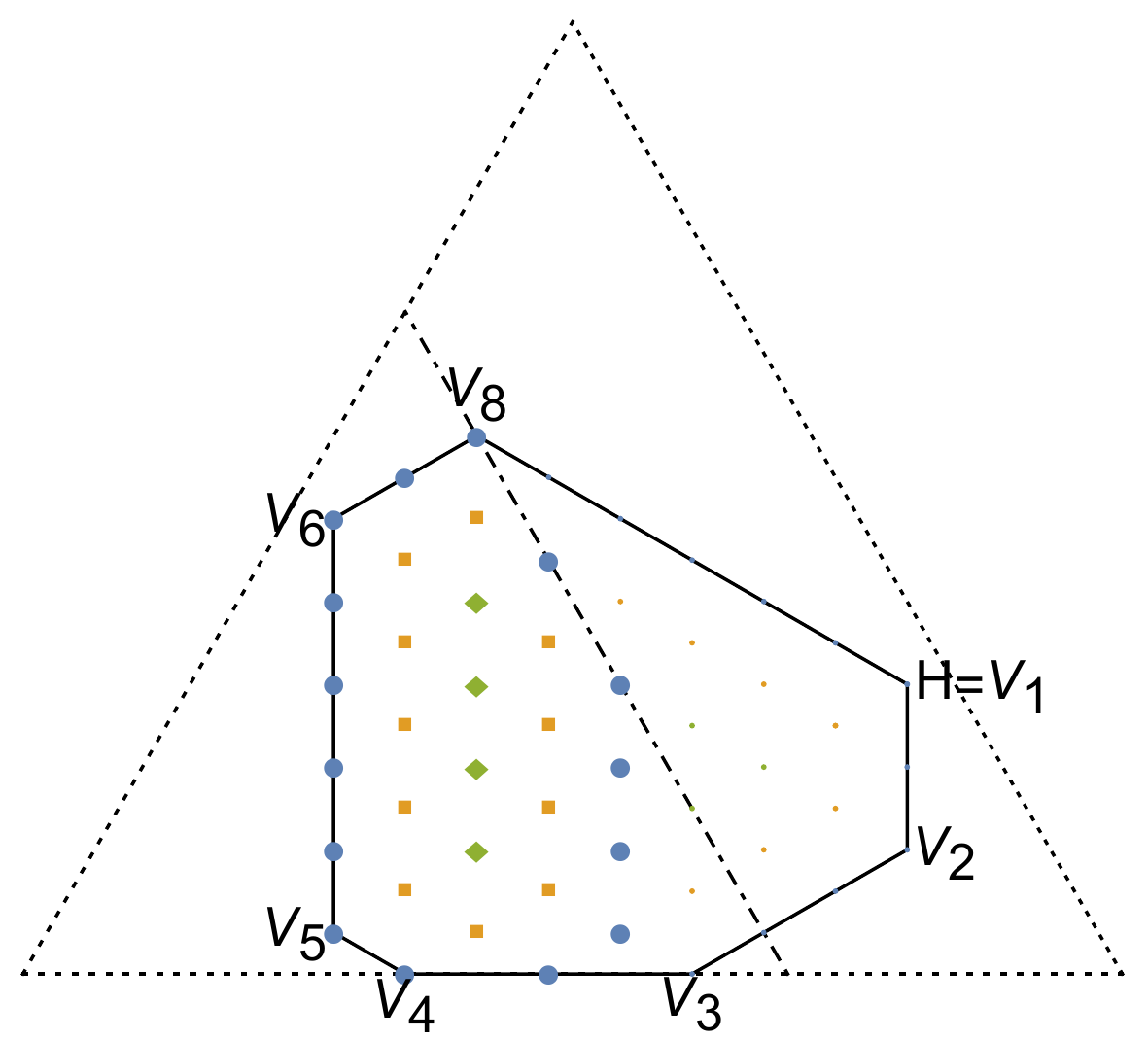}&\includegraphics[width=0.26\textwidth]{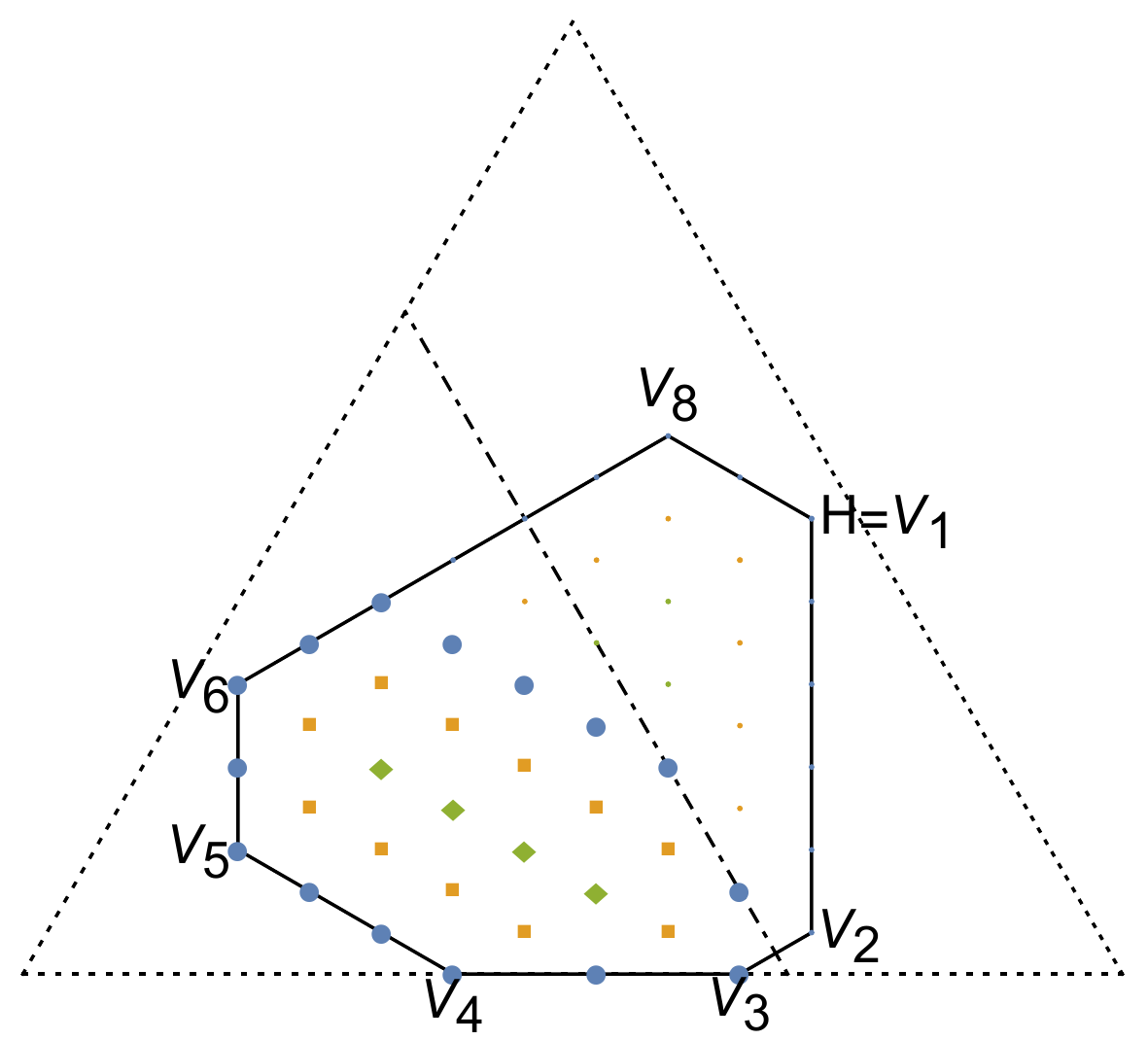}&\raisebox{25pt}{$\scriptstyle u_{20}=7\ u_{21}=6\ u_{22}=4$}\\
\raisebox{25pt}{$k=17$}&\includegraphics[width=0.26\textwidth]{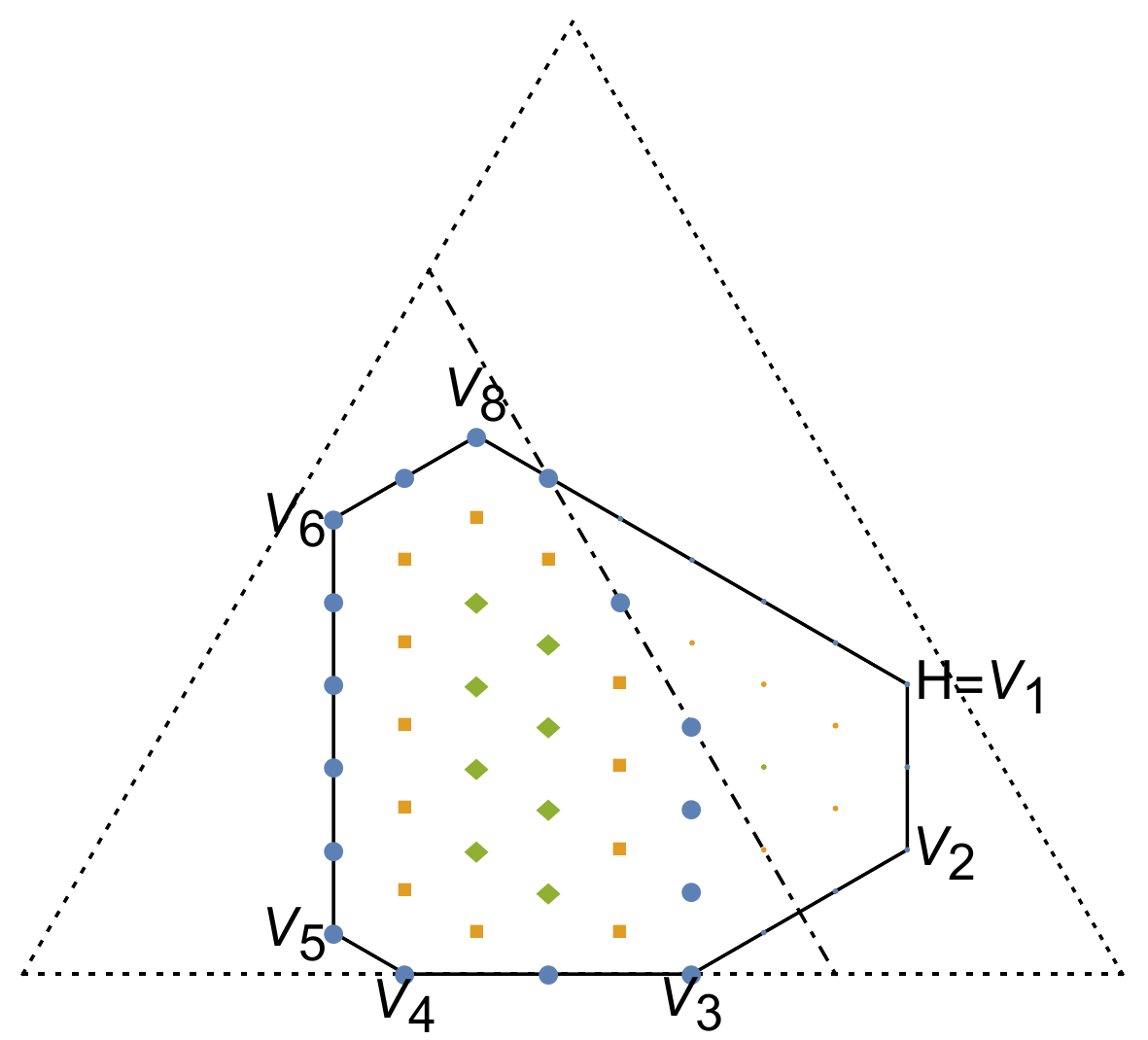}&\includegraphics[width=0.26\textwidth]{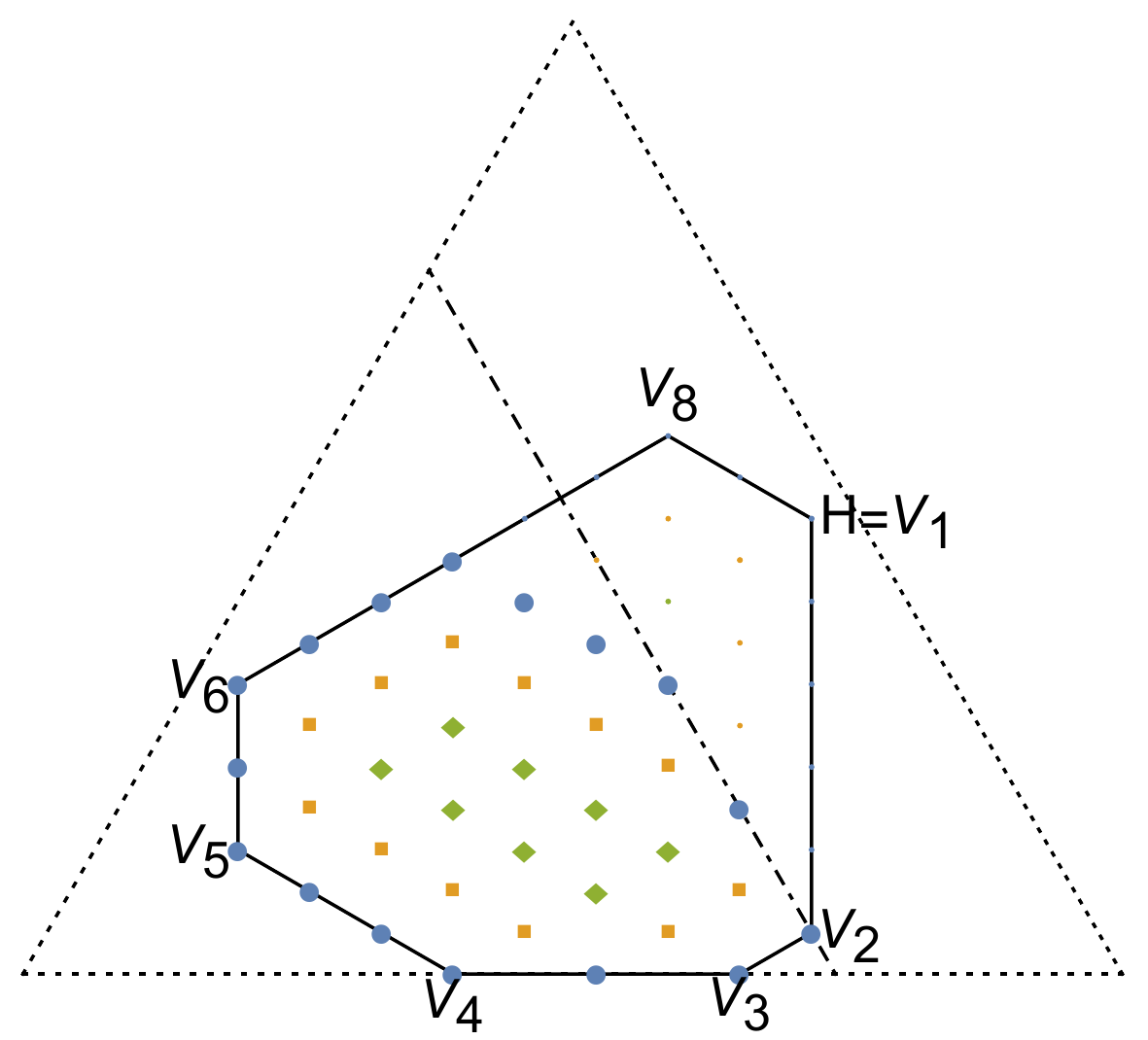}&\raisebox{25pt}{$\scriptstyle u_{30}=6\ u_{31}=5\ u_{32}=4$}
\end{tabular}
\end{table}

\begin{table}[H]
\begin{tabular}{l|c|c|c|}
\raisebox{25pt}{$k=18$}&\includegraphics[width=0.26\textwidth]{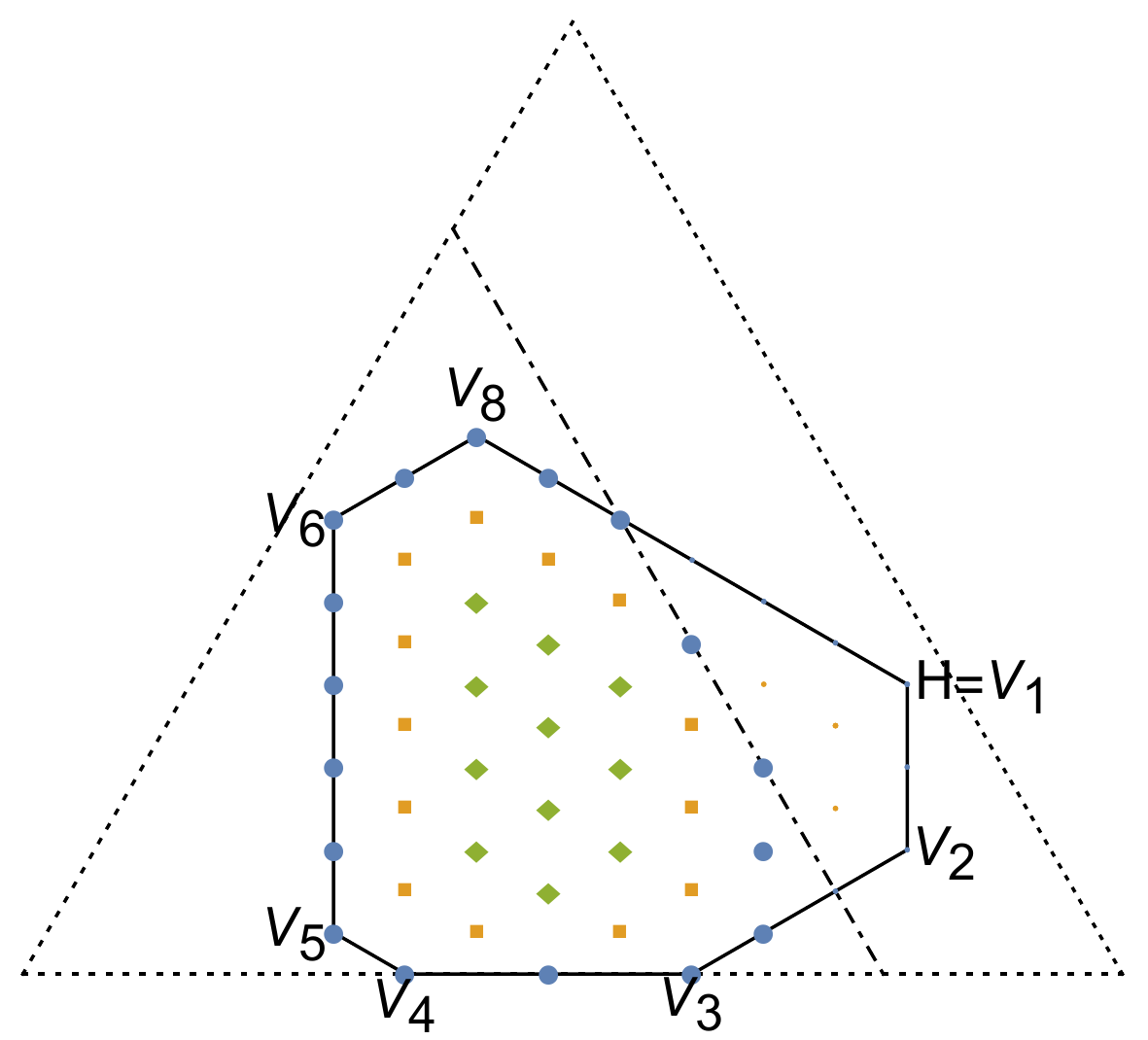}&\includegraphics[width=0.26\textwidth]{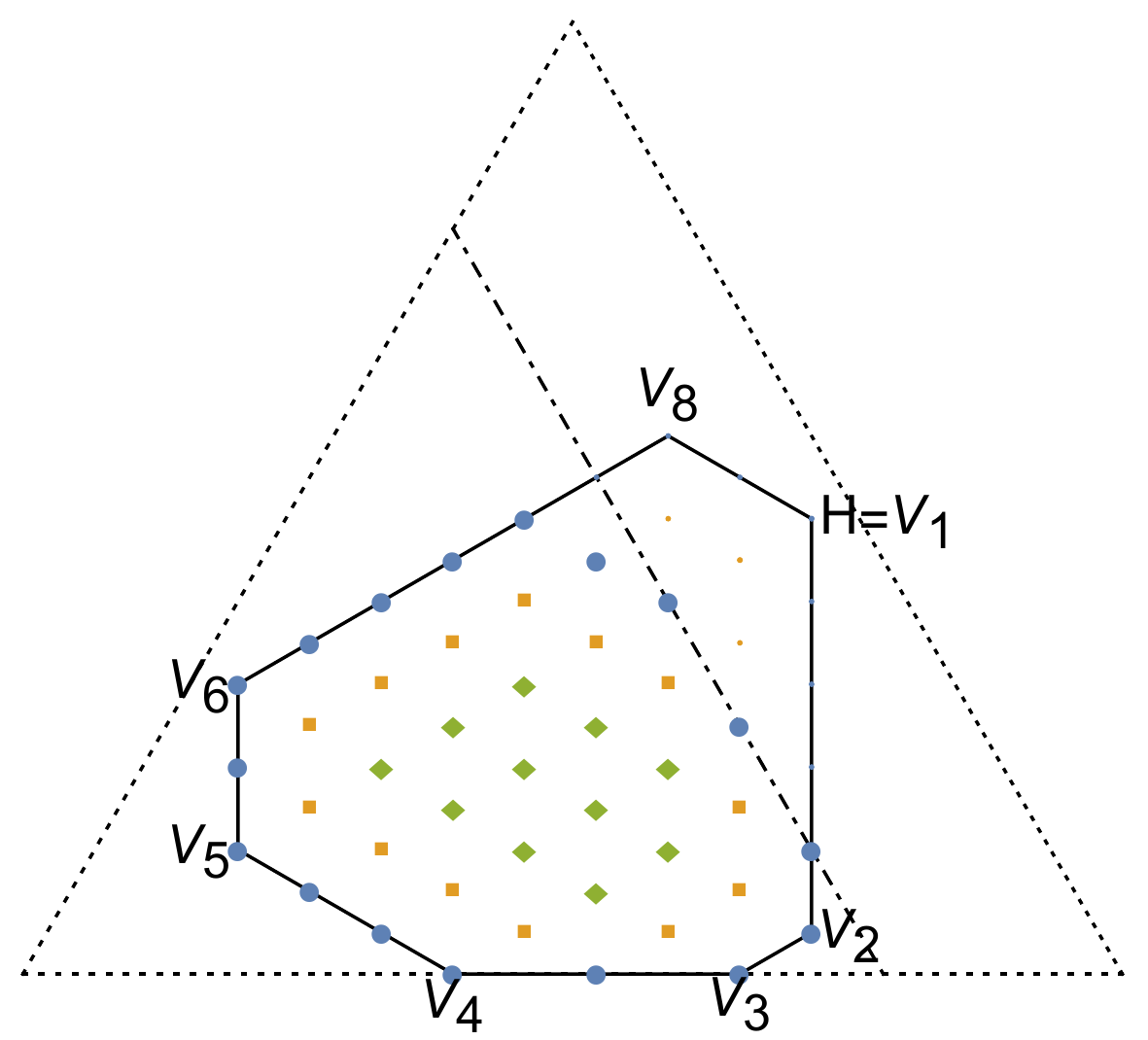}&\raisebox{25pt}{$\scriptstyle u_{40}=5\ u_{41}=4\ u_{42}=3$}
\\
\raisebox{25pt}{$k=19$}&\includegraphics[width=0.26\textwidth]{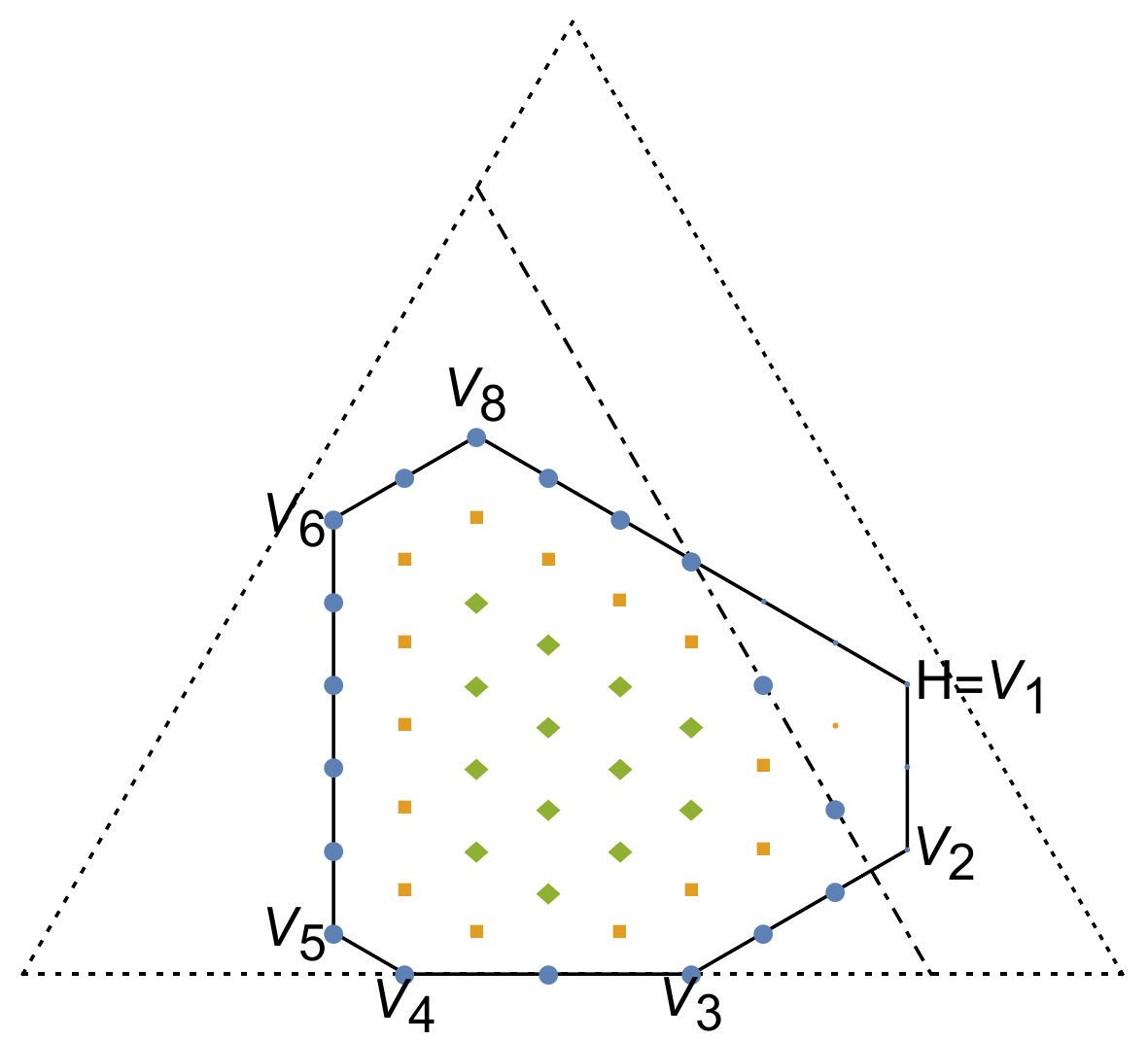}&\includegraphics[width=0.26\textwidth]{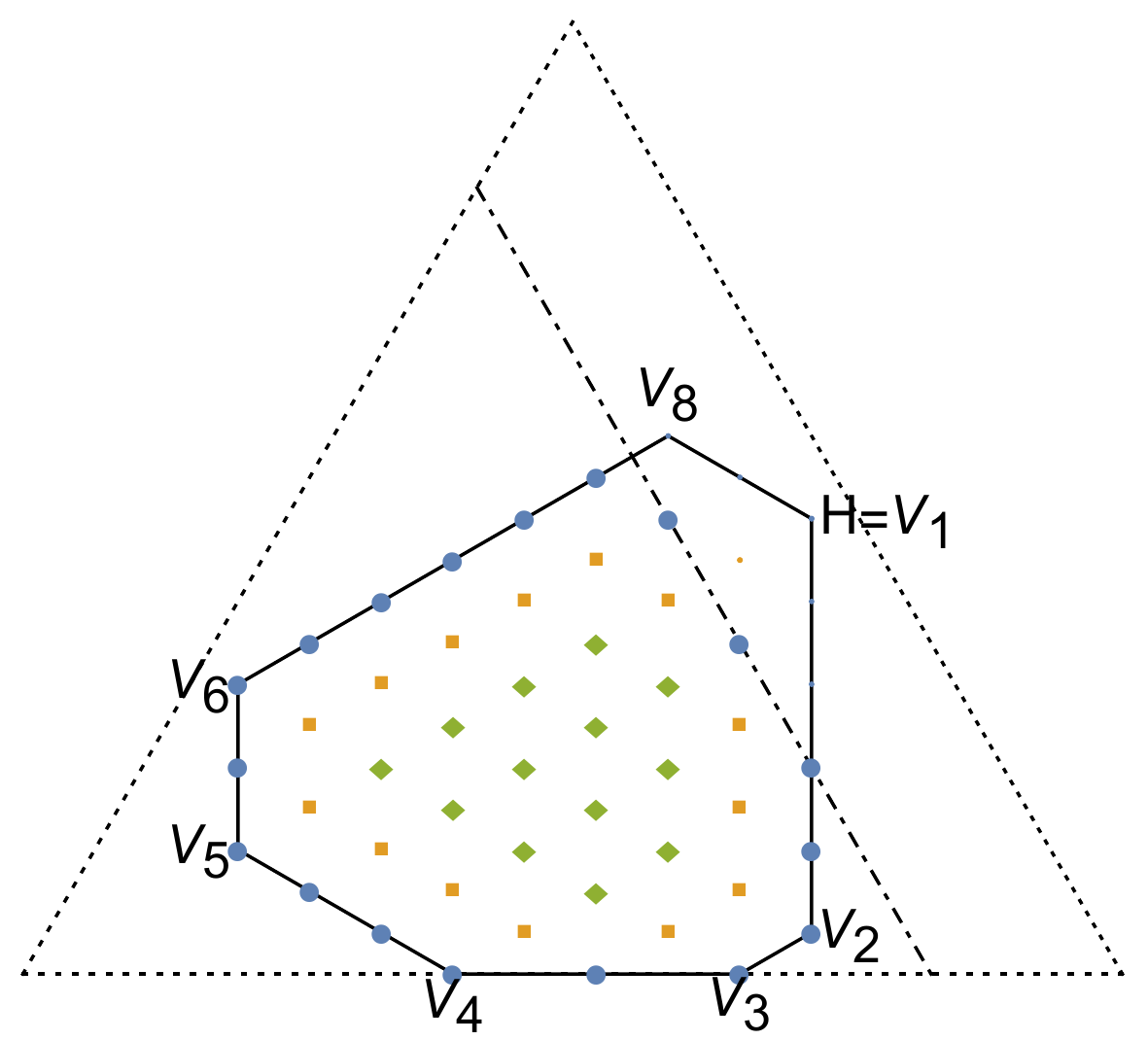}&\raisebox{25pt}{$\scriptstyle u_{50}=4\ u_{51}=3\ u_{52}=2$}
\\
\raisebox{25pt}{$k=20$}&\includegraphics[width=0.26\textwidth]{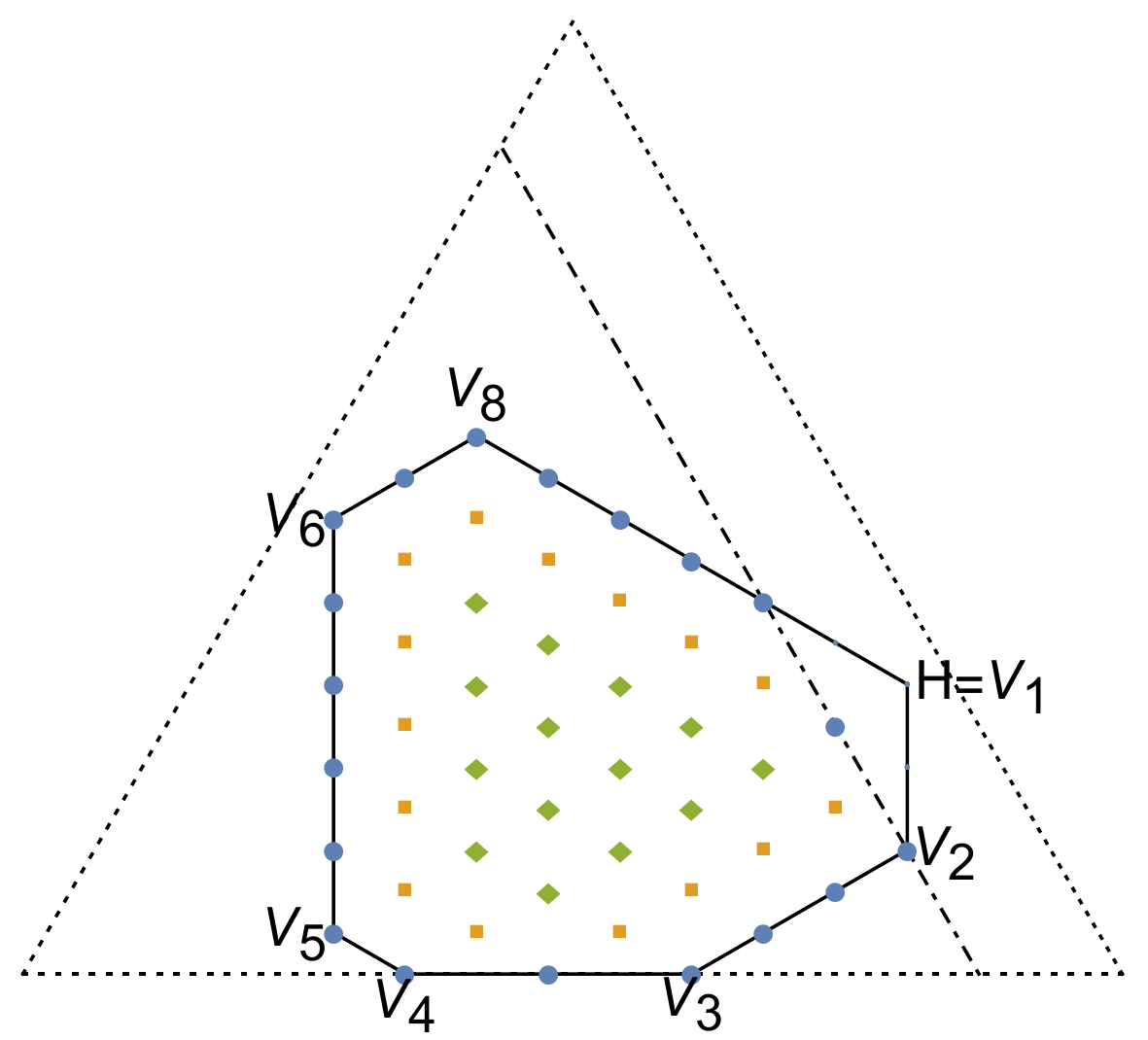}&\includegraphics[width=0.26\textwidth]{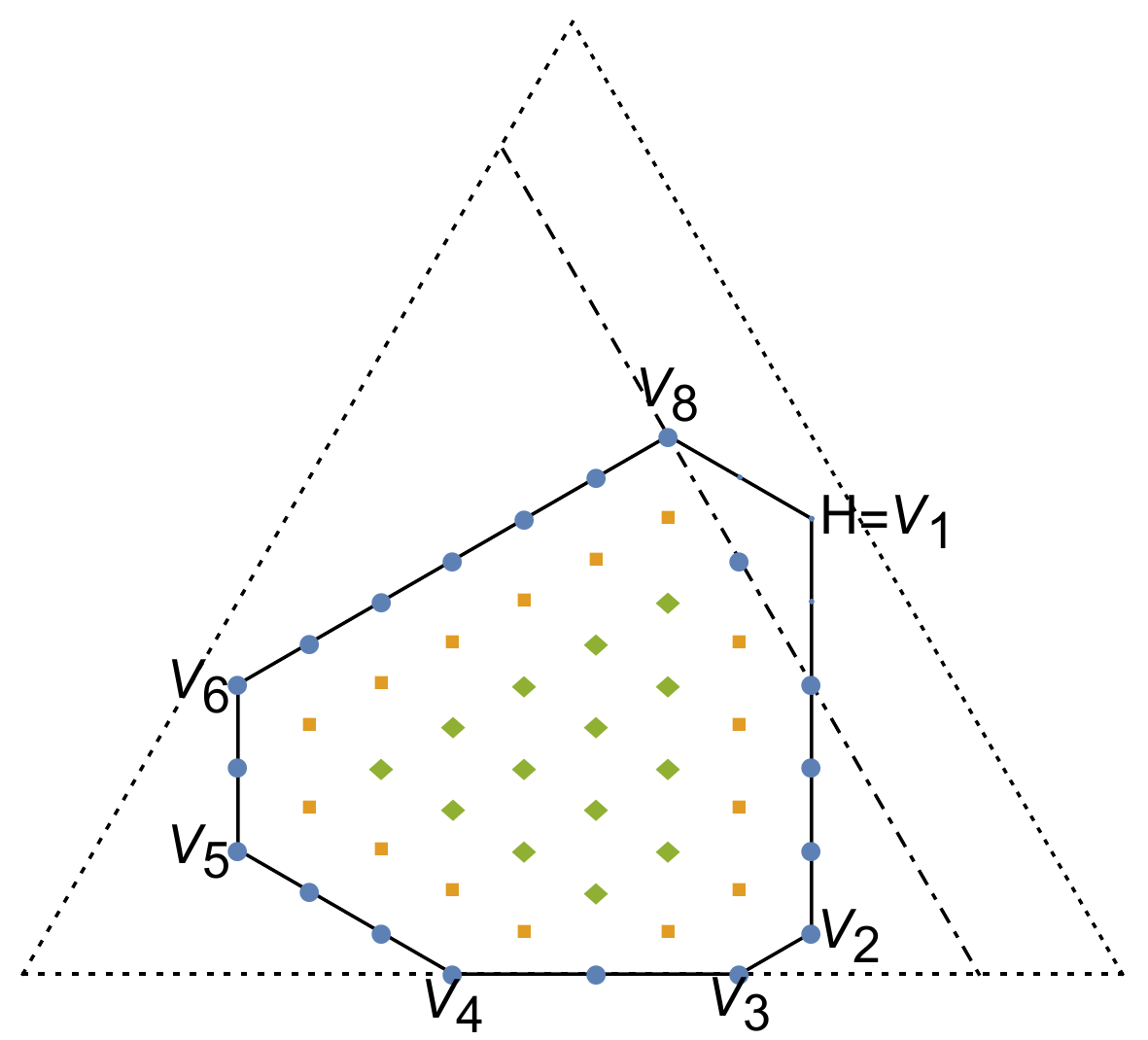}&\raisebox{25pt}{$\scriptstyle u_{60}=3\ u_{61}=2\ u_{62}=1$}
\\
\raisebox{25pt}{$k=21$}&\includegraphics[width=0.26\textwidth]{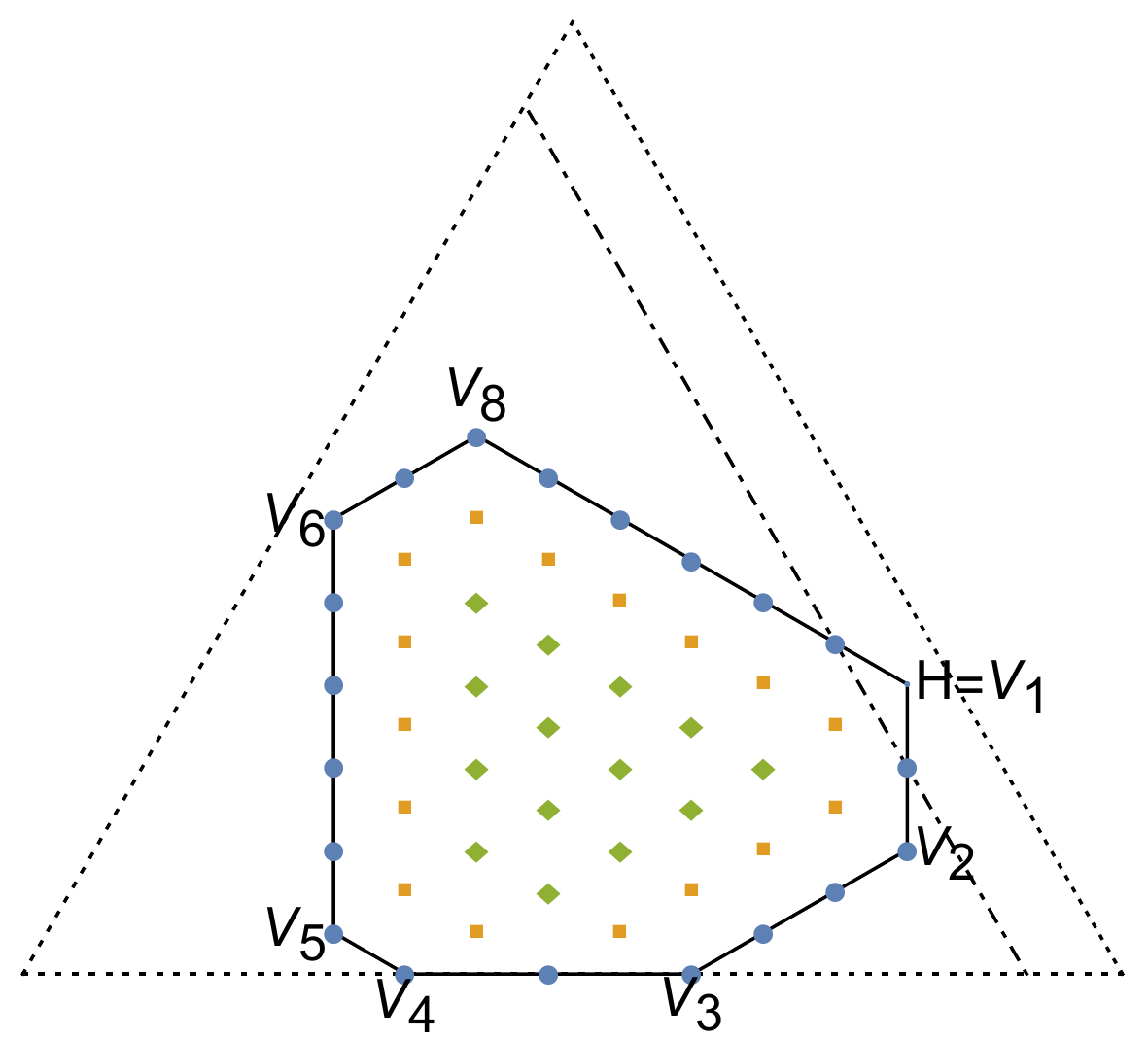}&\includegraphics[width=0.26\textwidth]{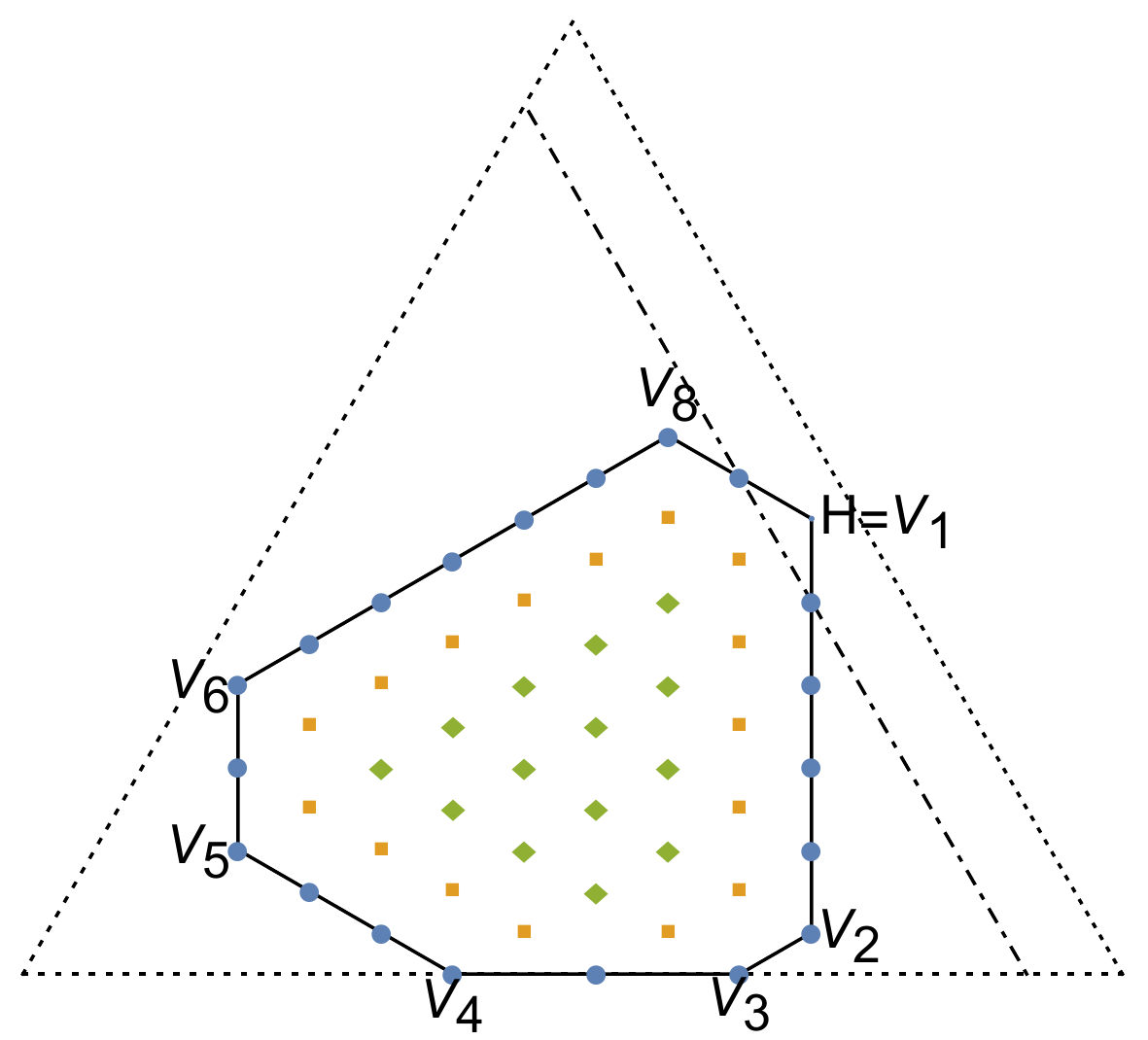}&\raisebox{25pt}{$\scriptstyle u_{70}=2\ u_{71}=1\ u_{72}=0$}
\\
\raisebox{25pt}{$k=22$}&\includegraphics[width=0.26\textwidth]{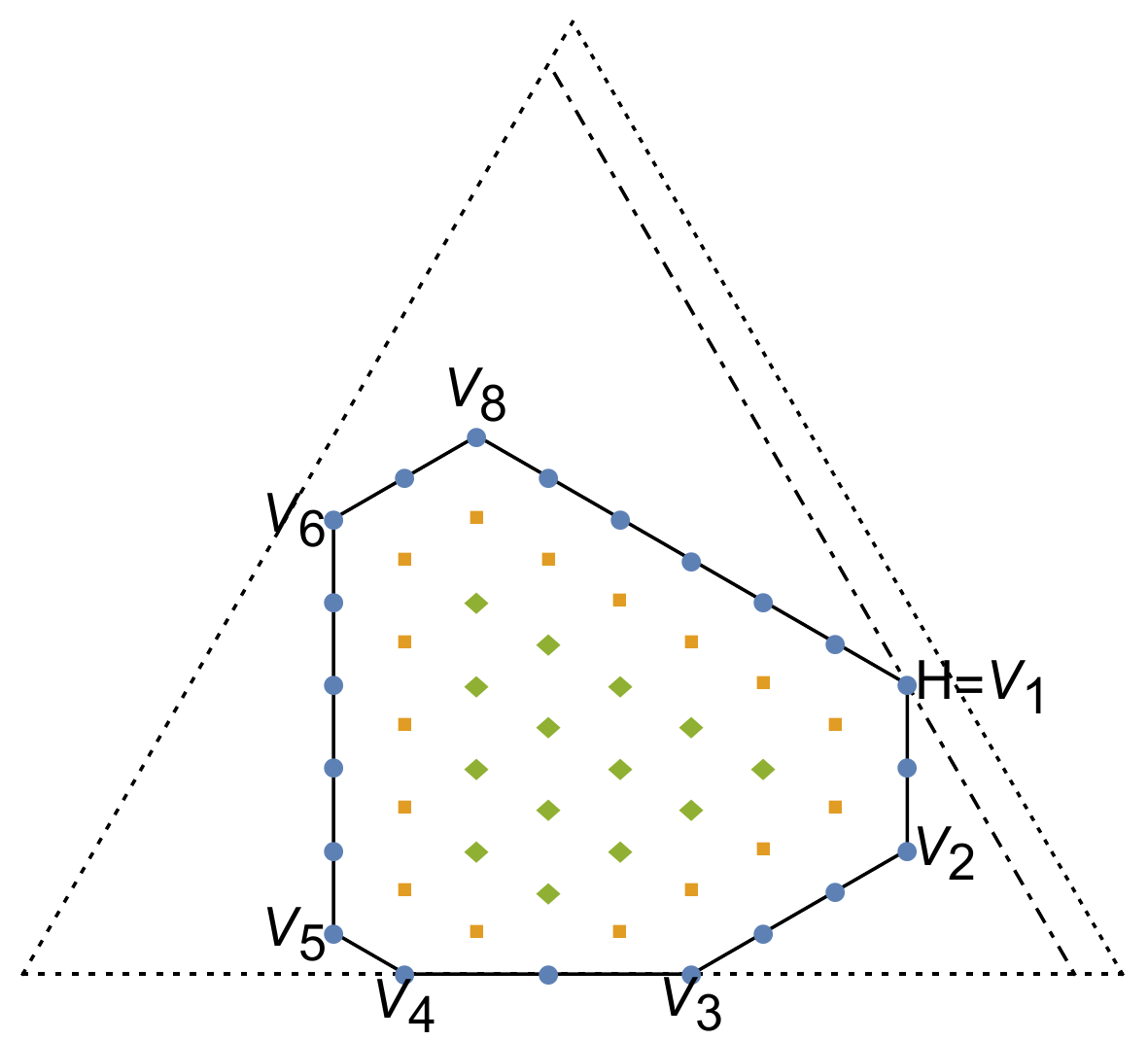}&\includegraphics[width=0.26\textwidth]{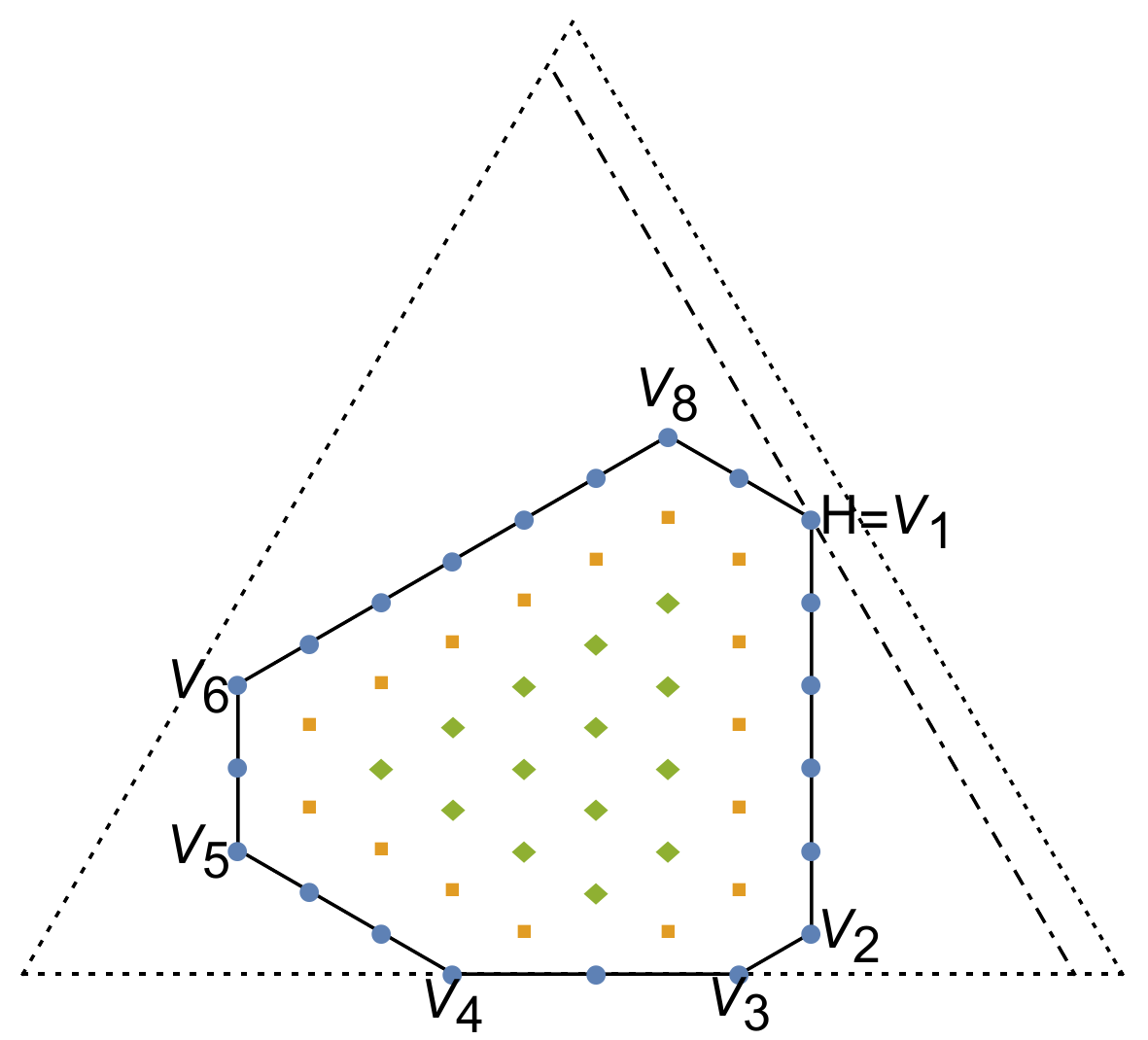}&\raisebox{25pt}{$\scriptstyle u_{80}=1\ u_{81}=0\ u_{82}=0$}

\end{tabular}
\caption{
{\small Truncated Wessl\'en domains in $\nu$ for the branchings $(9,5)\star(6,2)$ (left column) and $(9,5)\star(2,6)$ (right column),
 for increasing $k$. The solid polygon is the classical tensor polygon, drawn here in oblique axes. The dot-dashed line 
is the level line $\nu_1+\nu_2=k$. Points of multiplicity 1, 2 and 3 are displayed respectively in blue, orange and green.
The $u_{pj}$ of the last column are defined below in sect. \ref{proofofP} and can be read from the pattern of arrows in Fig.\;\ref{9562}.}}
\label{table:polygons}
\end{table}

Conversely as $k$ is decremented from $k^{max}$ by one unit, the inequality (\ref{BMW1}), and  possibly the
other inequalities,  start to exclude points from Wessl\'en's
domain. 
It is clear from Fig.\;\ref{wessl-tronqu} that inequalities (\ref{BMW2}),   (\ref{BMW3}),  (\ref{BMW4})$_L$,
 (\ref{BMW4})$_R$, are satisfied by all points of Wessl\'en's domain if they are satisfied respectively at points 
 $V_1=(\lambda_1+\mu_1, \lambda_2+\mu_2)$, $V_8=(\lambda_1-\mu_1, \lambda_2+\mu_1+\mu_2)$ 
 and $V_2=(\lambda_1+\mu_1+\min(\lambda_2,\mu_2), \lambda_2+\mu_2-2\min(\lambda_2,\mu_2)$.
 Thus 
  \bea\label{24p} { (\ref{BMW2})} {\rm \ is\ satisfied\ if \ }  && k\ge \lambda_1+\mu_1+\max(\lambda_2,\mu_2)\\
  \label{25p}{ (\ref{BMW3})} {\rm \ is\ satisfied\ if \ }  && k\ge 
 \lambda_1+ \lambda_2+\mu_2\\ 
 \label{26L}
 {(\ref{BMW4})_L} {\rm \ is\ satisfied\ if \ }  &&  k\ge \lambda_2+\mu_1+\mu_2\\ 
  \label{26R} {(\ref{BMW4})_R} {\rm \ is\ satisfied\ if \ }  &&   k\ge \lambda_1+\mu_1+\min(\lambda_2,\mu_2)\,.
  \eea
This introduces several successive boundary values of the level, as $k$ decreases from $k^{max}$.

\subsubsection{Proof of property $\mathfrak P$}
\label{proofofP}
As the level $k$ is decreased from $k^{min}+p$ to $k=k^{min}+p-1$, there are $u_{pj}$ terms in the decomposition of $\lambda\star\mu$  that have their multiplicity decreased from $j+1$ to $j$, for $j=0,\ldots, p$. In this section, we first give an explicit expression for $u_{p0}$, then use 
the recursion formula (\ref{recurN}) to extend it to all $u_{pj}$. We then observe that these expressions of $u_{pj}$ are invariant 
under $\mu\to \bar \mu$. Since we know from \cite{RCJBZ2014} that property  $\mathfrak P$ is satisfied for $k\ge k^{max}$, 
and since decrementing $k$ preserves it, it follows that it is satisfied for all $k$, qed.
\\[2pt]
\subsubsubsection{Computing the $u_{p0}$}\\[-5pt]
As level $k$ is decreased from $k^{min}+p$ to $k^{min}+p-1$, there are $u_{p0}$ terms that are excluded from the truncated 
Wessl\'en domain because they violate one of the inequalities.
According  to \cite{BMW}, they lie on the boundary of the truncated  Wessl\'en domain of level $k$. Which inequality is relevant for a
given $k$ depends on the relative values of $k, \lambda_1,\lambda_2,\mu_1$ and $\mu_2$. Recall that by convention, we
have $\lambda_1\ge \max(\lambda_2,\mu_1,\mu_2)$.\\
By a tedious analysis of all possibilities that is not be reproduced here, 
we have found the following expression for $u_{p0}$, $p=k-k^{min}$
\be \label{up0}\!\!\!\!\!\!\!\!\!\!\!\!\!\!\!\!\!\!\!\!\!\!\!\!\!\!\!\!\!
u_{p0}
=\begin{cases}  
\scriptstyle 0 &\scriptstyle {\rm if}\ \scriptstyle k > k^{max}\\
\scriptstyle  k^{max}-k+1  &\scriptstyle {\rm if}\ \max(\lambda_1 +\max(\mu_1,\mu_2),k^{min}+1)  \le k \le k^{max}=\lambda_1+\lambda_2+\mu_1+\mu_2  \\  
\scriptstyle \lambda_2+ \min(\mu_1,\mu_2)  +1 &\scriptstyle {\rm if}\ \scriptstyle
 {k^{min}+1 \le \lambda_1 +\max(\mu_1,\mu_2) \ {\rm and}}\ 
 {\max(\lambda_1 +\min(\mu_1,\mu_2),k^{min}+1)}\le k \le \lambda_1 +\max(\mu_1,\mu_2)\\
\scriptstyle k-\lambda_1+\lambda_2+1 &\scriptstyle {\rm if}\ \scriptstyle{k^{min}+1 \le \lambda_1 +\min(\mu_1,\mu_2) \ {\rm and}}\  k^{min}+1 \le k \le \lambda_1 +\min(\mu_1,\mu_2)\\
\cdots &\scriptstyle {\rm if}\ \scriptstyle k=k^{min}\\
\scriptstyle 0 &\scriptstyle {\rm if}\ \scriptstyle k<k^{min}
\end{cases}
\ee
Note the ``continuity" at the boundary values $\lambda_1 +\max(\mu_1,\mu_2)$ and $\lambda_1 +\min(\mu_1,\mu_2)$ of $k$.
\\
Note also that this formula {\it does not} yet determine the value of $u_{p0}$ for the lowest value $0$ of $p$ (viz $k=k^{min}$),
\ie the number of points of multiplicity 1 at level $k^{min}$. 
The latter, however,
is determined from the ``classical" expression given by Mandel'tsveig \cite{Mandeltsveig} and denoted $\sigma(\lambda,\mu; 1)$ in \cite{RCJBZ2014}
for the total number of $\nu$ of 
multiplicity $1$ (at levels higher than or equal to $k^{max}$), 
and from the successive $u_{p0}$:  
{\be u_{00}=\sigma(\lambda,\mu; 1) -\sum_{p=1}^{k^{max}-k^{min}} u_{p0}\,.\ee }
We observe that, as anticipated, the resulting expression for $u_{p0}$ is invariant under $\mu_1\leftrightarrow \mu_2$. \\[4pt]
\subsubsubsection{Computing the $u_{pj}$}\\ 
Computing the variation of the number of points of multiplicity $j+1\ge 2$ as $k$ decreases by one unit may look more
difficult. 
Fortunately, the recursive argument of sect. 5.2.1 
 comes to the rescue. For $k\ge 4$, the number of $\nu$ (with $\nu_i\ge 1$) 
in $\lambda\star\mu$ that are ``downgraded" from multiplicity $j>1$ to multiplicity $j-1$ as $k\to k-1$ equals the number of 
$\nu'$  in  $(\lambda-\rho)\star(\mu-\rho)$ that are downgraded from multiplicity $j-1$ to multiplicity $j-2$ as the level goes from 
$k-3$ to $k-4$. Whence an expression of $u_{p1}$ in terms of $u_{p-1\,0}$: 
indeed, under a shift of $\lambda$ and $\mu$ by $-\rho$ and of $k$ to $k-3$, one has $k^{max}\to k^{max}-4$ and $k^{min}\to k^{min}-2$, see eq.\;(\ref{defskmin-kmax}), 
thus $p=k-k^{min}\to p-1$)
\footnote{In order not to clutter the following equations with conditions involving $k^{min}$ (like in eq.\;(\ref{up0})) we read the 
consecutive lines as ``if clauses"  from top to bottom, assuming that they should be used only if $k^{min}$ obeys appropriate inequalities\dots}
\be\nonumber \!\!\!\!\!\!\!\!\!\!\!\!\!\!\!\!\!\!\!\!\!
\scriptstyle u_{p1}(\lambda,\mu)=u_{p-1,0}(\lambda-\rho,\mu-\rho)
=\begin{cases} \scriptstyle 0 & \scriptstyle {\rm if}\ k -3> k^{max}-4\\
 \scriptstyle (k^{max}-4)-(k-3)+1  & \scriptstyle {\rm if}\ \lambda_1 +\max(\mu_1,\mu_2)-2\le k-3 \le k^{max}-4  \\ 
\scriptstyle \lambda_2+ \min(\mu_1,\mu_2) -2+1  &\scriptstyle {\rm if}\ \lambda_1 +\min(\mu_1,\mu_2)-2\le k-3 \le \lambda_1 +\max(\mu_1,\mu_2)-2\\
\scriptstyle k-\lambda_1+\lambda_2-3+1 & \scriptstyle{\rm if}\ k^{min}-2+1 \le k-3 \le \lambda_1 +\min(\mu_1,\mu_2)-2\\ 
\scriptstyle \cdots &\scriptstyle {\rm if } \ \scriptstyle k-3=k^{min}-2 \\
\scriptstyle 0 & \scriptstyle {\rm if }\   k-3< k^{min}-2  
\end{cases}
\ee
or in other words
\be u_{p1}(\lambda,\mu)
=\begin{cases} 0 & {\rm if}\ k > k^{max}-1\\
 k^{max}-k  & {\rm if}\ \lambda_1 +\max(\mu_1,\mu_2) +1\le k \le k^{max}-1  \\ 
 \lambda_2+ \min(\mu_1,\mu_2) -1  & {\rm if}\ \lambda_1 +\min(\mu_1,\mu_2)+1\le k \le \lambda_1 +\max(\mu_1,\mu_2)+1\\
k-\lambda_1+\lambda_2-2 & {\rm if}\ k^{min}+2 \le k \le \lambda_1 +\min(\mu_1,\mu_2)+1\\ 
\cdots & {\rm if } \ k=k^{min}+1 \\
0 & {\rm if }\   k< k^{min}+1  
\end{cases}
\ee
and more generally
\be \!\!\!\!
u_{pj}(\lambda,\mu)
=\begin{cases} 0 & {\rm if}\ k > k^{max}-j\\
k^{max}-k-j+1  & {\rm if}\ \lambda_1 +\max(\mu_1,\mu_2) +j\le k \le k^{max}-j  \\ 
 \lambda_2+ \min(\mu_1,\mu_2) -2j+1  & {\rm if}\ \lambda_1 +\min(\mu_1,\mu_2)+j\le k \le \lambda_1 +\max(\mu_1,\mu_2)+j\\
k-\lambda_1+\lambda_2-3j+1 & {\rm if}\ k^{min}+j+1 \le k \le \lambda_1 +\min(\mu_1,\mu_2)+j\\ 
\cdots & {\rm if } \ k=k^{min}+j \\
0 & {\rm if }\   k< k^{min}+j  
\end{cases}
\ee
Again the missing value of $u_{pp}$ may be derived from Mandel'tsveig's formula  {(see \cite{Mandeltsveig}, \cite{RCJBZ2014})} 
{giving $\sigma(\lambda,\mu; j+1)$, 
the ``classical" number of $\nu$'s of
multiplicity $j+1$: 
$$u_{pp}=\sigma(\lambda,\mu; {j+1})-\sum_{p=j+1}^{k^{max} - k^{min}} u_{pj}\,.$$}
We observe that this expression of $u_{pj}$ is again invariant under $\mu\leftrightarrow \bar\mu$. As explained at the beginning of
this section, this completes the proof of property $\mathfrak P$.


\section{Miscellanea}
\label{miscellanea}
\subsection{From $\lambda\otimes\mu$ to $\lambda\otimes\overline{\mu}$ : other approaches}

Admittedly the previous proof of property $\mathfrak{P}$, with its brute force calculation of the number of points at a given level,
lacks elegance and simplicity. We thus attempted to explore other approaches\dots Although unsuccessful, as far as leading to another proof of the above property, some of these investigations lead to other results that may have a separate interest,  we gather them here.


\subsubsection{Proof of property $\mathfrak{P}$ in some particular cases (elementary approach)}

In addition to the ``linear sum rule" proved in \cite{RCJBZ2011} (equality of the total multiplicities for $\lambda\otimes\mu$ and $\lambda\otimes\overline{\mu}$ at each level),
we recall from sect. \ref{summaryonmutiplicities} the existence of a 
``quadratic sum rule": the sum of squares of multiplicities are also the same at each level.
These two sum rules put constraints on the two lists of multiplicities.
\\[8pt]
As in section \ref{proofofP},  and for $j=0,\ldots, p$, we call $u_{pj}$ (resp $v_{pj}$)
the number of terms in the decomposition of $\lambda\otimes\mu$ (resp $\lambda\otimes\overline{\mu}$)
that have their multiplicity increased from $j$ to $j+1$ when the level is increased from $k^{min}+p-1$ to $k^{min}+p$.
Keeping track of the increase in multiplicities at each level, the reader will have no difficulty to show that the two sum rules imply:
$$\sum_{j=0}^{p} u_{pj} = \sum_{j=0}^{p} v_{pj} \qquad  \text{and} \qquad \sum_{j=0}^{p} ((j+1)^2 - j^2)u_{pj} = \sum_{j=0}^{p} ((j+1)^2 - j^2) v_{pj}\,. $$
The general result obtained in sect. \ref{proofofP}, namely that $u_{pj} = v_{pj}$ for all $p$ and all $j$,  cannot be deduced from the two sum rules recalled above. However this equality can be immediately obtained at levels $k=k^{min}$, $k=k^{min}+1$, and $k\geq k^{max}$. 
Indeed, when $k=k^{min}+1$ \ie $p=1$, the two previous  equations read, equivalently,  $u_{10}-v_{10} = - (u_{11}-v_{11})$  and $u_{10}-v_{10} = - 3 (u_{11}-v_{11})$, which implies $u_{10}=v_{10}$ and $u_{11}=v_{11}$.  When $k=k^{min}$ (multiplicities are all equal to $1$)  and when $k\geq k^{max}$ (multiplicities have their classical values), the result is obvious since it is just a way to re-express already known results.

The approach described in the present section provides therefore an elementary proof of property $\mathfrak{P}$ when $k=k^{min}$, $k^{min}+1$, and $k=k^{max}$.
By the same token, it also gives a proof for all values of $k$ if the highest weights $\lambda$ and $\mu$ are such that $k^{max} - k^{min} = 0,1$ or $2$.


\subsubsection{Using automorphisms.}

The group of automorphisms of the affine version of the $A_2$ Dynkin diagram is $\ZZ_3$.
Let $(\lambda_1,\lambda_2)$ be the components of a weight $\lambda$ on the basis of fundamental weights.
If the level is $k$ one introduces an affine component $\lambda_0 = k - (\lambda_1+\lambda_2)$ and consider the affine weight $\hat\lambda$ with components $(\lambda_0, \lambda_1,\lambda_2)$. The generator $\zeta$ of $\ZZ_3$ acts on affine weights as follows: 
$\zeta (\lambda_0, \lambda_1,\lambda_2) = (\lambda_2,\lambda_0, \lambda_1)$. Obviously, $\zeta^3=\one$.
Existence of automorphisms $a,b$ imply $N_{a\lambda,\, b\mu}^{(k) \; ab\nu}=N_{\lambda, \mu}^{(k)\; \nu}$. In the present case, $a$ and $b$ can taken as $\zeta, \zeta^2$ or $\one$ and it is understood that automorphisms act on the affine extension of the weights $\lambda, \mu, \nu$, although their affine component (their first component) is usually dropped from the notation: in other words $\zeta (\lambda_1,\lambda_2) =  (\lambda_0, \lambda_1)$, $\zeta^2 (\lambda_1,\lambda_2) =  (\lambda_2, \lambda_0)$, \etc

We consider a given branching $\lambda \otimes \mu \rightarrow \nu$. 
The affine weights at level $k$ are 
$\hat \lambda=(k - (\lambda_1+\lambda_2),  \lambda_1, \lambda_2)$, 
$\hat \mu=(k - (\mu_1+\mu_2),  \mu_1, \mu_2)$, 
$\hat \nu=(k - (\nu_1+\nu_2),  \nu_1, \nu_2)$.
Let us choose the level $k$ in such a way that $k - (\mu_1+\mu_2) = \mu_1$, so we take $k=2\mu_1+\mu_2$. 
Notice that $\hat \mu = (\mu_1, \mu_1, \mu_2)$, therefore $\zeta^2 (\mu_1, \mu_1, \mu_2)=(\mu_1,\mu_2, \mu_1)$.
Using automorphisms $a=\one, b=\zeta^2$, 
one has  $N_{\lambda,\, \zeta^2 \mu}^{(k) \; \zeta^2 \nu}=N_{\lambda, \mu}^{(k)\; \nu}$, \ie
\be
\label{autom1}
k=2\mu_1+\mu_2, \qquad
N_{(\lambda_1,\lambda_2),\, (\mu_2,\mu_1)}^{(k) \; (\nu_2, 2\mu_1+\mu_2 - (\nu_1+\nu_2))}=N_{(\lambda_1,\lambda_2), (\mu_1, \mu_2)}^{(k)\; (\nu_1, \nu_2)}\,.
\ee
In order for the components of the weights (including the affine component) to stay non negative, one needs to assume 
$2\mu_1+\mu_2 \geq \lambda_1+\lambda_2$ and $2 \mu_1 + \mu_2 \geq \nu_1+\nu_2$.

In the same way, assuming $2\mu_2+\mu_1 \geq \lambda_1+\lambda_2$ and $2 \mu_2 + \mu_1 \geq \nu_1+\nu_2$, one chooses the level $k$ in such a way that 
 $k - (\mu_1+\mu_2) = \mu_2$, so $k=2\mu_2+\mu_1$.  
 With $\hat \mu = (\mu_2, \mu_1, \mu_2)$,  we have $\zeta (\mu_2, \mu_1, \mu_2)=(\mu_2,\mu_2, \mu_1)$.
 Using automorphisms $a=\one, b=\zeta$, one gets  $N_{\lambda,\, \zeta \mu}^{(k) \; \zeta \nu}=N_{\lambda, \mu}^{(k)\; \nu}$, explicitly: 
\be
\label{autom2}
k=2\mu_2+\mu_1, \qquad
N_{(\lambda_1,\lambda_2),\, (\mu_2,\mu_1)}^{(k) \; (2\mu_2+\mu_1 - (\nu_1+\nu_2), \nu_1)}=N_{(\lambda_1,\lambda_2), (\mu_1, \mu_2)}^{(k)\; (\nu_1, \nu_2)}
\ee
It is clear that the two above transformations are inverse of one another.

We now use $mult(\lambda,\mu; \nu) = mult(\mu,\lambda; \nu)$ and $mult(\mu, \overline{\lambda}; \nu^\prime) = mult(\overline{\mu}, \lambda; \overline{\nu^\prime}) = mult(\lambda, \overline \mu; \overline{\nu^\prime})$.\\
Assuming  $2\lambda_2+\lambda_1 \geq \mu_1+\mu_2$ and  $2\lambda_2+\lambda_1 \geq \nu_1+\nu_2$, one obtains in the same way 
\be
\label{autom3}
 k=2 \lambda_2+\lambda_1, \qquad
 N_{(\lambda_1,\lambda_2),\, (\mu_2,\mu_1)}^{(k) \; (\nu_1, 2\lambda_2+\lambda_1 - (\nu_1+\nu_2))}=N_{(\lambda_1,\lambda_2), (\mu_1, \mu_2)}^{(k)\; (\nu_1, \nu_2)}
\ee
And, assuming $2\lambda_1+\lambda_2 \geq \mu_1+\mu_2$ and  $2\lambda_1+\lambda_2 \geq \nu_1+\nu_2$
\be
\label{autom4}
 k=2 \lambda_1+\lambda_2, \qquad
 N_{(\lambda_1,\lambda_2),\, (\mu_2,\mu_1)}^{(k) \; (2\lambda_1+\lambda_2 - (\nu_1+\nu_2), \nu_2)}=N_{(\lambda_1,\lambda_2), (\mu_1, \mu_2)}^{(k)\; (\nu_1, \nu_2)}
\ee
 For illustration, we apply eq.\;(\ref{autom1})  to our former example $\lambda = (9,5),\; \mu=(6,2)$.
 One checks that the chosen automorphism applied to the list of weights $\nu$ appearing on the line\footnote{this is also equal to $k^{min}(\lambda, \mu)$, but it is an accident.} $k=2\mu_1+\mu_2=14$ of Fig.\;\ref{9562} gives, up to reordering, the line $k=14$ of table \ref{9526}. In the same way, the equality of multiplicities for the appropriate triples also holds if we use eq.\;(\ref{autom4})  (now, $k$ has to be chosen as $2\lambda_2+\lambda_1=19$).
 
 Although giving non-trivial results for particular values of the level, we do not see how to generalize this approach to handle the general case.


\subsubsection{Using a piece-wise linear map in the space of weights}

In reference \cite{RCJBZ2014} several proofs of property $\mathfrak P$ in the classical case, \ie for tensor products, were given. One of them was based on the construction of an involutive piece-wise linear map $T$ from the set of admissible couplings (or of their corresponding pictographs) associated with the various branchings  $\lambda\otimes\mu \rightarrow \nu$ to the set of couplings associated with the branchings $\lambda\otimes \overline{\mu}  \rightarrow \nu^\prime$.  The two weights $\lambda, \mu$ being fixed,  this particular transformation $T :  (\nu,\alpha) \mapsto (\nu^\prime, \alpha^\prime)$,  with $\alpha$ denoting some coupling (or some pictograph) for the triple $\lambda, \mu, \nu$, cannot be used in the present situation where we deal with fusion product at level $k$ because it does not respect the thresholds: 
in order to prove property $\mathfrak P$ for fusion products, one should exhibit a bijective piecewise-linear transformation $T$ such that the couplings defined by $\nu,\alpha$ and $\nu^\prime, \alpha^\prime$ have the same threshold.

Taking into account the specificities of SU(3), in particular  the fact that the multiplicities of a given branching increase by one unit when the level $k$ runs from $k_0^{\min}$ to $k_0^{\max}$, it is enough to look for an invertible map $T$ that 
is compatible with these two values of the level.
In other words, if $T$ is a one-to-one map from the set of admissible triples $(\lambda, \mu, \nu)$ to the set of admissible triples $(\lambda, \overline\mu, \nu^\prime)$  and is such that 
$k_0^{\min}(\lambda, \mu, \nu) = k_0^{\min}(\lambda, \overline\mu, \nu^\prime)$ and $k_0^{\max}(\lambda, \mu, \nu) = k_0^{\max}(\lambda, \overline\mu, \nu^\prime)$, the theorem is proved.

Unfortunately, many particular cases have to be considered, and the discussion leading to a definition of such a piece-wise linear map seems to be as complicated  as the one leading to the proof of property $\mathfrak P$ in sect. \ref{proofofP}.


\subsubsection{Using the matrix polynomial $X(s,t)$ to prove property $\mathfrak{P}$ (a failed attempt) }
The property $\mathfrak P$ can be rephrased as follows: for given weights $\lambda$ and $\mu$ and level $k$, the
  multisets\footnote{\ie sets with multiplicities, or lists, up to order. As we work with a fixed level $k$ in this section, we drop this index from the notation used for fusion matrices.} 
  $\{ N_{\lambda\mu}^\nu\}$ and  $\{N_{\lambda\bar\mu}^{\nu'}\} $  are identical
and we may use the conjugation properties of the fusion coefficients
$$ N_{\lambda\bar\mu}^{\nu'}=N_{\bar\lambda\mu}^{\bar\nu'}=N_{\lambda\bar\nu'}^\mu$$
to write this statement in  equivalent forms. 
Thus $\{N_{\lambda\mu}^\nu\}=\{N_{\lambda\bar\nu'}^{\mu}\}$ means that

{\sl For an arbitrary $\widehat{\su}(3)_k$ fusion matrix $N_{\lambda}$, 
the content of any row $\mu$ 
is the same (up to a permutation that  in general depends on the choice of the  label $\mu$) as that of the column
$\mu$ of the same matrix.
}

{Equality of the multisets} $\{N_{\lambda\mu}^\nu\}=\{N_{\bar\lambda\mu}^{\nu'}\}$ implies
that our generating function $X(s,t)$ satisfies the following property that we call  $\CP$:

{\bf  $\CP$}: {\sl For any row label $\mu$ and for any monomial $s^{\lambda_1} t^{\lambda_2} $ appearing\footnote{
\ie appearing in the decomposition into monomials of any matrix element (polynomials) of the row $\mu$
}
along the row $\mu$,
the monomial  $s^{\lambda_2} t^{\lambda_1}$ also appears the same {\rm total} number of times along the same row $\mu$}

Can one derive this result directly from properties of the generating functions?\\
The reader  will have no difficulty to prove the following lemmas L1-L5:

\textbf{L1} The matrix elements of the fusion matrices of type $N_{(\p,0)}$ are either $0$ or $1$. \\
\ommit{Proof. This is an immediate consequence of the Pieri formula (see for example \cite{DFMS}  page 695). 
Reminder. The terms appearing in the decomposition of the tensor product of a given irrep (associated with some Young diagram) by an irrep whose associated Young diagram only contains one row, are classically given  by the Pieri formula, with associated multiplicities at most equal to $1$. The fact of working, instead, with the fusion product at level $k$ can only reduce the multiplicity. In particular, multiplicities are still at most equal to 1. A fusion matrix of the kind $N_{(\p,0)}$ is associated with a Young diagram possessing $\p$ columns but only one row. The lemma follows.}
\textbf{L2} The number of  non-zero elements, in the row $\mu$ (arbitrary)  of the fusion matrix $N_{(\p,0)}$, is equal to the number of non-zero elements in the same row $\mu$ of the fusion matrix $N_{(0,\p)}$, or, equivalently, in the column $\mu$ of the same matrix $N_{(\p,0)}$.\\
\ommit{Proof. Because of theorem \ref{CZthm} (actually a very particular case of the latter),  we already know that for all  $\p$ and $\mu$,  $\sum_{\nu} (N_{(\p,0)})_{\mu\nu} = \sum_{\nu} (N_{(0,\p)})_{\mu\nu}$. Since the elements of those two matrices are all $1$'s or $0$'s (previous lemma), this implies the lemma.
  Since $N_{(0,\p)} = N_{(\p,0)}^T$, one can rephrase this property in terms of rows and columns of $N_{(\p,0)}$.}
\textbf{L3} The matrix elements of the fusion matrices of type  $N_{(\p,k-\p)}$ (\,\ie those belonging to the third edge of the Weyl alcove) are either $0$ or $1$. \\
\ommit{Proof. Remember that $P=N_{(k,0)}$ describes the rotation of angle $2 \pi/3$ of the Weyl alcove. As seen above, 
for all $0 \leq \p \leq  k$ we have $P\cdot N_{(\p,k-\p)} = N_{(0,\p)}$.
The result follows from the first lemma and from the fact that $P$ is a permutation matrix.}
\textbf{L4}  For every choice of an irrep $\mu$, the number of non-zero elements
 in the row $\mu$ of the matrix  $N_{(\p,k-\p)}$ is equal to the number of non-zero elements in the 
same row  $\mu$ of the matrix $N_{(k-\p,\p)}$ or equivalently, in the column $\mu$ of the same matrix $N_{(\p,k-\p)}$.
\\
\ommit{Proof. Again, this is a trivial consequence of theorem \ref{CZthm} and from the fact that all the non-zero elements are equal to $1$.
Warning: the non-zero elements of those matrices $N_{(\p,\q)}$ that do not belong to the boundary of the alcove can very well be bigger than $1$. For example for level $k=3$ and the middle point $\lambda=(1,1)$ of the diagram of Fig.\;\ref{Glevel3}, we have $(N_\lambda)_\lambda^\lambda=2$ . }
\textbf{L5}  Every row of the boundary matrix polynomial\footnote{$\Lambda(s,t)$ is defined as the rhs of eq(\ref{sumeqforX}):} $\Lambda(s,t)$ containing some {monomial $s^{\lambda_1} t^{\lambda_2}$} (in a matrix element) 
also contains the {monomial $s^{\lambda_2} t^{\lambda_1}$} the same number of times.
{Equivalently, $\Lambda(s,t)$ satisfies property $\CP$.}\\
 \\
\ommit{{Proof. 
By definition, the row $\mu$ of $s^j N_{(j,0)}$,  for every choice of $\mu$,
 contains as many monomials $s^j$ as there are $1$'s in the row $\mu$ of $N_{(j,0)}$.
From the first two lemmas, with the same choice of $\mu$, the matrix $t^j N_{(0,j)}$ contains an equal number of monomials $t^j$ but their positions (\ie columns) can be different.
For this reason, the row $\mu$ of $s^j N_{(j,0)} + t^j N_{(0,j)}$ contains either monomials $s^j$, or monomials $t^j$ (in equal number), or terms of the type $s^j+t^j$ in those cases where the positions of $1$'s (column indices) happen to be the same for the row $\mu$ of $N_{(j,0)}$ and the row $\mu$ of $N_{(0,j)}$.
\\
We now sum over the $j$ index and consider the edge polynomial matrices  $\Lambda_1(s)=\sum_{j=0}^k s^j N_{(j,0)}$ and $\Lambda_2(t) = \ \sum_{j=0}^k  t^j N_{(0,j)}$. The previous argument shows that an arbitrary row $\mu$ of $\Lambda_1(s)$ contains a number of monomials $s^j$ (for particular values of $j$'s) equal to the number of monomials $t^j$ (with the same values of $j$'s) in the row $\u$ of $\Lambda_2(t)$.  So, every row of the sum $\Lambda_1(s) + \Lambda_2(t)$ containing a number of copies of the monomial $s^j$ (coming from $\Lambda_1(s)$) also contains the same number of monomials $t^j$ (coming from $\Lambda_2(t)$). If it happens that the position of the matrix elements $s^j$ and $t^j$ coincide, we shall get a term $s^j + t^j$ in the sum. In any case, the conclusion is that an arbitrary row of the sum of these two edge matrices containing  one or several copies of some monomial $p(s,t)$ also contains the same number of copies of the monomial $p(t,s)$. 
What we have actually to consider is not the sum but the weighted sum  $(s^3 + s t)  \Lambda_1(s)+ (t^3 + s t)  \Lambda_2(t)$.  The introduction of the two prefactors modifies the matrix elements (one gets polynomials $(s^3 + s t) s^j$ or $(t^3 + s t) t^j$, or their sum), but as the two prefactors are interchanged if we permute $s$ and $t$, the previous conclusions are not modified.
\\
Finally we consider the third side of the alcove, the polynomial matrix $ \Lambda_3(s,t) = \sum_{j=0}^k  s^j  t^{k-j} \, N_{(j,k-j)}$. 
Lemma \ref{lemma4} shows that an arbitrary row of this matrix contains, for some values of $j$ (corresponding to the position of $1$'s in the $N_{(j,k-j)}$), an equal number of monomials $s^j t^{(k-j)}$ and of monomials $s^{(k-j)} t^j$, moreover,  for the particular value $j=k$, it contains a term $s^k+t^k$, which is symmetric in $s$ and $t$.  The symmetry property described in the lemma therefore holds separately for $\Lambda_3(s,t)$. Multiplying this matrix by a factor $2 s^2  t^2 $, which is symmetric in $s$ and $t$, does not modify the argument.
Since the lemma holds for the weighted sum of $\Lambda_1(s)$ and $\Lambda_2(t)$, and separately for $\Lambda_3(s,t)$ (with its symmetric prefactor), it also holds for the boundary polynomial matrix $\Lambda(s, t) = (s^3 + s t)  \Lambda_1(s)+ (t^3 + s t)  \Lambda_2(t) + 2 s^2  t^2  \Lambda_3(s,t)$. This completes the proof of Lemma \ref{lemma5}.
 Notice that the matrix elements 
of $\Lambda(s, t)$ are not usually symmetric in $s$ and $t$. 
{\red so far, so good}
At this stage, we observe that the previous lemma may be rephrased as: \\
{\sl The polynomial $\Lambda(s,t)$ satisfies the 
property $\CP$}.}} 
It is also clear that the polynomial $A(s,t) = [(s^3 + t^3 + 2 s t + 2 s^2 t^2) \one  -  s t (s G + t G^T)]$
satisfies the same property $\CP$. If we could assert that this property 
is preserved by matrix multiplication and inversion, we would conclude that
$K(s,t)=A^{-1}(s,t)$  and $X(s,t)=K(s,t).\Lambda(s,t)$ also satisfy the same property, 
thus establishing property $\CP$, which is equivalent to $\mathfrak P$. 
{Unfortunately, this is not the case in general and we could not prove property $\mathfrak P$ in this way. It is nevertheless interesting to see how this property is translated when read in terms of generating functions: this justifies the inclusion of the above discussion in the present paper.}


\subsection{An application of O-blades: a new relation between fusion coefficients}
\label{applOblade}
\def\a{\alpha}
\def\oblade{O-blade}
\def\pictograph{pictograph} 

\begin{figure}[ht]
\centering{\includegraphics[width=14pc]{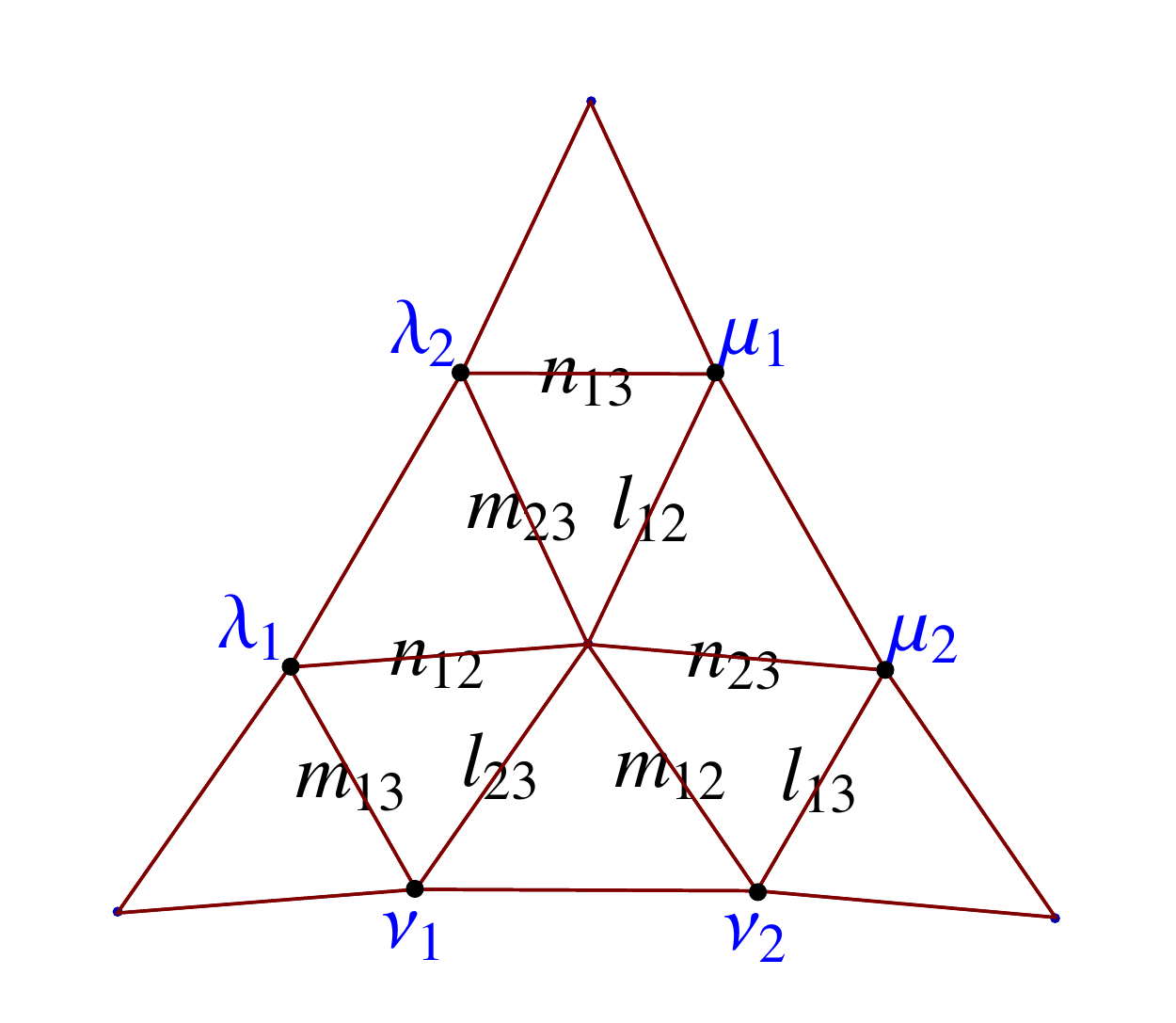} \hskip10mm\includegraphics[width=14pc]{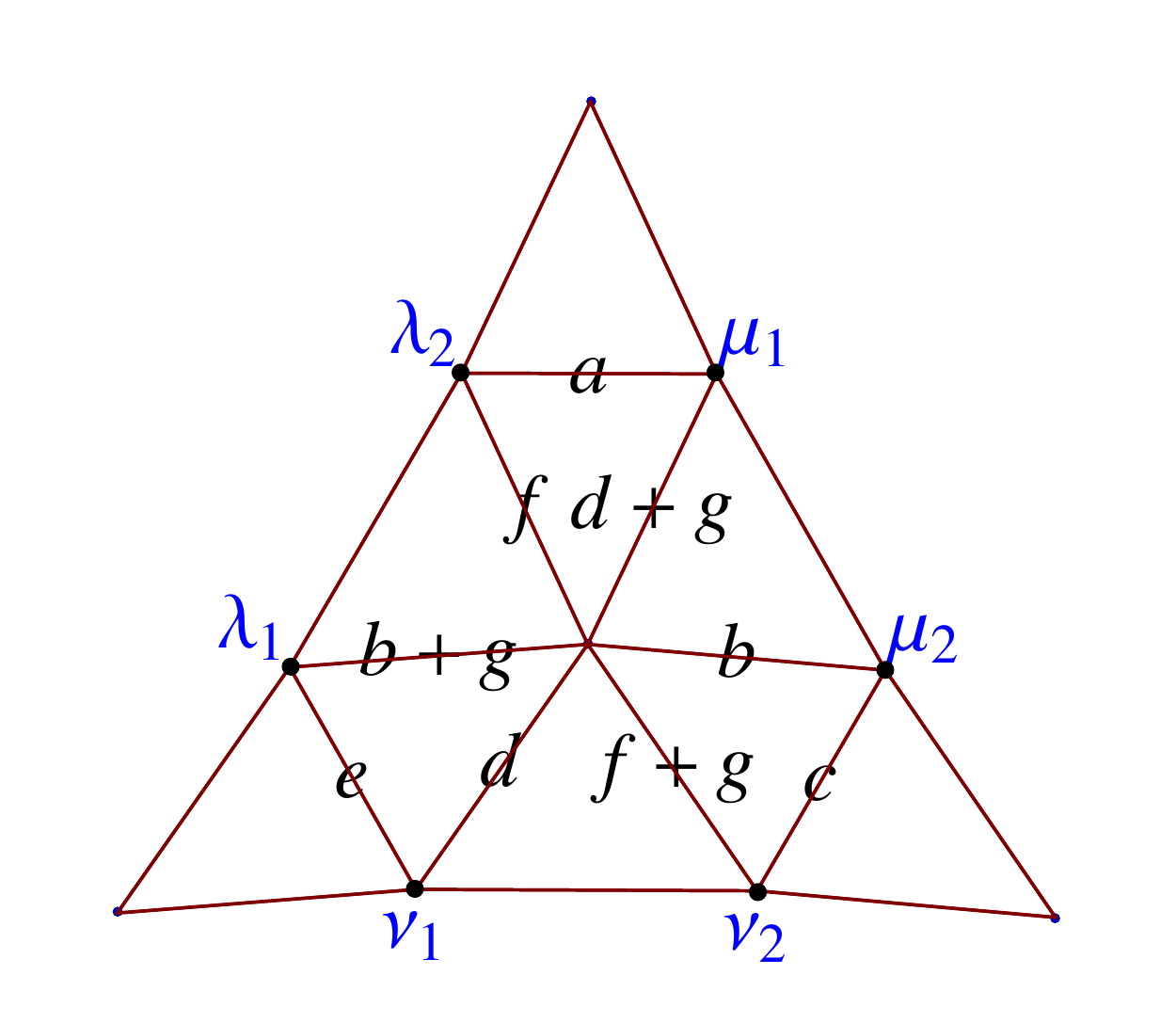}}\\
\caption{\label{genericObladeSU3}(a): A generic \oblade\  for SU(3) and its nine internal edges.\\ (b): Its seven components $(a,b,c,d,e,f;g)$ along the left fundamental basis}
\end{figure}

\begin{figure}[tbp]
\centering{\includegraphics[width=45pc]{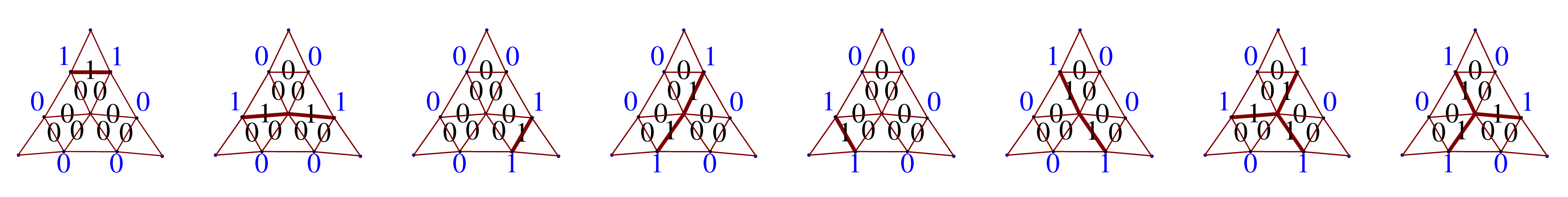}}
\caption{\label{fundamentalIntertwinersSU3} Fundamental \oblade s for SU(3)}
\end{figure}

\begin{figure}[tbp]
\centering{\includegraphics[width=35pc]{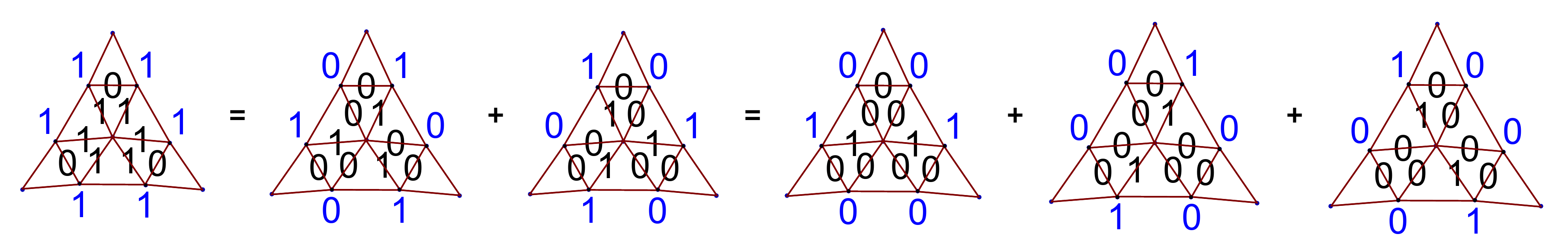}}
\centering{\includegraphics[width=35pc]{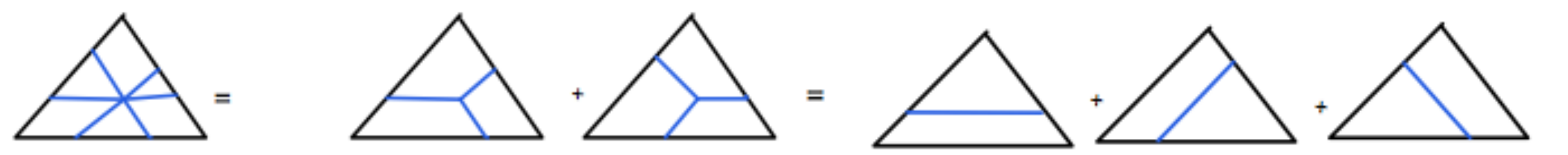}}
\caption{\label{su3RelationOblades} The SU(3) intertwiner relation in terms of \oblade s }
\end{figure}

{\bf Proposition.}{\ }
The involutive transformation defined by  eq.\;(\ref{forktransfo}) is involutive, and, for arbitrary levels $k$, one has: 
 $$ N^{(k)\;\nu}_{\lambda\,\mu}= N^{(k)\;\nu'}_{\lambda'\,\mu'}$$
A generic \oblade\  for SU(3) is displayed in Fig.\;\ref{genericObladeSU3}(a).\\
It  contains nine edges
$m_{12}, m_{23}, m_{13}, n_{12}, n_{23}, n_{13}, l_{12}, l_{23}, l_{13}$,
 labelled with non-negative integers.  \\
 It has a single inner vertex, and  obeys one single constraint:
the three pairs of opposite ``angles" defined by the lines intersecting at the inner vertex should be equal, the value of an angle being defined as the sum of its sides.
The three weights  appearing in the branching rule $\lambda, \mu \rightarrow {\nu}$ follow each other as indicated in Fig. ~\ref{genericObladeSU3}.
Their components, in terms of edges,  read (``integer conservation at the external vertices''):
\begin{eqnarray*}
\lambda_1 &=& m_{13} + n_{12}  \qquad  \mu_1 = n_{13} + l_{12}  \qquad   \nu_1 = m_{13} + l_{23}\\
\lambda_2 &=& m_{23} + n_{13}  \qquad  \mu_2 = n_{23} + l_{13}   \qquad  \nu_2 = m_{12} + l_{13} \,.
\end{eqnarray*}
The reader will have no difficulty to re-express the ``Wessl\'en inequalities'' (\ref{W1}-\ref{W3}) as positivity constraints for the nine edges (``blades'') of the \oblade s.
As in \cite{RCJBZ2014} (see this reference for a more complete discussion), 
we call {\sl fundamental \pictograph s} (here fundamental \oblade s) the combinatorial models associated with intertwiners  of the type $f_1 \otimes f_2 \otimes f_3 \rightarrow \one$ where  the $f_i$'s denote either a fundamental representation or the trivial one. 
The dimension of the corresponding spaces of intertwiners being equal to $0$ or $1$, there is only one such {\pictograph} or none.
For $\SU(3)$, we have six fundamental \oblade s obtained by permuting the factors of $f \otimes f^* \otimes \one$ where $f$ is the three-dimensional vector representation (of highest weight $(1,0)$), and we have two other fundamental \oblade s respectively associated with the cubes of $(1,0)$ and of $(0,1)$; the last two (resp. the first six) are called ``primitive fundamental'' (resp. ``non-primitive fundamental '') in \cite{RCJBZ2014}.
 The eight fundamental \oblade s are given in Fig.\ref{fundamentalIntertwinersSU3} {-- edges carrying a ``$1$'' label have been thickened}. 
Obviously, \oblade s, thought of as characteristic functions associated with their set of edges, can be added or multiplied by scalars (they may be called virtual whenever some inner edges carry values that do not belong to the set of 
non negative integers. 
One could think that this family of eight fundamental \pictograph s can be chosen as a basis in the space of characteristic functions, but this is not so because they are not independent: there is also one relation which is displayed in Fig.\;\ref{su3RelationOblades}.
The two primitive fundamental \oblade s appear as Y-shapes (forks) with a horizontal tail to the left or to the right of the  inner vertex;  the relation displayed  in Fig.\;\ref{su3RelationOblades} exhibits two ways of making a star out of its components.
Notice  that the space of {\pictograph s} comes with a distinguished generating family -- {those that are fundamental} --  but not with a distinguished basis.
Up to permutations, we may consider two interesting particular basis, each  contains seven fundamental intertwiners: the six non-primitive ones, and a last one chosen among the two that are primitive.

\begin{figure}[tbp]
 \centering{\includegraphics[width=42pc]{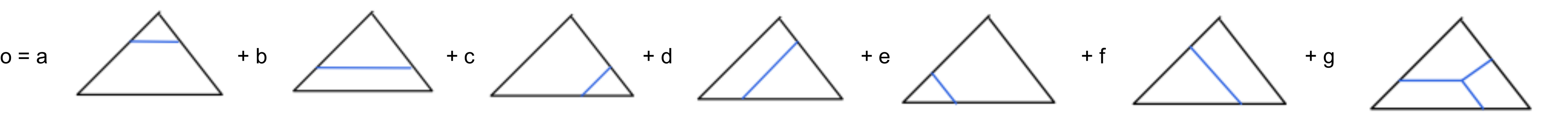}}
\centering{\includegraphics[width=42pc]{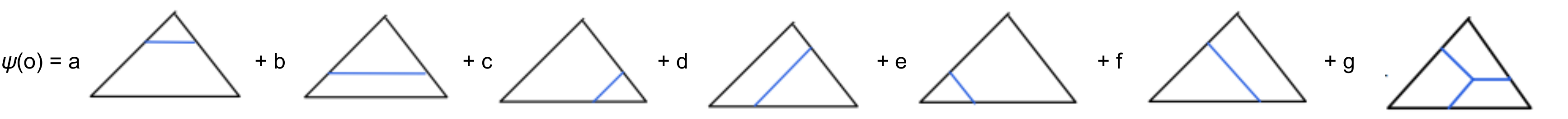}}
\centering{\includegraphics[width=42pc]{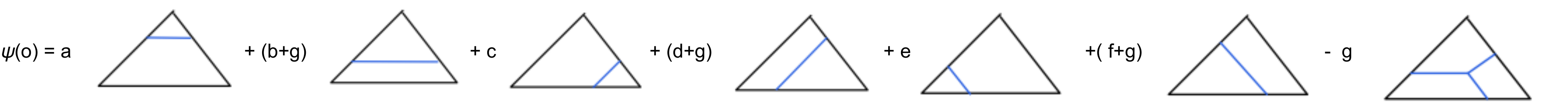}}
\caption{\label{arbitraryobladeSU3decomp} (a) Decomposition of an \oblade\ along the left fundamental basis, (b): its image par $\Psi$, and (c): its decomposition along the same basis using the SU(3) intertwiner relation.}
\end{figure}

Chosing  the ``left'' basis (\ie the last basis element being Y-shaped,  with a horizontal tail to the left of the inner vertex), 
we see in Fig.\;\ref{arbitraryobladeSU3decomp}(a) or in Fig.\;\ref{genericObladeSU3}(b) how an arbitrary \oblade\  $o$ can be defined as a superposition of its seven components and write\footnote{One can easily show that the component $g$ coincides with the parameter $x=(S_1-S_2)/3$ of section \ref{formulaeformultiplicities}.}  $o_L = (a,b,c,d,e,f,g)$. The associated weights are $\lambda_1=b+e+g, \lambda_2 = a+f, \mu_1 = a+d+g, \mu_2 = b+c, \nu_1=d+e, \nu_2=c+f+g$.

Call $\Psi$ the linear map that leaves invariant the six non-primitive O-blades, and  permutes the two that are Y-shaped. 
Under this map, an arbitrary element $o$ (see Fig.\;\ref{arbitraryobladeSU3decomp}(a)) becomes $\Psi(o)$ (see Fig.\;\ref{arbitraryobladeSU3decomp}(b)), which, using the relation between fundamental \oblade s displayed in Fig.\;\ref{su3RelationOblades}, can be expanded on the left basis as in Fig.\;\ref{arbitraryobladeSU3decomp}(c).
We have therefore\footnote{One should notice that positivity of edges of a (non virtual) \oblade\   implies positivity of its components along thes first six intertwiners, those that we called  non-primitive fundamental, but not along the last one (the left fork).} $\Psi(o)_L = (a,b+g,c,d+g,e,f+g,-g)$, with associated weights $\lambda^\prime, \mu^\prime, \nu^\prime$ and $\lambda_1^\prime = e+b, \lambda_2^\prime = a+f+g, \mu_1=a+d, \mu_2=b+g+c, \nu_1=f+g+e, \nu_2=f+c$. 
The linear map $\Psi$, defined on the space of couplings, induces a transformation $(\lambda, \mu; \nu) \rightarrow (\lambda^\prime, \mu^\prime; \nu^\prime)  $ on branchings, that we also call $\Psi$, with
\\
\begin{equation}
\label{forktransfo}
\begin{aligned}[c]
\lambda_1^\prime =& \frac{1}{3}(2 \lambda_1 + \lambda_2 - \mu_1 + \mu_2 + \nu_1 - \nu_2)\\
\mu_1^\prime =& \frac{1}{3}(- \lambda_1 + \lambda_2 +2 \mu_1 + \mu_2 + \nu_1 - \nu_2)\\
\nu_1^\prime =& \frac{1}{3}( \lambda_1 - \lambda_2 + \mu_1 - \mu_2 +2 \nu_1 + \nu_2)
\end{aligned}
\qquad
\begin{aligned}[c]
\lambda_2^\prime =& \frac{1}{3}(\lambda_1 + 2 \lambda_2 + \mu_1 - \mu_2 - \nu_1 + \nu_2)\\
\mu_2^\prime =& \frac{1}{3}( \lambda_1 - \lambda_2 + \mu_1 +2 \mu_2 - \nu_1 + \nu_2)\\
\nu_2^\prime =& \frac{1}{3}(- \lambda_1 + \lambda_2 - \mu_1 + \mu_2 + \nu_1 +2 \nu_2)
\end{aligned}
\end{equation}
%
\begin{figure}[tbp]
\centering{\includegraphics[width=14pc]{arbitraryobladeSU3.pdf} \hskip10mm \includegraphics[width=14pc]{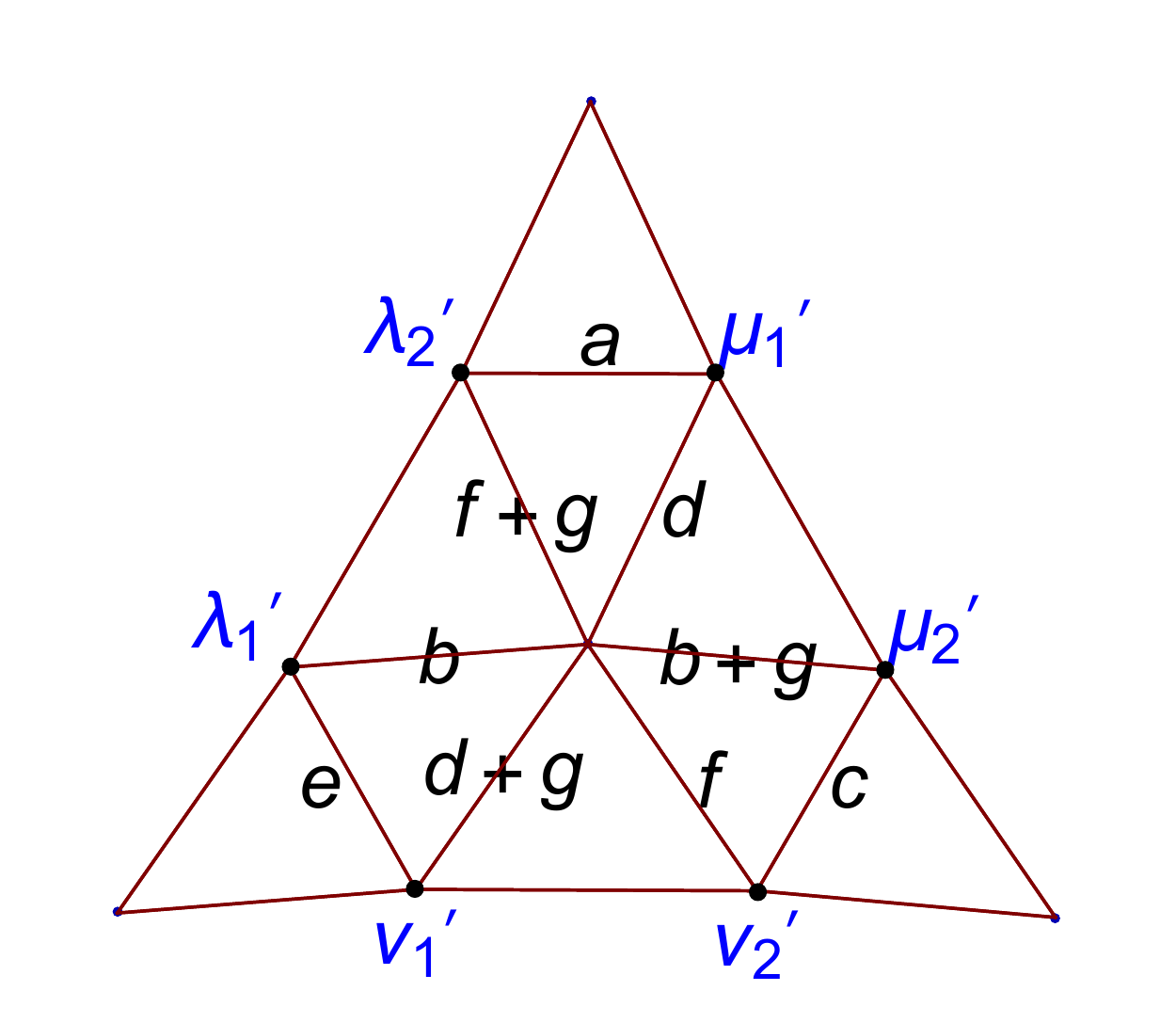}}
\caption{\label{permutedforks} Permuting forks: an  \oblade\ $o$ and its image $\Psi(o)$.}
\end{figure}
The nine internal edges and the three highest weights associated with $\Psi(o)$ are displayed in Fig.\;\ref{permutedforks}(b).
From its geometrical definition (permutation of left and right forks), it is already clear that $\Psi$ is an involution, but this can also be checked from the equations given previously. In particular $\Psi$ is injective, both for branchings and for \oblade s (or couplings).
This implies that if a given branching has classical multiplicity $\mathfrak m$, and can therefore be associated with $\mathfrak m$ distinct \oblade s describing its corresponding couplings, its image by $\Psi$ is automatically admissible and will have the same multiplicity, the \oblade s describing the couplings of the latter being obtained as the image by $\Psi$ of the \oblade s of the former.

We now show that an \oblade\ $o$ (characterizing a coupling), and its image $\Psi (o)$ have the same threshold.
In terms of components of highest weights and internal edges, we know (see Fig.\;\ref{genericObladeSU3}(a)) that $k_0(o) = \max(\lambda_1+\lambda_2+l_{13}, \mu_1+\mu_2+m_{13},\nu_1+\nu_2+n_{13})$, or, equivalently, in terms of  the (left)  fundamental components $o_L=(a,b,c,d,e,f,g)$ of $o$, that $k_0(o)=\max(a+b+c+e+f+g, a+c+d+e+f+g,a+b+c+d+e+g)$. Using $\Psi(o)_L = (a,b+g,c,d+g,e,f+g,-g)$, one finds immediately $k(\Psi(o))=k_0(o)$. In particular this also implies that branchings exchanged by $\Psi$ have the same minimal threshold, so that this transformation also holds in the affine case and that it is compatible with the levels.

Example. $\Psi((9,5), (6,2); (10,5))=((8, 6), (5, 3); (11, 4))$. In both cases, multiplicity and threshold are the same: $mult((9,5), (6,2); (10,5))=mult((8,6), (5,3); (11,4))=3$ and $k_0^{min} = 16$.

Remark. It is easy to see that the map $\Psi$ is compatible with the transposition of the first two arguments and with the six Frobenius transformations.

\section*{Acknowledgments} 
R.C. thanks hospitality of the Instituto Nacional de Matem\'atica Pura e Aplicada (IMPA, Rio de Janeiro), where part of this work was done.
\\
J.-B. Z. gratefully acknowledges support from the Simons Center for Geometry and
Physics, Stony Brook University, at which part of the research of this paper was completed. He thanks Pavel Bleher, Vladimir
Korepin and Bernard Nienhuis, the organizors of
the workshop ``Statistical Physics and Combinatorics", for their invitation.


\newpage


\begin{thebibliography}{99}

\bibitem{RCJBZ2014} R.Coquereaux and J.-B. Zuber, \emph{Conjugation properties of tensor product multiplicities}, J. Phys. {\bf A 47} (2014) 455202; \url{http://arxiv.org/abs/1405.4887}

\bibitem{BCM-tensorpro} L. B\'egin, C. Cummins and P. Mathieu,
\emph{Generating functions for tensor products}, 
J. Math. Phys. {\bf 41} (11) (2000) 7611-7639;
\url{http://arxiv.org/abs/hep-th/9811113}

\bibitem{BCM-fusionpro} L. B\'egin, C. Cummins and P. Mathieu, \emph{Generating-function method for fusion rules}, 
J. Math. Phys. {\bf 41} (11) (2000) 7640-7674; 
\url{http://arxiv.org/abs/math-ph/0005002}

\bibitem{DFMS} P. Di Francesco, P. Mathieu and D. S\'en\'echal, \emph{Conformal Field Theory} (Springer, 1997)

\bibitem{Dyn} E.B. Dynkin, \emph{ Semisimple subalgebras of semisimple Lie algebras} Mat. Sb. (N.S.), {\bf 30} (72) (1952) 349-462

\bibitem{Ocneanu:paths} A. Ocneanu, \emph{ Paths on Coxeter diagrams: from 
Platonic solids and singularities to minimal models and subfactors.}  Notes taken by S. Goto,
 Fields Institute Monographs, AMS 1999, Rajarama Bhat et al, eds.
 
  \bibitem{CIZ}  A. Cappelli, C. Itzykson and J.-B. Zuber, \emph{The ADE classification of minimal and $A_{1}^{(1)}$ conformal invariant theories.} Commun. Math. Phys. {\bf 13} (1987) 1-26
 
 \bibitem{Ostrik}V. Ostrik, \emph{Module categories, weak Hopf algebras and modular invariants}, Transform. groups, {\bf 8} (2)  (2003) 177-206;  \url{http://arxiv.org/abs/math.QA/0111139}
 
 \bibitem{PZ} V. Petkova and J.-B. Zuber, \emph{The many faces of Ocneanu cells}, Nucl. Phys. {\bf B 603} (2001) 449-496;
\url{http://arxiv.org/abs/hep-th/0101151}

\bibitem{MOV} A. Malkin, V. Ostrik and M. Vybornov, \emph{Quiver varieties and Lusztig's algebra}, 
Advances in Maths {\bf 203} (2) (2006) 514-536; \url{http://arxiv.org/abs/math.RT/0403222}

\bibitem{CoquereauxSchieberRio} R. Coquereaux and G. Schieber, \emph{Quantum Symmetries of sl (2) and sl (3) graphs}, Proceedings of the Fifth International Conference on Mathematical Methods in Physics (Rio de Janeiro), Proceedings of Science : PoS (IC2006) 001;
\url{http://pos.sissa.it/archive/conferences/031/001/IC2006_001.pdf}

\bibitem{CoquereauxSchieberJMP} R. Coquereaux and G. Schieber, \emph{Orders and dimensions for sl2 or sl3 module-categories and boundary conformal field theories on a torus},  J. of Mathematical Physics {\bf 48} (2007) 043511; 
\url{http://arxiv.org/abs/math-ph/0610073}

\bibitem{RCJBZ2011} R. Coquereaux and J.-B. Zuber, \emph{On sums of tensor and fusion multiplicities}, J. Phys. {\bf A  44} (2011) 295208;  \url{http://arxiv.org/abs/1103.2943}

\bibitem{KacPeterson}   V. Kac  and   D. Peterson, 
\emph{Infinite dimensional Lie algebras,  theta functions, and modular forms},   Adv.  Math.  {\bf 53} (1984), 125-264

\bibitem{ZeierZimboras} R.Zeier and Z.Zimbor\'as, \emph{ On squares of representations of compact Lie algebras}, J. Math. Phys. {\bf 56} (2015) 081702;  \url{http://arxiv.org/abs/1504.07732}

\bibitem{BMW} L. B\'egin, P. Mathieu and M.A. Walton, \emph{$\widehat{su}(3)_k$ fusion coefficients}, 
\url{http://arxiv.org/abs/hepth/9206032}

\bibitem{KMSW} A.N. Kirillov, P. Mathieu, D. S\'en\'echal and M. Walton, \emph{Can fusion coefficients be calculated from the depth rule?} LAVAL-PHY-20/92, LETH-PHY-2/92; \url{http://arxiv.org/abs/hep-th/9203004}

\bibitem{LK}S. Lu, PhD thesis, MIT (1990). A.N. Kirillov, unpublished

\bibitem{Mandeltsveig} V.B. Mandel'tsveig, \emph{ Irreducible representations of the $SU_3$ group}, Soviet Physics JETP 
{\bf 20} (1965) 1237-1243

\bibitem{Suciu} L. Suciu. \textit{The SU(3) wire model}, PhD thesis, The Pennsylvania State University, 1997, AAT 9802757.
A. Ocneanu, unpublished

\bibitem{Wesslen} M. Wessl\'en, \emph{A geometric description of tensor product decompositions in $\mathfrak{su}$(3)},  
J. Math. Phys. {\bf 49} (2008) 073506 

\end{thebibliography}
\end{document}